\documentclass[11pt]{article}
\pdfoutput=1

\usepackage[normalem]{ulem}

\usepackage{epsfig}
\usepackage{psfrag}
\usepackage{latexsym}
\usepackage{indentfirst}
\usepackage{fancyhdr}
\usepackage{dsfont}
\usepackage{amsmath}
\usepackage{amssymb}
\usepackage{amsfonts}
\usepackage{mathrsfs}
\usepackage{amsthm}
\usepackage{pifont}
\usepackage{dsfont}
\usepackage{multirow}
\usepackage{array}
\usepackage{chngpage}
\usepackage{longtable}
\usepackage{cite}
\usepackage{bbold}
\usepackage{color}
\usepackage{braket}
\usepackage{colordvi}
\usepackage{fancybox}
\usepackage[footnotesize]{caption2}
\usepackage{graphicx}
\usepackage[center,footnotesize,hang]{subfigure}
\usepackage{bbm}
\usepackage[colorlinks, linkcolor=red, anchorcolor=black, citecolor=green]{hyperref}
\usepackage{multirow}
\usepackage{array}
\usepackage{textcomp}  
\usepackage{enumitem}  
\usepackage{mathrsfs}  
\newcommand{\PreserveBackslash}[1]{\let\temp=\\#1\let\\=\temp}
\newcolumntype{C}[1]{>{\PreserveBackslash\centering}p{#1}}
\newcolumntype{R}[1]{>{\PreserveBackslash\raggedleft}p{#1}}
\newcolumntype{L}[1]{>{\PreserveBackslash\raggedright}p{#1}}
\allowdisplaybreaks

\newcommand{\bq}{\begin{eqnarray}}
\newcommand{\nq}{\end{eqnarray}}

\begin{document}
\title{
\begin{flushright}
\hfill\mbox{\small USTC-ICTS-19-33} \\[5mm]
\begin{minipage}{0.2\linewidth}
\normalsize
\end{minipage}
\end{flushright}
{\Large \bf
Modular symmetry origin of texture zeros and quark lepton unification
\\[2mm]}}
\date{}

\author{Jun-Nan Lu$^{a,b}$\footnote{E-mail: {\tt
hitman@mail.ustc.edu.cn}},  \
Xiang-Gan Liu$^{a}$\footnote{E-mail: {\tt
hepliuxg@mail.ustc.edu.cn}},\
Gui-Jun Ding$^{a}$\footnote{E-mail: {\tt
dinggj@ustc.edu.cn}} \
\\*[20pt]
\centerline{
\begin{minipage}{\linewidth}
\begin{center}
$^a${\it \small
Interdisciplinary Center for Theoretical Study and  Department of Modern Physics,\\
University of Science and Technology of China, Hefei, Anhui 230026, China}\\[2mm]
$^b${\it \small
AHEP Group, Institut de F\'{i}sica Corpuscular --
  CSIC/Universitat de Val\`{e}ncia, Parc Cient\'ific de Paterna.\\
  C/ Catedr\'atico Jos\'e Beltr\'an, 2 E-46980 Paterna (Valencia) - SPAIN}\\
\end{center}
\end{minipage}}
\\[10mm]}
\maketitle
\thispagestyle{empty}

\begin{abstract}

The even weight modular forms of level $N$ can be arranged into the common irreducible representations of the inhomogeneous finite modular group $\Gamma_N$ and the homogeneous finite modular group $\Gamma'_N$ which is the double covering of $\Gamma_N$, and the odd weight modular forms of level $N$ transform in the new representations of $\Gamma'_N$. We find that the above structure of modular forms can naturally generate texture zeros of the fermion mass matrices if we properly assign the representations and weights of the matter fields under the modular group.
We perform a comprehensive analysis for the $\Gamma'_3\cong T'$ modular symmetry. The three generations of left-handed quarks are assumed to transform as a doublet and a singlet of $T'$, we find six possible texture zeros structures of quark mass matrix up to row and column permutations. We present five benchmark quark models which can produce very good fit to the experimental data. These quark models are further extended to include lepton sector, the resulting models can give a unified description of both quark and lepton masses and flavor mixing simultaneously although they contain less number of free parameters than the observables.

\end{abstract}
\newpage

\section{Introduction}

The standard model (SM) of particle physics has been precisely tested so far. However, neither significant evidence for the departures from the SM nor convincing hints for the presence of new physics has been found. The masses of quarks and charged leptons are parameterized by the Yukawa coupling constants in SM. The Yukawa sector is still poorly understood, and the SM itself can not predict the exact values of quark masses and CKM mixing matrix. The fundamental principle which rules the hierarchical charged fermion mass spectra, tiny neutrino masses, flavor mixing and CP violation, is still elusive. Neutrino oscillation experiments have made enormous progress in past years. The three lepton mixing angles and neutrino mass squared differences have been precisely measured. The latest global fit of neutrino oscillation data gives the best fit values and $1\sigma$ errors of the lepton mixing angels as $\theta_{12}/^\circ=33.82^{+0.78}_{-0.76}$, $\theta_{23}/^{\circ}=49.6^{+1.0}_{-1.2}$ and $\theta_{13}/^{\circ}=8.61\pm0.13$~\cite{Esteban:2018azc}. The present neutrino oscillation data favor normal ordering (NO) neutrino mass spectrum over inverted ordering (IO), CP conservation in neutrino oscillation is disfavored at $2\sigma$ confidence level, and the leptonic Dirac CP phase $\delta_{CP}$ around $3\pi/2$ is preferred~\cite{Abe:2017vif,Abe:2018wpn}.

Neutrino oscillation provides us new insight to understanding the flavor puzzle. It is found that the neutrino mixing angles can be reproduced by extending the SM with a finite discrete non-abelian flavor symmetry~\cite{Altarelli:2010gt,Ishimori:2010au,King:2013eh,King:2014nza,King:2015aea,King:2017guk,Feruglio:2019ktm}.
The observed neutrino mixing pattern arises as the result of particular
vacuum alignment of scalar fields called flavons which spontaneously break certain discrete flavor symmetry. The flavons are SM singlet and they transform nontrivially under the flavor symmetry group. Usually a number of flavons are necessary, and the scalar potential as well as additional shaping symmetries has to be cleverly designed to obtain the desired vacuum alignment. Another drawback of this approach is that the predictability of a discrete flavor symmetry model could be degraded by the possible higher dimensional operators. Moreover, flavor symmetry is usually used to constrain the neutrino mixing angles while the neutrino masses are undetermined except in some specific models.

In order to overcome the above mentioned drawbacks of the conventional finite discrete flavor symmetry, a new approach of modular invariance playing the role of flavor symmetry was recently proposed in~\cite{Feruglio:2017spp}. Modular invariance has a long history both in string and in field theories. The idea of modular invariance as flavor symmetry to constrain the Yukawa couplings has been naturally realized in string theory~\cite{Dixon:1986qv,Hamidi:1986vh,Lauer:1989ax}, D-brane compactification~\cite{Cremades:2003qj,Blumenhagen:2005mu}, magnetized extra dimensions~\cite{Cremades:2004wa,Abe:2009vi} and in orbifold compactification~\cite{Ibanez:1986ka,Lebedev:2001qg}. Modular invariance has been used to address several aspects of the flavour problem in model building~\cite{Brax:1994kv,Binetruy:1995nt,Dudas:1995eq}. In the present paper, we shall follow Ref.~\cite{Feruglio:2017spp} to take a bottom-up approach based on supersymmetric modular invariant theories. In the most economical version of the modular invariant models, the flavon fields other than the modulus are not needed and the flavor symmetry is uniquely broken by the vacuum expectation value (VEV) of the complex modulus $\tau$. The complicated vacuum alignment is not required although one needs some mechanism to fix the value of $\tau$. The Yukawa couplings are functions of modular forms which are holomorphic functions of $\tau$, and all higher dimensional operators in the superpotential are completely determined by modular invariance in the limit of supersymmetry~\cite{Feruglio:2017spp}. It is notable that the neutrino mass models based on modular invariance could involve only few coupling constants such that neutrino masses and mixing parameters are correlated. This formalism has been extended to consistently combine with the generalized CP symmetry, and the consistency conditions requires that the modulus transforms as $\tau\rightarrow-\tau^{*}$ up to modular transformations under the action of CP symemtry~\cite{Novichkov:2019sqv,Baur:2019kwi,Acharya:1995ag,Dent:2001cc,Giedt:2002ns,Baur:2019iai}. In a symmetric basis where the representation matrices of both $S$ and $T$ are symmetric, the multiplets of modular forms become complex conjugated under CP transformation if they are properly normalized. As a consequence, the generalized CP symmetry would constrain all the couplings in a modular invariant model to be real~~\cite{Novichkov:2019sqv}, and thus the predictive power of such models is enhanced. It was noted that the K$\ddot{\mathrm{a}}$hler potential is not at all fixed by the symmetries and transformation properties of the models, although the superpotential is completely fixed by the modular transformations. The corrections from the most general K$\ddot{\mathrm{a}}$hler potential consistent with the symmetries of the model could potentially reduce the predictive power of this formalism~\cite{Chen:2019ewa}.

The crucial element of this new approach is the modular forms of even weights and level $N$ which can be arranged into irreducible representations of inhomogeneous finite modular group $\Gamma_N$. Several models of lepton masses and mixing have been constructed based on the finite modular groups $\Gamma_2\cong S_3$~\cite{Kobayashi:2018vbk,Kobayashi:2018wkl,Kobayashi:2019rzp,Okada:2019xqk}, $\Gamma_3\cong A_4$~\cite{Feruglio:2017spp,Criado:2018thu,Kobayashi:2018vbk,Kobayashi:2018scp,deAnda:2018ecu,
Okada:2018yrn,Kobayashi:2018wkl,Novichkov:2018yse,Nomura:2019jxj,Okada:2019uoy,Nomura:2019yft,
Ding:2019zxk,Okada:2019mjf,Nomura:2019lnr,Kobayashi:2019xvz,Asaka:2019vev,Gui-JunDing:2019wap,
Zhang:2019ngf,Nomura:2019xsb,Kobayashi:2019gtp}, $\Gamma_4\cong S_4$~\cite{Penedo:2018nmg,Novichkov:2018ovf,deMedeirosVarzielas:2019cyj,Kobayashi:2019mna,King:2019vhv,Criado:2019tzk,Wang:2019ovr,Gui-JunDing:2019wap} and $\Gamma_5\cong A_5$~\cite{Novichkov:2018nkm,Ding:2019xna,Criado:2019tzk}. This new approach has been extended to modular forms of general integer weights which can be arranged into irreducible representations of the homogeneous finite modular group $\Gamma'_N$~\cite{Liu:2019khw}. Notice that $\Gamma'_N$ is the double covering of $\Gamma_N$. The modular forms of weight 1 and level 3 have been explicitly constructed and shown to furnish a two dimensional irreducible representation of $\Gamma'_3\cong T'$~\cite{Liu:2019khw}. The phenomenological predictions of modular symmetry models for leptogenesis have been discussed in~\cite{Asaka:2019vev,Wang:2019ovr}. The SU(5) grand unified models with modular symmetry have been constructed~\cite{deAnda:2018ecu,Kobayashi:2019rzp}. The modular symmetry has also been applied to dark matter and radiative neutrino mass models~\cite{Nomura:2019jxj,Okada:2019mjf,Nomura:2019lnr}.

An interesting attempt to understand the dynamics of fermion mass generation and flavor mixing is to assume some entries of the fermion mass matrices are vanishing~\cite{Fritzsch:1977za,Weinberg:1977hb,Wilczek:1977uh}, and such scenarios are more popularly known as texture-zero models. Usually abelian flavor symmetry is used to realize texture zero structure exactly or approximately~\cite{Grimus:2004hf,Felipe:2014vka}. Systematical and complete studies of all possibilities have been performed for both the lepton sector~\cite{Ludl:2014axa} and quark sector~\cite{Ludl:2015lta}. For recent review on texture zeros we refer the reader to~\cite{Fritzsch:1999ee,Gupta:2013yha}. The homogeneous finite modular group $\Gamma'_N$ is the double covering of the inhomogeneous finite modular group $\Gamma_N$, and $\Gamma'_N$ has twice as many elements as $\Gamma_N$.
Besides the irreducible representations of $\Gamma_N$, $\Gamma'_N$ has other new representations~\cite{Liu:2019khw}. The even weight modular forms of level $N$ can be arranged into the irreducible representations of $\Gamma_N$ up to the automorphy factor~\cite{Feruglio:2017spp}, while the odd weight modular forms of level $N$ are arranged into the new representations of $\Gamma'_N$~\cite{Liu:2019khw}. In this work, we shall show that the above structure of modular forms can naturally produce texture zeros of the fermion mass matrices if we properly assign the representations and weights of the matter fields under the modular group. In this sense, the modular invariance approach has the merits of both abelian flavor symmetry and discrete non-abelian flavor symmetry.

In the present paper, we shall use the modular forms of level 3 to show concrete examples. In addition to the representations $\mathbf{1}$, $\mathbf{1}'$, $\mathbf{1}''$ and $\mathbf{3}$ of $\Gamma_3\cong A_4$,
$\Gamma'_3\cong T'$ has three doublet representations $\mathbf{2}$, $\mathbf{2}'$ and $\mathbf{2}''$. The even weight modular forms of level 3 transform as singlets and triplets of $\Gamma'_3$ while the odd weight  modular forms of level 3 arrange into $\Gamma'_3$ doublets. We will perform a comprehensive analysis of possible texture zeros of the quark mass matrices with the introduction of the $T'$ modular symmetry. Since the mass of the third generation quark is much heavier than the first two generations, we shall assume that the three generations of the left-handed quark fields transform as a doublet and a singlet under $T'$ modular symmetry, while the right-handed quark fields can be assigned to a direct sum of $T'$ doublet and singlet or they could be three singlets of $T'$.

This paper is structured as follows: in section~\ref{sec:ModSym_rev} we briefly review the basic aspects of modular symmetry and give the expressions of the modular forms of level 3. In section~\ref{sec:QMM_ds}, we present the possible structures and texture zeros of the quark mass matrices if both left-handed and right-handed quark fields are assigned to direct sum of singlet and double representations of $T'$. The texture zero structures of the quark mass matrices for singlet assignments of right-handed quark fields are given in section~\ref{sec:QMM_s}. Furthermore, in section~\ref{sec:model_quark} we give five benchmark quark models which
leads to up and down quark mass matrices with texture zeros. These models contain only ten or eleven independent real parameters, and they produce excellent fit to the data of quark masses and CKM mixing matrix. In section~\ref{sec:QLU} we include the lepton sector to give a unified description of both quark and lepton masses and flavor mixing simultaneously. The non-minimal K\"ahler potential and its contribution in the quark sector are discussed in section~\ref{sec:non-minimal-Kahler}, and we give most general form of K\"ahler potential consistent with modular symmetry. Finally we conclude and draw our conclusions in section~\ref{sec:conclusion}. We give the representation matrices of the generators of $T'$ group and the Clebsch-Gordan (CG) coefficients in our basis in the Appendix~\ref{app:Tp_group}. The modular forms of higher weights $k=5, 6, 7, 8$ and level 3 are collected in the Appendix~\ref{app:High_MDF}. The non-minimal K\"ahler potential in lepton sector is discussed in the Appendix~\ref{app:non-minimal-Kahler-lepton}.

\section{\label{sec:ModSym_rev}Modular symmetry and modular forms of level $N=3$}

The modular group $\overline{\Gamma}$ can be regarded as the group of linear fraction transformations acting on the complex modulus $\tau$ with $\mathrm{Im}\tau>0$,
\begin{equation}
\tau\rightarrow\gamma\tau=\frac{a\tau+b}{c\tau+d},~~a,b,c, d\in\mathbb{Z},~~ad-bc=1\,.
\end{equation}
Obviously the linear fraction transformation $\frac{a\tau+b}{c\tau+d}$ is identical to $\frac{-a\tau-b}{-c\tau-d}$. Therefore the modular group $\overline{\Gamma}$ is isomorphic to $PSL(2, \mathbb{Z})=SL(2, Z)/\{I, -I\}$, where $SL(2, \mathbb{Z})$ is the group of $2\times2$ matrices with integer entries and determinant 1, and $I$ refers to the two-dimensional unit matrix. The modular group has infinite elements, and it can be generated by two transformations $S$ and $T$,
\begin{equation}
S:\tau\rightarrow-\frac{1}{\tau},~~~~~~T:\tau\rightarrow\tau+1\,,
\end{equation}
which fulfill the following relations
\begin{equation}
S^2=(ST)^3=1
\end{equation}
and also $(TS)^3=1$ which is equivalent to $(ST)^3=1$ if $S^2=1$. Let $N$ be a positive integer. The principal congruence subgroup of level $N$ is
\begin{equation}
\Gamma(N)=\left\{\left(\begin{array}{cc}
a  & b \\
c  & d
\end{array}\right)\in SL(2, \mathbb{Z}),~~ \left(\begin{array}{cc}
a  & b \\
c  & d
\end{array}\right)=\left(
\begin{array}{cc}
1 ~&~0 \\
0  ~&~ 1
\end{array}
\right)\,(\mathrm{mod~N})
\right\}\,,
\end{equation}
which is infinite normal subgroup of $SL(2, \mathbb{Z})$. It is obvious that $\Gamma(1)\cong SL(2, \mathbb{Z})$. We define $\overline{\Gamma}(N)=\Gamma(N)/\{I, -I\}$ for $N=1, 2$ and
$\overline{\Gamma}(N)=\Gamma(N)$ for $N>2$ since the element $-I$ doesn't belong to $\Gamma(N)$ for $N>2$. Note $\overline{\Gamma}(1)\cong PSL(2, \mathbb{Z})\equiv\overline{\Gamma}$. The inhomogeneous finite modular groups are defined as the quotient groups $\Gamma_N\equiv\overline{\Gamma}/\overline{\Gamma}(N)$. Because the element $T^{N}$ belongs to $\Gamma(N)$, the finite modular group $\Gamma_N$ can be generated by $S$ and $T$ obeying the relations~\cite{deAdelhartToorop:2011re}
\begin{equation}
\label{eq:Gamma_N}S^2=(ST)^3=T^N=1\,.
\end{equation}
The groups $\Gamma_N$ for $N=2, 3, 4, 5$ are isomorphic to $S_3$, $A_4$, $S_4$ and $A_5$ respectively. Additional relations besides these in Eq.~\eqref{eq:Gamma_N} are necessary in order to render the group finite for $N>5$. The homogeneous finite modular groups are the quotient groups $\Gamma'_N\equiv SL(2, \mathbb{Z})/\Gamma(N)$, and they can generated by three generators $S$, $T$ and $\mathbb{R}$ which satisfy~\cite{Liu:2019khw}
\begin{equation}
\label{eq:Gammap_N}S^{2}=\mathbb{R},~~(ST)^{3}=T^{N}=\mathbb{R}^{2}=1,~~\mathbb{R}T = T\mathbb{R}
\end{equation}
and additional constraints should be imposed for $N>5$. The group $\Gamma'_2$ is isomorphic to $S_3$, and $\Gamma'_N$ is the double covering of $\Gamma_N$, and it has twice as many elements as $\Gamma_N$. In the present work, we shall focus on $\Gamma'_3$ which is the double covering of $A_4$ and is isomorphic to the binary tetrahedral group $T'$.

Modular forms of weight $k$ and level $N$ are holomorphic functions $f(\tau)$ which transforms under the action of $\Gamma(N)$ in the following way,
\begin{equation}
f\left(\gamma\tau\right)=(c\tau+d)^kf(\tau),~~~~\gamma=\left(\begin{array}{cc}
a  &  b \\
c  &  d
\end{array}
\right)\in\Gamma(N)\,,
\end{equation}
where $k$ is a generic non-negative integer. The modular forms of weight $k$ and level $N$ span a linear space of finite dimension. As shown in~\cite{Liu:2019khw}, the modular forms can be organized into some modular multiplets $f_{\mathbf{r}}\equiv\left(f_1(\tau), f_{2}(\tau),...\right)^{T}$ which transform as certain irreducible representation $\mathbf{r}$ of the finite modular group $\Gamma'_N$~\cite{Feruglio:2017spp,Liu:2019khw}, i.e.
\begin{equation}
f_{\mathbf{r}}(\gamma\tau)=(c\tau+d)^k\rho_{\mathbf{r}}(\gamma)f_{\mathbf{r}}(\tau)~~~\mathrm{for}~~\forall~\gamma\in SL(2,\mathbb{Z})\,,
\end{equation}
where $\gamma$ is the representative element of the coset
$\gamma\Gamma(N)$ in $\Gamma'_N$, and $\rho_{\mathbf{r}}(\gamma)$ is the representation matrix of the element $\gamma$ in the irreducible representation $\mathbf{r}$.

\subsection{\label{subsec:MFormN3} Modular forms of level 3 }

The linear space of modular forms of weight $k$ and level 3 has dimension $k+1$. The modular forms of level 3 has been constructed in terms of the
Dedekind eta function $\eta(\tau)$~\cite{Liu:2019khw}. There are two linearly independent modular forms of the lowest weight 1 with $k=1$, and they can arrange into a doublet $\mathbf{2}$ of $T'$,
\begin{equation}
Y^{(1)}_{\mathbf{2}}(\tau)\equiv\left(\begin{array}{c}
Y^{(1)}_{\mathbf{2},1}(\tau) \\
Y^{(1)}_{\mathbf{2}, 2}(\tau)
\end{array}\right)
=\Big(
\sqrt{2}e^{i7\pi/12}\dfrac{\eta^3(3\tau)}{\eta(\tau)} \,,\quad
\dfrac{\eta^3(3\tau)}{\eta(\tau)} - \dfrac{1}{3}\dfrac{\eta^3(\tau/3)}{\eta(\tau)} \Big)^T\,.
\end{equation}
Note the overall coefficient of $Y^{(1)}_{\mathbf{2}}$ can not be uniquely determined. We shall also denote $Y^{(1)}_{\mathbf{2},1}(\tau)=Y_1(\tau)$ and $Y^{(1)}_{\mathbf{2},2}(\tau)=Y_2(\tau)$ for notation simplicity in the following. The doublet modular forms $Y^{(1)}_{\mathbf{2},1}$ and $Y^{(1)}_{\mathbf{2}, 2}$ have the following $q-$expansions,
\begin{eqnarray}
\nonumber Y^{(1)}_{\mathbf{2},1}(\tau)&=& \sqrt{2} e^{7\pi i/12} q^{1/3} (1 + q + 2 q^2 + 2 q^4 + q^5 + 2 q^6 +q^8+2q^9+2q^{10}+\ldots) ,\\
\label{eq:qexpans}Y^{(1)}_{\mathbf{2}, 2}(\tau)&=& 1/3 + 2 q + 2 q^3 + 2 q^4 + 4 q^7+2q^9+2q^{12}+ \ldots\,,
\end{eqnarray}
with $q=e^{2\pi i\tau}$. The modular forms of higher weights can be constructed from the tensor products of $Y^{(1)}_{\mathbf{2}}$. There are three linearly independent weight 2 modular forms which can be arranged into a $T'$ triplet,
\begin{equation}
Y^{(2)}_{\mathbf{3}} \equiv \begin{pmatrix}
Y^{(2)}_{\mathbf{3},1}(\tau)\\Y^{(2)}_{\mathbf{3},2}(\tau)\\Y^{(2)}_{3,3}(\tau)
\end{pmatrix}= \begin{pmatrix}
e^{i\pi/6}Y^2_2 \\ \sqrt{2}e^{i7\pi /12}Y_1Y_2 \\ Y^2_1  \end{pmatrix}
\end{equation}
At weight 3, we have four independent modular forms which can be decomposed into two doublets transforming in the representations $\mathbf{2}$ and $\mathbf{2}''$ of $T'$,
\begin{eqnarray}
\nonumber&&Y^{(3)}_{\mathbf{2}} \equiv \begin{pmatrix}
Y^{(3)}_{\mathbf{2},1}\\Y^{(3)}_{\mathbf{2},2} \end{pmatrix} = \begin{pmatrix} 3e^{i\pi/6}Y_1Y^2_2 \\ \sqrt{2}e^{i5\pi/12}Y^3_1-e^{i\pi/6}Y^3_2 \end{pmatrix} \,, \\
&&Y^{(3)}_{\mathbf{2}''} \equiv \begin{pmatrix}
Y^{(3)}_{\mathbf{2}'',1}\\Y^{(3)}_{\mathbf{2}'',2} \end{pmatrix} = \begin{pmatrix} Y^3_1+(1-i)Y^3_2 \\ -3Y_2Y^2_1 \end{pmatrix} \,.
\end{eqnarray}
The modular forms of weight 4 can be obtained from the contractions of $Y^{(1)}_{\mathbf{2}}$ and $Y^{(3)}_{\mathbf{2}}$,  $Y^{(3)}_{\mathbf{2}''}$, and they decompose as $\mathbf{3}\oplus\mathbf{1}\oplus\mathbf{1}'$ under $T'$,
\begin{eqnarray}
\nonumber&&Y^{(4)}_{\mathbf{1}}=-4Y^3_1Y_2-(1-i)Y^4_2\,,\\
\nonumber&&Y^{(4)}_{\mathbf{1}'}=\sqrt{2}e^{i5\pi/12}Y^4_1-4e^{i\pi/6}Y_1Y^3_2\,,\\
&&Y^{(4)}_{\mathbf{3}} \equiv \begin{pmatrix}
Y^{(4)}_{\mathbf{3},1}\\Y^{(4)}_{\mathbf{3},2} \\ Y^{(4)}_{\mathbf{3},3} \end{pmatrix} = \begin{pmatrix} \sqrt{2}e^{i7\pi/12}Y^3_1Y_2-e^{i\pi/3}Y^4_2 \\ -Y^4_1-(1-i)Y_1Y^3_2 \\ 3e^{i\pi/6}Y^2_1Y^2_2 \end{pmatrix} \,.
\end{eqnarray}
The analytical expressions of modular forms of weights 5, weight 6, weight 7 and weight 8 are reported in the Appendix~\ref{app:High_MDF}. The modular multiplets of level 3 at different weights are summarized in table~\ref{Tab:Level3_MM}. It is remarkable that the odd weight modular forms always transform in the doublet irreducible representations $\mathbf{2}$, $\mathbf{2}'$ and $\mathbf{2}''$ of $T'$, while the even weight modular forms arrange themselves into $T'$ triplet $\mathbf{3}$ and singlets $\mathbf{1}$, $\mathbf{1}'$ and $\mathbf{1}''$ which
are identical with the representation matrices of $A_4$ in our working basis. This interesting structure of modular forms can help to produce texture zeros of the quark mass matrix, as we shall show in the following.

We shall formulate our models in the framework of $\mathcal{N}=1$ global supersymmetry. In modular invariant theory, it is generally assumed that the chiral supermultiplet $\Phi_I$ carries a modular weight $-k_I$ and transforms according to certain representation $\rho_I$ of $\Gamma'_N$~\cite{Ferrara:1989bc,Ferrara:1989qb,Feruglio:2017spp},
\begin{equation}
\label{eq:MDTrans}\tau\rightarrow\gamma\tau=\frac{a\tau+b}{c\tau+d},~~~\Phi_I\rightarrow(c\tau+d)^{-k_I}\rho_I(\gamma)\Phi_I
\end{equation}
In the present paper, we choose a minimal form of the K\"{a}hler potential,
\begin{equation}
\label{eq:kahler_min}\mathcal{K}(\tau, \overline{\tau}, \Phi_I, \overline{\Phi}_I)=-h\Lambda^2\log(-i\tau+i\overline{\tau})+\sum_I\frac{|\Phi_I|^2}{(-i\tau+i\overline{\tau})^{k_I}}
\end{equation}
which is invariant up to a K\"{a}hler transformation under the modular transformations given in Eq.~\eqref{eq:MDTrans}, and $h$ is a positive constant. It is the leading order K\"ahler potential of general matter field $\Phi_I$ with modular weight $-k_I$ in the context of string compactifications on abelian orbifolds~\cite{Dixon:1989fj}. The K\"ahler potential is less constrained by modular symmetry, at least the operators $(-i\tau+i\overline{\tau})^{k-k_I}\left(\Phi^{\dagger}_IY^{(k)\dagger}_{\mathbf{r}}Y^{(k)}_{\mathbf{r}}\Phi_I\right)_{\mathbf{1}}$ for any integer $k$ and irreducible representation $\mathbf{r}$ can not be avoided, they can lead to off-diagonal elements of the K\"ahler metric and the predictions for neutrino masses and mixing parameters can be changed considerably~\cite{Chen:2019ewa}. As a consequence, in the bottom-up model construction based on modular symmetry models~\cite{Feruglio:2017spp}, it is necessary to constrain the K\"ahler potential better to enhance the predictive power. Motivated by top-down model building in string theory, the modular flavor symmetry is extended to combine with traditional flavor symmetry~\cite{Baur:2019kwi,Baur:2019iai,Nilles:2020nnc,Nilles:2020kgo}. In this new scheme, both superpotential and K\"ahler potentials are strongly constrained, the non-diagonal contributions
to the K\"ahler metric are forbidden by all symmetries of the theory. Therefore the extra terms depending on modular forms in the K\"ahler potential don't considerably alter the phenomenological predictions which have been obtained by using just the standard K\"ahler potential in Eq.~\eqref{eq:kahler_min}~\cite{Nilles:2020kgo}. Hence we shall mainly focus on the minimal canonical K\"ahler potential of Eq.~\eqref{eq:kahler_min} in the present work, and the non-minimal K\"ahler potential would be discussed in section~\ref{sec:non-minimal-Kahler} and Appendix~\ref{app:non-minimal-Kahler-lepton}. The superpotential $\mathcal{W}(\tau, \Phi_I)$ can be expanded in power series of the supermultiplets $\Phi_I$ as follows,
\begin{equation}
\mathcal{W}(\tau, \Phi_I)=\sum_{n}Y_{I_1\ldots I_n}(\tau)\Phi_{I_1}\ldots\Phi_{I_n}\,,
\end{equation}
where the function $Y_{I_1\ldots I_n}(\tau)$ should be modular forms of weight $k_Y$ and level $N$, and should transform in the representation $\rho_Y$ of $\Gamma'_N$,
\begin{equation}
\tau\rightarrow\gamma\tau=\frac{a\tau+b}{c\tau+d},~~~~~Y_{I_1\ldots I_n}(\tau)\rightarrow Y_{I_1\ldots I_n}(\gamma\tau)=(c\tau+d)^{k_Y}\rho_{Y}(\gamma)Y_{I_1\ldots I_n}(\tau)\,.
\end{equation}
The requirement that $\mathcal{W}$ is invariant under the modular transformation entails $k_Y$ and $\rho_Y$ should fulfill the following constraints
\begin{equation}
k_Y=k_{I_1}+k_{I_1}+\ldots+k_{I_n},~~~\rho_Y\otimes\rho_{I_1}\otimes\ldots\rho_{I_n}\supset\mathbf{1}\,,
\end{equation}
where $\mathbf{1}$ denotes the invariant singlet representation of $\Gamma'_{N}$. In the following, we shall investigate the possible texture zero structures of the quark mass matrix for the assignments that the three generations of the right-handed quark fields transform as a doublet and a singlet under $T'$ modular symmetry or they are three singlets of $T'$.

\begin{table}[t!]
\centering
\begin{tabular}{|c|c|}
\hline  \hline

Modular weight $k$ & Modular form $Y^{(k)}_{\mathbf{r}}$ \\ \hline

$k=1$ & $Y^{(1)}_{\mathbf{2}}$\\  \hline

$k=2$ & $Y^{(2)}_{\mathbf{3}}$\\ \hline

$k=3$ & $Y^{(3)}_{\mathbf{2}}, Y^{(3)}_{\mathbf{2}''}$\\ \hline

$k=4$ & $Y^{(4)}_{\mathbf{1}}, Y^{(4)}_{\mathbf{1}'}, Y^{(4)}_{\mathbf{3}}$\\ \hline

$k=5$ & $Y^{(5)}_{\mathbf{2}}, Y^{(5)}_{\mathbf{2}'}, Y^{(5)}_{\mathbf{2}''}$\\ \hline

$k=6$ & $Y^{(6)}_{\mathbf{1}}, Y^{(6)}_{\mathbf{3}I}, Y^{(6)}_{\mathbf{3}II}$\\ \hline

$k=7$ & $Y^{(7)}_{\mathbf{2}I}, Y^{(7)}_{\mathbf{2}II}, Y^{(7)}_{\mathbf{2}'}, Y^{(7)}_{\mathbf{2}''}$\\ \hline

$k=8$ & $Y^{(8)}_{\mathbf{1}}, Y^{(8)}_{\mathbf{1}'}, Y^{(8)}_{\mathbf{1}''}, Y^{(8)}_{\mathbf{3}I}, Y^{(8)}_{\mathbf{3}II}$\\ \hline \hline
\end{tabular}
\caption{\label{Tab:Level3_MM}Summary of modular forms of level 3 up to weight 8, the subscript $\mathbf{r}$ denote the transformation property under $T'$ modular symmetry. Here $Y^{(6)}_{\mathbf{3}I}$ and $Y^{(6)}_{\mathbf{3}II}$ stand for two weight 6 modular forms transforming in the representation $\mathbf{3}$ of $T'$. Similar conventions are adopted for $Y^{(7)}_{\mathbf{2}I}$, $Y^{(7)}_{\mathbf{2}II}$ and $Y^{(8)}_{\mathbf{3}I}$, $Y^{(8)}_{\mathbf{3}II}$. }
\end{table}

\section{\label{sec:QMM_ds}Quark mass matrices for doublet plus singlet assignments of the right-handed quarks }

In this case, both left-handed and right-handed quark fields transform as a direct sum of one-dimensional and two-dimensional representations of the $T'$ modular group, i.e.
\begin{equation}
\label{eq:assign_1st}
Q_D\equiv\begin{pmatrix}
Q_1 \\
Q_2
\end{pmatrix}\sim\mathbf{2}^{i},\quad Q_3\sim \mathbf{1}^{j}\,,~~~
q^c_D\equiv\begin{pmatrix}
q^c_1 \\
q^c_2
\end{pmatrix}\sim\mathbf{2}^{k},\quad q^c_3\sim \mathbf{1}^{l}\,,
\end{equation}
where $i,j,k,l=0, 1, 2$ and we have denoted $\mathbf{1}\equiv \mathbf{1}^{0}$, $\mathbf{1}^{'}\equiv \mathbf{1}^{1}$, $\mathbf{1}^{''}\equiv \mathbf{1}^{2}$ for singlet representations and $\mathbf{2}\equiv \mathbf{2}^{0}$, $\mathbf{2}^{'}\equiv \mathbf{2}^{1}$, $\mathbf{2}^{''}\equiv \mathbf{2}^{2}$ for the doublet representations.
The notations $Q_i$ and $q^c_i$ $(i=1, 2, 3)$ stand for the left-handed and right-handed quark fields respectively, and $q^c_i$ can be either up type quark fields $u^c_i$ or down type quark fields $d^c_i$. We denote the modular weights of $Q_D$, $Q_3$, $q^c_D$ and $q^c_3$ as $k_{Q_D}$, $k_{Q_3}$, $k_{q^c_D}$ and $k_{q^c_3}$ respectively in the following, and the modular weights of the Higgs doublets $H_{u,d}$ are assumed to be vanishing. Thus the most general form of the superpotential for the quark masses is given by \begin{equation}
\label{eq:wq}\mathcal{W}_q=q^c_DQ_DH_{u/d}f_{DD}(Y)+q^c_DQ_3H_{u/d}f_{D3}(Y)+q^c_3Q_DH_{u/d}f_{3D}(Y)+q^c_3Q_3H_{u/d}f_{33}(Y)\,,
\end{equation}
where $f_{DD}(Y)$, $f_{D3}(Y)$, $f_{3D}(Y)$ and $f_{33}(Y)$ are general functions of modular forms, and their explicit forms depend on the group indices $i$, $j$, $k$, $l$ and the modular weights of the quark fields. The coupling constants in front of each term are neglected in Eq.~\eqref{eq:wq}. The Higgs field in Eq.~\eqref{eq:wq} is $H_u$ for up type quark fields $u^c$ and it is $H_d$ for down type quark fields $d^c$. Each term of
Eq.~\eqref{eq:wq} should be a singlet under $T'$, and its modular weight should be zero. Moreover, from Eq.~\eqref{eq:wq} we see that the quark mass matrix can be divided into four blocks as follow,
\begin{equation}
M_q=\left(\begin{array}{ccc}
 ~S~ &   \vdots & ~C~\\
\cdots & \cdots  & \cdots \\
 ~R ~ &   \vdots & ~T~\\
\end{array}\right)v_{u/d}\,,
\end{equation}
where $v_{u/d}$ is the vacuum expectation value (VEV) of the Higgs field $H_{u/d}$, and the quark mass matrix $M_q$ is given in the right-left basis, $q^{c}_i(M_q)_{ij}Q_j$, $S$, $C$, $R$ and $T$ are $2\times 2$, $2\times 1$, $1\times 2$ and $1\times 1$ sub-matrices respectively. Notice that we can assign the first generation or the second generation quark field instead of the third generation to be a singlet under $T'$ while the remaining two generations transform as a $T'$ doublet, the corresponding quark mass matrix can be obtained by multiplying certain permutation matrices from both sides. As a consequence, the results for the quark masses and mixing matrix are not changed. Using the Kronecker products and the CG coefficients of $T'$ given in Appendix~\ref{app:Tp_group}, we can find out the explicit forms of $S$, $C$, $R$ and $T$, and each matrix element can be expressed in terms of modular form. We shall be concerned with modular forms up to weight 6 in the present work, higher weight modular forms can be discussed in a similar way, nevertheless more modular invariant interactions as well as more free parameters would be involved.

\subsection{General structure of $S$}

This $2\times 2$ submatrix is determined by the representations $\mathbf{2}^{i}$ and $\mathbf{2}^{k}$ as well as the modular weights
$k_{Q_D}$ and $k_{q^c_D}$. Modular invariance requires that $f_{DD}(Y)$ should be a modular form in the triplet representation $\mathbf{3}$ or singlet representations $\mathbf{1}$, $\mathbf{1}'$, $\mathbf{1}''$ of $T'$. Hence $S$ would be vanishing if $k_{q^c_D}+k_{Q_D}$ is odd. As a consequence, the rank of the quark mass matrix would be less than three and at least one quark is massless which is disfavored by experimental data.
For all possible assignments of $\mathbf{2}^{i}$ and $\mathbf{2}^{k}$, we can find out the following possible structures for the $2\times 2$ submatrix $S$.
\begin{itemize}[labelindent=-0.8em, leftmargin=1.2em]
\item{$\mathbf{2}^{i}\otimes\mathbf{2}^{k}=\mathbf{2}\otimes\mathbf{2}=\mathbf{2}'\otimes \mathbf{2}''=\mathbf{3}\oplus\mathbf{1}'$}

In this case, if the summation of the modular weights $k_{Q_D}$ and $k_{q^c_D}$ is equal to 2, i.e. $k_{q^c_D}+k_{Q_D}=2$, $f_{DD}(Y)$ would be uniquely proportional to the weight two modular form $Y^{(2)}_{\mathbf{3}}=(Y^{(2)}_{\mathbf{3},1},Y^{(2)}_{\mathbf{3},2},Y^{(2)}_{\mathbf{3},3})^{T}$ which couples to $q^c_D$ and $Q_D$ to form a modular invariant singlet. From the CG coefficients of $T'$ given in Appendix~\ref{app:Tp_group}, we can straightforwardly read out the general form of $S$ as follows,
\begin{equation}
S_{1}=\left( \begin{array}{cc} \kappa \sqrt{2}e^{\frac{i5\pi}{12}}Y^{(2)}_{\mathbf{3},2} & -\kappa Y^{(2)}_{\mathbf{3},3} \\ -\kappa Y^{(2)}_{\mathbf{3},3} & \kappa \sqrt{2}e^{\frac{i7\pi}{12}}Y^{(2)}_{\mathbf{3},1} \\ \end{array} \right)\,,
\end{equation}
which is a symmetric matrix, and $\kappa$ is in general a complex Yukawa coupling parameter. In the case of $k_{q^c_D}+k_{Q_D}=4$, $f_{DD}(Y)$ would be the weight four modular form $Y^{(4)}_{\mathbf{3}}=(Y^{(4)}_{\mathbf{3}, 1},Y^{(4)}_{\mathbf{3}, 2},Y^{(4)}_{\mathbf{3}, 3})^{T}$, and the submatrix $S$ reads as
\begin{equation}
S_{2}=\left( \begin{array}{cc} \kappa \sqrt{2}e^{\frac{i5\pi}{12}}Y^{(4)}_{\mathbf{3}, 2} & -\kappa Y^{(4)}_{\mathbf{3}, 3} \\ -\kappa Y^{(4)}_{\mathbf{3}, 3} & \kappa \sqrt{2}e^{\frac{i7\pi}{12}}Y^{(4)}_{\mathbf{3}, 1} \\ \end{array} \right)\,.
\end{equation}
We have two independent modular forms $Y^{(6)}_{\mathbf{3}I}$ and $Y^{(6)}_{\mathbf{3}II}$ at weight six, as shown in table~\ref{Tab:Level3_MM}. Thus $f_{DD}(Y)$ is a linear combination of $Y^{(6)}_{\mathbf{3}I}$ and $Y^{(6)}_{\mathbf{3}II}$ for $k_{q^c_D}+k_{Q_D}=6$, and we have
\begin{equation}
S_{3}=\left( \begin{array}{cc} \sqrt{2}e^{\frac{i5\pi}{12}}(\kappa_{1} Y^{(6)}_{\mathbf{3}I,2} + \kappa_{2} Y^{(6)}_{\mathbf{3}II,2}) & -(\kappa_{1} Y^{(6)}_{\mathbf{3}I,3} + \kappa_{2} Y^{(6)}_{\mathbf{3}II,3}) \\ -(\kappa_{1} Y^{(6)}_{\mathbf{3}I,3} + \kappa_{2} Y^{(6)}_{\mathbf{3}II,3}) & \sqrt{2}e^{\frac{i7\pi}{12}}(\kappa_{1} Y^{(6)}_{\mathbf{3}I,1} + \kappa_{2} Y^{(6)}_{\mathbf{3}II,1}) \\ \end{array} \right)\,,
\end{equation}
where $\kappa_1$ and $\kappa_2$ are complex free parameters.

\item{$\mathbf{2}^{i}\otimes \mathbf{2}^{k}=\mathbf{2}\otimes \mathbf{2}'=\mathbf{2}''\otimes \mathbf{2}''=\mathbf{3}\oplus \mathbf{1}''$}

In this case, $f_{DD}(Y)$ has to be modular form transforming as $\mathbf{3}$ or $\mathbf{1}'$ in order to fulfill modular invariance.
Hence $f_{DD}(Y)$ is proportional to $Y^{(2)}_{\mathbf{3}}$ for the value $k_{q^c_D}+k_{Q_D}=2$, and $S$ is of the following form
\begin{equation}
S_{4}=\left( \begin{array}{cc} \kappa \sqrt{2}e^{\frac{i5\pi}{12}}Y^{(2)}_{\mathbf{3},1} & -\kappa Y^{(2)}_{\mathbf{3},2} \\ -\kappa Y^{(2)}_{\mathbf{3},2} & \kappa \sqrt{2}e^{\frac{i7\pi}{12}}Y^{(2)}_{\mathbf{3},3} \\ \end{array} \right)\,.
\end{equation}
For the value of $k_{q^c_D}+k_{Q_D}=4$, both $Y^{(4)}_{\mathbf{3}}$ and $Y^{(4)}_{\mathbf{1}'}$ can contribute to the quark mass terms, and the submatrix $S$ is given by
\begin{equation}
S_{5}=\left( \begin{array}{cc} \kappa_{1} \sqrt{2}e^{\frac{i5\pi}{12}}Y^{(4)}_{\mathbf{3}, 1} & \kappa_{2}Y^{(4)}_{\mathbf{1}'}-\kappa_{1} Y^{(4)}_{\mathbf{3}, 2} \\ -\kappa_{2}Y^{(4)}_{\mathbf{1}'}-\kappa_{1} Y^{(4)}_{\mathbf{3}, 2} & \kappa_{1} \sqrt{2}e^{\frac{i7\pi}{12}}Y^{(4)}_{\mathbf{3}, 3} \\ \end{array} \right)\,,
\end{equation}
which is not a symmetric matrix because of the term $\kappa_2(q^c_DQ_D)_{\mathbf{1}''}Y^{(4)}_{\mathbf{1}'}H_{u/d}$. Similarly for $k_{q^c_D}+k_{Q_D}=6$, both $Y^{(6)}_{\mathbf{3}I}$ and $Y^{(6)}_{\mathbf{3}II}$ are relevant, and we find
\begin{equation}
S_{6}=\left( \begin{array}{cc} \sqrt{2}e^{\frac{i5\pi}{12}}(\kappa_{1} Y^{(6)}_{\mathbf{3}I,1}+\kappa_{2} Y^{(6)}_{\mathbf{3}II,1}) & -(\kappa_{1} Y^{(6)}_{\mathbf{3}I,2} + \kappa_{2} Y^{(6)}_{\mathbf{3}II,2}) \\ -(\kappa_{1} Y^{(6)}_{\mathbf{3}I,2} + \kappa_{2} Y^{(6)}_{\mathbf{3}II,2}) & \sqrt{2}e^{\frac{i7\pi}{12}}(\kappa_{1} Y^{(6)}_{\mathbf{3}I,3} + \kappa_{2} Y^{(6)}_{\mathbf{3}II,3})\\ \end{array} \right)\,.
\end{equation}

\item{$\mathbf{2}^{i}\otimes \mathbf{2}^{k}=\mathbf{2}\otimes \mathbf{2}''=\mathbf{2}'\otimes \mathbf{2}'=\mathbf{3}\oplus \mathbf{1}$}

If the modular weights satisfy $k_{q^c_D}+k_{Q_D}=2$, the mass term of the first two generation quarks is given by $\kappa(q^c_DQ_DY^{(2)}_{\mathbf{3}})_{\mathbf{1}}H_{u/d}$ which gives rise to
\begin{equation}
S_{7}=\left( \begin{array}{cc} \kappa \sqrt{2}e^{\frac{i5\pi}{12}}Y^{(2)}_{\mathbf{3},3} & -\kappa Y^{(2)}_{\mathbf{3},1} \\ -\kappa Y^{(2)}_{\mathbf{3},1} & \kappa \sqrt{2}e^{\frac{i7\pi}{12}}Y^{(2)}_{\mathbf{3},2} \\ \end{array} \right)\,.
\end{equation}
For $k_{q^c_D}+k_{Q_D}=4$, $f_{DD}(Y)$ can be the weight 4 modular forms $Y^{(4)}_{\mathbf{3}}$ and $Y^{(4)}_{\mathbf{1}}$. We can read out the general form of the submatrix $S$ as follow,
\begin{equation}
S_{8}=\left( \begin{array}{cc} \kappa_{1} \sqrt{2}e^{\frac{i5\pi}{12}}Y^{(4)}_{\mathbf{3}, 3} & \kappa_{2}Y^{(4)}_{\mathbf{1}}-\kappa_{1} Y^{(4)}_{\mathbf{3}, 1} \\ -\kappa_{2}Y^{(4)}_{\mathbf{1}}-\kappa_{1} Y^{(4)}_{\mathbf{3}, 1} & \kappa_{1} \sqrt{2}e^{\frac{i7\pi}{12}}Y^{(4)}_{\mathbf{3}, 2} \\ \end{array} \right)\,.
\end{equation}
For the value of $k_{q^c_D}+k_{Q_D}=6$, all the three weight 6 modular forms $Y^{(6)}_{\mathbf{3}I}$, $Y^{(6)}_{\mathbf{3}II}$ and  $Y^{(6)}_{\mathbf{1}}$ are relevant, and $S$ is given by
\begin{equation}
S_{9}=\left( \begin{array}{cc} \sqrt{2}e^{\frac{i5\pi}{12}}(\kappa_{1} Y^{(6)}_{\mathbf{3}I,3} + \kappa_{2} Y^{(6)}_{\mathbf{3}II,3}) & -(\kappa_{1} Y^{(6)}_{\mathbf{3}I,1}+\kappa_{2} Y^{(6)}_{\mathbf{3}II,1}) + \kappa_{3}Y^{(6)}_{\mathbf{1}} \\ -(\kappa_{1} Y^{(6)}_{\mathbf{3}I,1}+\kappa_{2} Y^{(6)}_{\mathbf{3}II,1}) - \kappa_{3}Y^{(6)}_{\mathbf{1}} & \sqrt{2}e^{\frac{i7\pi}{12}}(\kappa_{1} Y^{(6)}_{\mathbf{3}I,2} + \kappa_{2} Y^{(6)}_{\mathbf{3}II,2}) \\ \end{array} \right)\,.
\end{equation}
We summarized all the above possible forms of the submatrix $S$ for different assignments of $\mathbf{2}^i$, $\mathbf{2}^{k}$ and the modular weights $k_{Q_D}$, $k_{q^c_D}$ in table~\ref{Tab:S}.

\end{itemize}

\begin{table}[t!]
\centering
\small
\begin{tabular}{|c|c|c|}
\hline  \hline
   & Expressions of $S$ & Constraints \\
  \hline
 \multirow{2}{*}{$S_0$} &  \multirow{2}{*}{$\left( \begin{array}{cc} 0 & 0 \\ 0 & 0 \\ \end{array} \right)$}  &  $k_{q^c_D}+k_{Q_D}<0$ \\
 & & $k_{q^c_D}+k_{Q_D}=0,1,3,5,\ldots$ \\ \hline
$S_{1}$ & $\left( \begin{array}{cc} \kappa \sqrt{2}e^{\frac{i5\pi}{12}}Y^{(2)}_{\mathbf{3},2} & -\kappa Y^{(2)}_{\mathbf{3},3} \\ -\kappa Y^{(2)}_{\mathbf{3},3} & \kappa \sqrt{2}e^{\frac{i7\pi}{12}}Y^{(2)}_{\mathbf{3},1} \\ \end{array} \right) $ & $k_{q^c_D}+k_{Q_D}=2,~k+i=0\,(\text{mod}~3)$\\ \hline

$S_{2}$ & $\left( \begin{array}{cc} \kappa \sqrt{2}e^{\frac{i5\pi}{12}}Y^{(4)}_{\mathbf{3}, 2} & -\kappa Y^{(4)}_{\mathbf{3}, 3} \\ -\kappa Y^{(4)}_{\mathbf{3}, 3} & \kappa \sqrt{2}e^{\frac{i7\pi}{12}}Y^{(4)}_{\mathbf{3}, 1} \\ \end{array} \right) $ & $k_{q^c_D}+k_{Q_D}=4,~k+i=0\,(\text{mod}~3)$\\\hline

$S_{3}$ & $\left( \begin{array}{cc} \sqrt{2}e^{\frac{i5\pi}{12}}(\kappa_{1} Y^{(6)}_{\mathbf{3}I,2} + \kappa_{2} Y^{(6)}_{\mathbf{3}II,2}) & -(\kappa_{1} Y^{(6)}_{\mathbf{3}I,3} + \kappa_{2} Y^{(6)}_{\mathbf{3}II,3}) \\ -(\kappa_{1} Y^{(6)}_{\mathbf{3}I,3} + \kappa_{2} Y^{(6)}_{\mathbf{3}II,3}) & \sqrt{2}e^{\frac{i7\pi}{12}}(\kappa_{1} Y^{(6)}_{\mathbf{3}I,1} + \kappa_{2} Y^{(6)}_{\mathbf{3}II,1}) \\ \end{array} \right) $ & $k_{q^c_D}+k_{Q_D}=6,~k+i=0\,(\text{mod}~3)$\\\hline

$S_{4}$ & $\left( \begin{array}{cc} \kappa \sqrt{2}e^{\frac{i5\pi}{12}}Y^{(2)}_{\mathbf{3},1} & -\kappa Y^{(2)}_{\mathbf{3},2} \\ -\kappa Y^{(2)}_{\mathbf{3},2} & \kappa \sqrt{2}e^{\frac{i7\pi}{12}}Y^{(2)}_{\mathbf{3},3} \\ \end{array} \right)$ & $k_{q^c_D}+k_{Q_D}=2,~k+i=1\,(\text{mod}~3)$\\\hline

$S_{5}$ & $\left( \begin{array}{cc} \kappa_{1} \sqrt{2}e^{\frac{i5\pi}{12}}Y^{(4)}_{\mathbf{3}, 1} & \kappa_{2}Y^{(4)}_{\mathbf{1}'}-\kappa_{1} Y^{(4)}_{\mathbf{3}, 2} \\ -\kappa_{2}Y^{(4)}_{\mathbf{1}'}-\kappa_{1} Y^{(4)}_{\mathbf{3}, 2} & \kappa_{1} \sqrt{2}e^{\frac{i7\pi}{12}}Y^{(4)}_{\mathbf{3}, 3} \\ \end{array} \right)$ & $k_{q^c_D}+k_{Q_D}=4,~k+i=1\,(\text{mod}~3)$\\\hline

$S_{6}$ & $\left( \begin{array}{cc} \sqrt{2}e^{\frac{i5\pi}{12}}(\kappa_{1} Y^{(6)}_{\mathbf{3}I,1}+\kappa_{2} Y^{(6)}_{\mathbf{3}II,1}) & -(\kappa_{1} Y^{(6)}_{\mathbf{3}I,2} + \kappa_{2} Y^{(6)}_{\mathbf{3}II,2}) \\ -(\kappa_{1} Y^{(6)}_{\mathbf{3}I,2} + \kappa_{2} Y^{(6)}_{\mathbf{3}II,2}) & \sqrt{2}e^{\frac{i7\pi}{12}}(\kappa_{1} Y^{(6)}_{\mathbf{3}I,3} + \kappa_{2} Y^{(6)}_{\mathbf{3}II,3})\\ \end{array} \right)$ & $k_{q^c_D}+k_{Q_D}=6,~k+i=1\,(\text{mod}~3)$\\\hline

$S_{7}$ & $\left( \begin{array}{cc} \kappa \sqrt{2}e^{\frac{i5\pi}{12}}Y^{(2)}_{\mathbf{3},3} & -\kappa Y^{(2)}_{\mathbf{3},1} \\ -\kappa Y^{(2)}_{\mathbf{3},1} & \kappa \sqrt{2}e^{\frac{i7\pi}{12}}Y^{(2)}_{\mathbf{3},2} \\ \end{array} \right)$ & $k_{q^c_D}+k_{Q_D}=2,~k+i=2\,(\text{mod}~3)$\\\hline

$S_{8}$ & $\left( \begin{array}{cc} \kappa_{1} \sqrt{2}e^{\frac{i5\pi}{12}}Y^{(4)}_{\mathbf{3}, 3} & \kappa_{2}Y^{(4)}_{\mathbf{1}}-\kappa_{1} Y^{(4)}_{\mathbf{3}, 1} \\ -\kappa_{2}Y^{(4)}_{\mathbf{1}}-\kappa_{1} Y^{(4)}_{\mathbf{3}, 1} & \kappa_{1} \sqrt{2}e^{\frac{i7\pi}{12}}Y^{(4)}_{\mathbf{3}, 2} \\ \end{array} \right)$ & $k_{q^c_D}+k_{Q_D}=4,~k+i=2\,(\text{mod}~3)$\\\hline

$S_{9}$ & $\left( \begin{array}{cc} \sqrt{2}e^{\frac{i5\pi}{12}}(\kappa_{1} Y^{(6)}_{\mathbf{3}I,3} + \kappa_{2} Y^{(6)}_{\mathbf{3}II,3}) & -(\kappa_{1} Y^{(6)}_{\mathbf{3}I,1}+\kappa_{2} Y^{(6)}_{\mathbf{3}II,1}) + \kappa_{3}Y^{(6)}_{\mathbf{1}} \\ -(\kappa_{1} Y^{(6)}_{\mathbf{3}I,1}+\kappa_{2} Y^{(6)}_{\mathbf{3}II,1}) - \kappa_{3}Y^{(6)}_{\mathbf{1}} & \sqrt{2}e^{\frac{i7\pi}{12}}(\kappa_{1} Y^{(6)}_{\mathbf{3}I,2} + \kappa_{2} Y^{(6)}_{\mathbf{3}II,2}) \\ \end{array} \right)$ & $k_{q^c_D}+k_{Q_D}=6,~k+i=2\,(\text{mod}~3)$\\
\hline \hline
\end{tabular}
\caption{\label{Tab:S}The possible structures of the submatrix $S$, where the left-handed quark doublet $Q_D$ and the right-handed quark field $q^c_D$ are assigned to transformed as $\mathbf{2}^i$ and $\mathbf{2}^k$ respectively under $T'$ modular symmetry, and their modular weights are denoted as $k_{Q_D}$ and $k_{q^c_D}$ respectively. }
\end{table}

\subsection{General structures of $C$ and $R$}

The submatrix $C$ comprises the (13) and (23) entries of the quark mass matrix $M_q$, and it is determined by the modular form $f_{D3}(Y)$. Once the assignments for $\mathbf{2}^{k}$, $\mathbf{1}^{j}$ and the modular weights $k_{q^c_D}$ and $k_{Q_3}$ are specified, we can easily read out $f_{D3}(Y)$ and the explicit form of $C$. The modular invariance requires that $q^c_D$, $Q_3$ and $f_{D3}(Y)$ should contract into a $T'$ singlet. From the Kronecker products $\mathbf{1}^a\otimes\mathbf{2}^b = \mathbf{2}^{a+b~(\text{mod}~3)}$ in Eq.~\eqref{eq:mult}, we know that $f_{D3}(Y)$ has to be a modular form transforming as $\mathbf{2}^{2-a-b~(\text{mod}~3)}$. As shown in table~\ref{Tab:Level3_MM}, the modular weights of the doublet modular forms must be odd. Therefore the sum $k_{q^c_D}+k_{Q_3}$ should be an odd integer otherwise the submatrix $C$ would be zero. We find that $C$ is vanishing exactly if any of the following conditions are fulfilled,
\begin{eqnarray}
\nonumber&&k_{q^c_D}+k_{Q_3}=0, 2, 4, 6,\ldots\,,\\
\nonumber\text{or}~&&k_{q^c_D}+k_{Q_3}=1~\text{with}~\mathbf{2}^{k}\otimes \mathbf{1}^{j}=\mathbf{2}~\text{or}~\mathbf{2}'\,,\\
\text{or}~&&k_{q^c_D}+k_{Q_3}=3~\text{with}~\mathbf{2}^{k}\otimes \mathbf{1}^{j}=\mathbf{2}'\,.
\end{eqnarray}
In the case that $C$ is non-vanishing, it can take the following nontrivial forms.
\begin{itemize}[labelindent=-0.8em, leftmargin=1.2em]
\item{$\mathbf{2}^{k}\otimes \mathbf{1}^{j}=\mathbf{2}$ for $k+j=0\,(\text{mod}~3)$ }

In this case, $f_{D3}(Y)$ should be modular form transforming as $\mathbf{2}''$ under $T'$. From table~\ref{Tab:Level3_MM}, we see that modular forms in the representation $\mathbf{2}''$ appear at weight 3 and weight 5. For $k_{q^c_D}+k_{Q_3}=3$, $f_{D3}(Y)$ is proportional to $Y^{(3)}_{\mathbf{2}''}\equiv(Y^{(3)}_{\mathbf{2}'', 1}, Y^{(3)}_{\mathbf{2}'', 2})^{T}$, and the corresponding structure of $C$ is given by
\begin{equation}
C_{1}=(\kappa Y^{(3)}_{\mathbf{2}'',2},~
-\kappa Y^{(3)}_{\mathbf{2}'',1})^{T}\,.
\end{equation}
For $k_{q^c_D}+k_{Q_3}=5$, we can easily read out the submatrix $C$ as follow,
\begin{equation}
C_{2}=(\kappa Y^{(5)}_{\mathbf{2}'',2},
-\kappa Y^{(5)}_{\mathbf{2}'',1})^{T}\,.
\end{equation}

\item{$\mathbf{2}^{k}\otimes \mathbf{1}^{j}=\mathbf{2}'$ for $k+j=1\,(\text{mod}~3)$}

The modular form $f_{D3}(Y)$ should transform as $\mathbf{2}'$ for this assignment. As shown in table~\ref{Tab:Level3_MM}, the lowest weight modular form in the doublet representation $\mathbf{2}'$ is $Y^{(5)}_{\mathbf{2}'}=(Y^{(5)}_{\mathbf{2}',1},Y^{(5)}_{\mathbf{2}',2})^{T}$. Accordingly the relevant quark mass term is $(q^c_DQ_3Y^{(5)}_{\mathbf{2}'})_{\mathbf{1}}H_{u/d}$ for $k_{q^c_D}+k_{Q_3}=5$, and the block $C$ is of the following form,
\begin{equation}
C_{3}=(\kappa Y^{(5)}_{\mathbf{2}',2},
-\kappa Y^{(5)}_{\mathbf{2}',1})^{T}\,.
\end{equation}

\item{$\mathbf{2}^{k}\otimes \mathbf{1}^{j}=\mathbf{2}''$ for $k+j=2\,(\text{mod}~3)$ }

A modular form transforming in the doublet representation $\mathbf{2}$ of $T'$ is necessary in order to form an invariant singlet in this case. There are three modular forms $Y^{(1)}_{\mathbf{2}}=(Y^{(1)}_{\mathbf{2},1},Y^{(1)}_{\mathbf{2},2})^{T}$, $Y^{(3)}_{\mathbf{2}}=(Y^{(3)}_{\mathbf{2},1},Y^{(3)}_{\mathbf{2},2})^{T}$ and $Y^{(5)}_{\mathbf{2}}=(Y^{(5)}_{\mathbf{2}, 1},Y^{(5)}_{\mathbf{2}, 2})^{T}$ transforming as $\mathbf{2}$. For the value of $k_{q^c_D}+k_{Q_3}=1$, $Y^{(1)}_{\mathbf{2}}$ is involved, and we find the submatrix $C$ is
\begin{equation}
C_{4}=(\kappa Y^{(1)}_{\mathbf{2},2},
-\kappa Y^{(1)}_{\mathbf{2},1})^{T}\,.
\end{equation}
For $k_{q^c_D}+k_{Q_3}=3$, the modular form $f_{D3}(Y)$ is $Y^{(3)}_{\mathbf{2}}$, and $C$ take the form
\begin{equation}
C_{5}=(\kappa Y^{(3)}_{\mathbf{2},2},
-\kappa Y^{(3)}_{\mathbf{2},1})^{T}\,.
\end{equation}
Similarly for $k_{q^c_D}+k_{Q_3}=5$, the submatrix $C$ reads as
\begin{equation}
C_{6}=(\kappa Y^{(5)}_{\mathbf{2}, 2},
-\kappa Y^{(5)}_{\mathbf{2}, 1})^{T}\,.
\end{equation}

\end{itemize}

In the exactly  same fashion, we can fix the possible structures of the submatrix $R$, and the results are summarized in table~\ref{Tab:C&R}.

\begin{table}[t!]
\centering
\resizebox{1.0\textwidth}{!}{
\begin{tabular}{|c|c|c||c|c|c|}
\hline  \hline
 & Expressions of $C$ & Constraints &  &  Expressions of $R$  & Constraints \\\hline
\multirow{4}{*}{$C_0$} & \multirow{4}{*}{$(0,0)^{T}$}  &  $ \textbf{i):}~k_{q^c_D}+k_{Q_3}<0,$ &   \multirow{4}{*}{$R_0$} & \multirow{4}{*}{$(0,0)$}  &  $ \textbf{i):}~k_{q^c_3}+k_{Q_D}<0,$ \\
 & &  $ \textbf{ii):}~k_{q^c_D}+k_{Q_3}=0,2,4,6,\ldots,$ &   & &  $ \textbf{ii):}~k_{q^c_3}+k_{Q_D}=0,2,4,6,\ldots,$ \\
& & $\textbf{iii):}~k_{q^c_D}+k_{Q_3}=1,~k+j=0,1\,(\text{mod}~3)\,,$ &  & & $\textbf{iii):}~k_{q^c_3}+k_{Q_D}=1,~l+i=0,1\,(\text{mod}~3)\,,$ \\
& & $\textbf{iv):}~k_{q^c_D}+k_{Q_3}=3,~k+j=1\,(\text{mod}~3)\,.$ &  & & $\textbf{iv):}~k_{q^c_3}+k_{Q_D}=3,~l+i=1\,(\text{mod}~3)\,.$ \\ \hline
$C_{1}$ & $(\kappa Y^{(3)}_{\mathbf{2}'',2}, - \kappa Y^{(3)}_{\mathbf{2}'',1})^{T}$ & $k_{q^c_D}+k_{Q_3}=3,~~k+j=0\,(\text{mod}~3)$ & $R_{1}$ & $(\kappa Y^{(3)}_{\mathbf{2}'',2}, - \kappa Y^{(3)}_{\mathbf{2}'',1})$ & $k_{q^c_3}+k_{Q_D}=3,~~l+i=0\,(\text{mod}~3)$ \\ \hline
$C_{2}$ & $(\kappa Y^{(5)}_{\mathbf{2}'',2}, - \kappa Y^{(5)}_{\mathbf{2}'',1})^{T}$ & $k_{q^c_D}+k_{Q_3}=5,~~k+j=0\,(\text{mod}~3)$ &  $R_{2}$ & $(\kappa Y^{(5)}_{\mathbf{2}'',2}, - \kappa Y^{(5)}_{\mathbf{2}'',1})$ & $k_{q^c_3}+k_{Q_D}=5,~~l+i=0\,(\text{mod}~3)$ \\ \hline
$C_{3}$ & $(\kappa Y^{(5)}_{\mathbf{2}',2}, - \kappa Y^{(5)}_{\mathbf{2}',1})^{T}$ & $k_{q^c_D}+k_{Q_3}=5,~~k+j=1\,(\text{mod}~3)$ &   $R_{3}$ & $(\kappa Y^{(5)}_{\mathbf{2}',2}, - \kappa Y^{(5)}_{\mathbf{2}',1})$ & $k_{q^c_3}+k_{Q_D}=5,~~l+i=1\,(\text{mod}~3)$\\ \hline
$C_{4}$ & $(\kappa Y^{(1)}_{\mathbf{2},2}, - \kappa Y^{(1)}_{\mathbf{2},1})^{T}$ & $k_{q^c_D}+k_{Q_3}=1,~~k+j=2\,(\text{mod}~3)$ &  $R_{4}$ & $(\kappa Y^{(1)}_{\mathbf{2},2}, - \kappa Y^{(1)}_{\mathbf{2},1})$ & $k_{q^c_3}+k_{Q_D}=1,~~l+i=2\,(\text{mod}~3)$ \\ \hline
$C_{5}$ & $(\kappa Y^{(3)}_{\mathbf{2},2}, - \kappa Y^{(3)}_{\mathbf{2},1})^{T}$ & $k_{q^c_D}+k_{Q_3}=3,~~k+j=2\,(\text{mod}~3)
$ &   $R_{5}$ & $(\kappa Y^{(3)}_{\mathbf{2}, 2}, - \kappa Y^{(3)}_{\mathbf{2}, 1})$ & $k_{q^c_3}+k_{Q_D}=3,~~l+i=2\,(\text{mod}~3)$ \\ \hline
$C_{6}$ & $(\kappa Y^{(5)}_{\mathbf{2}, 2}, - \kappa Y^{(5)}_{\mathbf{2}, 1})^{T}$ & $k_{q^c_D}+k_{Q_3}=5,~~k+j=2\,(\text{mod}~3)$ &   $R_{6}$ & $(\kappa Y^{(5)}_{\mathbf{2}, 2}, - \kappa Y^{(5)}_{\mathbf{2}, 1})$ & $k_{q^c_3}+k_{Q_D}=5,~~l+i=2\,(\text{mod}~3)$ \\
\hline \hline
\end{tabular}}
\caption{\label{Tab:C&R}The structures of the submatrix $C$ and $R$ for different possible values of modular weights and the assignments of the quark fields $q^c_D$, $q^c_3$, $Q_D$, $Q_3$ under the finite modular group $T'$. }
\end{table}

\subsection{General structure of $T$ }

$T$ is the (33) entry of the quark mass matrix, it would be non-zero if any of the following three conditions are satisfied,
\begin{equation}
\begin{aligned}
&\quad~~ k_{q^c_{3}}+k_{Q_{3}}=0\,,~\text{with}~\mathbf{1}^{l}\otimes \mathbf{1}^{j}=\mathbf{1}\,,\\
&\text{or}~~k_{q^c_{3}}+k_{Q_{3}}=4\,,~\text{with}~\mathbf{1}^{l}\otimes \mathbf{1}^{j}=\mathbf{1}~\text{or}~\mathbf{1}''\,,\\
&\text{or}~~k_{q^c_{3}}+k_{Q_{3}}=6\,,~\text{with}~\mathbf{1}^{l}\otimes \mathbf{1}^{j}=\mathbf{1}\,.
\end{aligned}
\end{equation}
Considering the free coupling constant associated with the term  $q^c_3Q_3H_{u/d}f_{33}(Y)$, the above three cases essentially give the same prediction for $T$. Moreover, the element $T$ would be exactly vanishing for
\begin{equation}
\begin{aligned}
  & \quad ~~ k_{q^c_{3}}+k_{Q_{3}}<0 \,, \\
  &\text{or}~~k_{q^c_{3}}+k_{Q_{3}}=1,2,3,5\,,\\
&\text{or}~~k_{q^c_{3}}+k_{Q_{3}}=4\,,~\text{with}~\mathbf{1}^{l}\otimes \mathbf{1}^{j}=\mathbf{1}'\,,\\
&\text{or}~~k_{q^c_{3}}+k_{Q_{3}}=0,6\,,~\text{with}~\mathbf{1}^{l}\otimes \mathbf{1}^{j}=\mathbf{1}'~\text{or}~\mathbf{1}''\,,
\end{aligned}
\end{equation}
up to weight 6 modular forms.

\subsection{\label{subsec:quark_mass_matrix_doublet}Possible structures of quark mass matrix}

Combining the possible forms of the submatrices $S$, $C$, $R$ and $T$ summarized in table~\ref{Tab:S} and table~\ref{Tab:C&R} , we can straightforwardly obtain the quark mass matrix. It is well established that no quark is massless, therefore we only consider the quark mass matrices with non-vanishing determinant in this paper. As a result, we find that
the quark mass matrix can take the following four possible structures with texture zeros,
\begin{equation}
\label{eq:texture1}
\begin{aligned}
&\text{Case }\mathcal{A}:~~\left( \begin{array}{ccc} \times & \times & 0 \\
\times & \times & 0 \\
0 & 0 & \times \\ \end{array} \right)\,,\quad
\text{Case }\mathcal{B}:~~\left( \begin{array}{ccc} \times & \times & \times\\
\times & \times & \times \\
0 & 0 & \times \\ \end{array} \right)\,,\\
& \text{Case }\mathcal{C}:~~ \left( \begin{array}{ccc} \times & \times & 0\\
\times & \times & 0 \\
\times & \times & \times \\ \end{array} \right)\,,\quad
\text{Case }\mathcal{D}:~~\left( \begin{array}{ccc} \times & \times & \times\\
\times & \times & \times \\
\times & \times & 0 \\ \end{array} \right)\,,
\end{aligned}
\end{equation}
where a cross denotes a non-vanishing entry. It is remarkable that the non-vanishing elements are correlated with each other in the present approach.

\section{\label{sec:QMM_s}Quark mass matrices for singlet assignments of the right-handed quarks }

In this section, we shall consider another case in which the three generations of left-handed quarks are assigned to a direct sum of doublet and singlet of $T'$ while the three generations of right-handed quarks are assumed to transform as one-dimensional representations of the $T'$ modular group, i.e.
\begin{equation}
\label{eq:assign_2nd}
Q_D\equiv\begin{pmatrix}
Q_1 \\
Q_2
\end{pmatrix}\sim\mathbf{2}^{i},\quad Q_3\sim \mathbf{1}^{j}\,, \quad
q^c_a\sim \mathbf{1}^{l_a}~~\textrm{with}~~a=1, 2, 3\,,
\end{equation}
where $i,j,l_{1,2,3}=0, 1, 2$ with $\mathbf{1}\equiv \mathbf{1}^{0}$, $\mathbf{1}^{'}\equiv \mathbf{1}^{1}$, $\mathbf{1}^{''}\equiv \mathbf{1}^{2}$ for singlet representations and $\mathbf{2}\equiv \mathbf{2}^{0}$, $\mathbf{2}^{'}\equiv \mathbf{2}^{1}$, $\mathbf{2}^{''}\equiv \mathbf{2}^{2}$ for the doublet representations. Thus the most general superpotential for the quark masses is given by
\begin{equation}
\label{eq:wq_singlet_assm}\mathcal{W}_q=\sum^{3}_{a=1}q^c_aQ_DH_{u/d}f_{aD}(Y)+q^c_aQ_3H_{u/d}f_{a3}(Y)\,,
\end{equation}
where we have suppressed all coupling constants. As a consequence, we can divide the quark mass matrix $M_q$ into six parts as follow,
\begin{equation}
\label{eq:mq_sinlets}M_q=\left( \begin{array}{cc}
R'_{1} ~&~ C'_{1} \\
R'_{2} ~&~ C'_{2} \\
R'_{3} ~&~ C'_{3}
\end{array} \right)\,,
\end{equation}
where $R'_{1,2,3}$ and $C'_{1,2,3}$ are $1\times 2$ and $1\times 1$ sub-matrices respectively, and they are determined by the modular forms $f_{aD}(Y)$ and $f_{a3}(Y)$. We shall not discuss explicitly the case where the left-handed quarks transform as three one-dimensional representations of $T'$ with the right-handed quark fields assigned to a singlet and a doublet under $T'$, since we only need to transpose the mass matrix in Eq.~\eqref{eq:mq_sinlets} to switch the transformation properties of left-handed and right-handed quarks.

\subsection{General structures of $R'_{a}$ and $C'_{a}$ }

It is easy to say that modular invariance requires the the modular form $f_{aD}(Y)$ should in the doublet representations $\mathbf{2}^{2-l_a-i~(\text{mod}~3)}$ of $T'$, and its modular weight should be $k_{q^c_a}+k_{Q_D}$. The submatrix $R'_{a}$ would be vanishing exactly, if any of the following relations is satisfied
\begin{equation}
\begin{aligned}
&\quad~~k_{q^c_{a}}+k_{Q_{D}}=0,2,4,\ldots\,,\\
&\text{or}~~k_{q^c_{a}}+k_{Q_{D}}=1,~~i+l_{a}=1, 2\,(\text{mod}~3)\,,\\
&\text{or}~~k_{q^c_{a}}+k_{Q_{D}}=3,~~i+l_{a}=1\,(\text{mod}~3)\,.
\end{aligned}
\end{equation}
Depending on the assignments $\mathbf{2}^{i}$, $\mathbf{1}^{l_a}$ and the modular weights $k_{q^c_a}$, $k_{Q_D}$, $R'_{a}$ can take the following nontrivial forms.

\begin{itemize}[labelindent=-0.8em, leftmargin=1.2em]
\item{$\mathbf{2}^{i}\otimes \mathbf{1}^{l_{a}}=\mathbf{2}$ with $i+l_{a}=0\,(\text{mod}~3)$ }

In this case, $f_{aD}(Y)$ should be in the representation $\mathbf{2}''$, and it can be either $Y^{(3)}_{\mathbf{2}''}$ or $Y^{(5)}_{\mathbf{2}''}$ up to weight 6. For $k_{q^c_{a}}+k_{Q_{D}}=3$, we can read out the submatrix $R'_{a}$ as
\begin{equation}
R'_{a,1}=(\kappa Y^{(3)}_{\mathbf{2}'',2}, -\kappa Y^{(3)}_{\mathbf{2}'',1})\,.
\end{equation}
For $k_{q^c_{a}}+k_{Q_{D}}=5$, $R'_{a}$ is of the form,
\begin{equation}
R'_{a,2}=(\kappa Y^{(5)}_{\mathbf{2}'',2}, -\kappa Y^{(5)}_{\mathbf{2}'',1})\,.
\end{equation}

\item{$\mathbf{2}^{i}\otimes \mathbf{1}^{l_{a}}=\mathbf{2}'$ with $i+l_{a}=1~(\text{mod}~3)$ }

If we only consider the modular forms with modular weight less than seven, then $f_{aD}(Y)$ would be proportional to $Y^{(5)}_{\mathbf{2}'}$, and  $R'_{a}$ takes the following form,
\begin{equation}
R'_{a,3}=(\kappa Y^{(5)}_{\mathbf{2}',2}, -\kappa Y^{(5)}_{\mathbf{2}',1})\,.
\end{equation}

\item{$\mathbf{2}^{i}\otimes \mathbf{1}^{l_{a}}=\mathbf{2}''$ with $i+l_{a}=2~(\text{mod}~3)$ }

The modular invariance requires that $f_{aD}(Y)$ should transform as the doublet representation $\mathbf{2}$ of $T'$. For $k_{q^c_{a}}+k_{Q_{D}}=1$, the quark mass Yukawa term is $\kappa q^c_aQ_DY^{(1)}_{\mathbf{2}}H_{u/d}$ such that the submatrix $R'_{a}$ is given by
\begin{equation}
R'_{a,4}=(\kappa Y^{(1)}_{\mathbf{2},2}, - \kappa Y^{(1)}_{\mathbf{2},1})\,.
\end{equation}
For $k_{q^c_{a}}+k_{Q_{D}}=3$, we can read out $R'_{a}$ as
\begin{equation}
R'_{a,5}=(\kappa Y^{(3)}_{\mathbf{2},2}, - \kappa Y^{(3)}_{\mathbf{2},1})\,.
\end{equation}
For $k_{q^c_{a}}+k_{Q_{D}}=5$, the weight 5 modular form $Y^{(5)}_{\mathbf{2}}=(Y^{(5)}_{\mathbf{2},1},Y^{(5)}_{\mathbf{2},2})^{T}$ is involved, and the submatrix $R'_{a}$ is
\begin{equation}
R'_{a,6}=(\kappa Y^{(5)}_{\mathbf{2}, 2}, - \kappa Y^{(5)}_{\mathbf{2}, 1})\,.
\end{equation}
All the above possible forms of $R'_{a}$ are summarized in table~\ref{Tab:Rap}.
\end{itemize}
Since both $q^c_a$ and $Q_3$ are assigned to singlet representations $\mathbf{1}^{l_a}$ and $\mathbf{1}^{j}$ respectively, the modular form $f_{a3}(Y)$ in the last term of Eq.~\eqref{eq:wq_singlet_assm} should transform as singlet $\mathbf{1}^{3-l_a-j~(\text{mod}~3)}$ under the $T'$ modular group. Analogous to previous cases, we find the element $C'_{a}$ would be non-zero for
\begin{equation}
\begin{aligned}
&\quad~~k_{q^c_{a}}+k_{Q_{3}}=0,~~j+l_a=0~(\text{mod}~3)\,,\\
&\text{or}~~k_{q^c_{a}}+k_{Q_{3}}=4,~~j+l_a=0, 2~(\text{mod}~3)\,,\\
&\text{or}~~k_{q^c_{a}}+k_{Q_{3}}=6,~~j+l_a=0~(\text{mod}~3)\,,
\end{aligned}
\end{equation}
otherwise $C'_{a}$ would be vanishing exactly in particular when the summation of modular weights $k_{q^c_{a}}+k_{Q_{3}}$ is an odd integer or $k_{q^c_{a}}+k_{Q_{3}}<0$.

\begin{table}[t!]
\centering
\begin{tabular}{|c|c|c|}
\hline  \hline
  & Expressions of $R'_{a}$ & Constraints \\ \hline
\multirow{4}{*}{$R'_{a,0}$} & \multirow{4}{*}{$(0,0)$}  & $\textbf{i):}~k_{q^c_{a}}+k_{Q_{D}}<0,$ \\
& &  $ \textbf{ii):}~k_{q^c_{a}}+k_{Q_{D}}=0, 2, 4, 6,\ldots,$ \\
& & $\textbf{iii):}~k_{q^c_{a}}+k_{Q_{D}}=1,~i+l_{a}=0, 1\,(\text{mod}~3)\,,$ \\
& & $\textbf{iv):}~k_{q^c_{a}}+k_{Q_{D}}=3,~i+l_{a}=1\,(\text{mod}~3)\,.$ \\ \hline
$R'_{a,1}$ & $(\kappa Y^{(3)}_{\mathbf{2}'',2}, - \kappa Y^{(3)}_{\mathbf{2}'',1})$ & $k_{q^c_{a}}+k_{Q_{D}}=3,~i+l_{a}=0\,(\text{mod}~3)$ \\ \hline
$R'_{a,2}$ & $(\kappa Y^{(5)}_{\mathbf{2}'',2}, - \kappa Y^{(5)}_{\mathbf{2}'',1})$ & $k_{q^c_{a}}+k_{Q_{D}}=5,~i+l_{a}=0\,(\text{mod}~3)$ \\ \hline
$R'_{a,3}$ & $(\kappa Y^{(5)}_{\mathbf{2}',2}, - \kappa Y^{(5)}_{\mathbf{2}',1})$ & $k_{q^c_{a}}+k_{Q_{D}}=5,~i+l_{a}=1\,(\text{mod}~3)$\\ \hline
$R'_{a,4}$ & $(\kappa Y^{(1)}_{\mathbf{2},2}, - \kappa Y^{(1)}_{\mathbf{2},1})$ & $k_{q^c_{a}}+k_{Q_{D}}=1,~i+l_{a}=2\,(\text{mod}~3)$ \\ \hline
$R'_{a,5}$ & $(\kappa Y^{(3)}_{\mathbf{2},2}, - \kappa Y^{(3)}_{\mathbf{2},1})$ & $k_{q^c_{a}}+k_{Q_{D}}=3,~i+l_{a}=2\,(\text{mod}~3)$ \\ \hline
$R'_{a,6}$ & $(\kappa Y^{(5)}_{\mathbf{2}, 2}, - \kappa Y^{(5)}_{\mathbf{2}, 1})$ & $k_{q^c_{a}}+k_{Q_{D}}=5,~i+l_{a}=2\, (\text{mod}~3)$ \\
\hline \hline
\end{tabular}
\caption{\label{Tab:Rap} The structures of the submatrix $R'_{a}$ for different possible values of modular weights and the assignments of the quark fields $q^c_a$, $Q_D$ under the finite modular group $T'$. }
\end{table}

\subsection{\label{subsec:quark_mass_matrix_singlet}Possible structures of quark mass matrix}

Given the possible forms of $R'_a$ listed in table~\ref{Tab:Rap}, we find that the quark mass matrix can take the following five possible structures with texture zeros,
\begin{eqnarray}
\nonumber &&\text{Case}~\mathcal{A}:~~\left( \begin{array}{ccc} \times & \times & 0 \\
\times & \times & 0 \\
0 & 0 & \times \\ \end{array} \right)\,,~~\quad~~
\text{Case}~\mathcal{B}:~~\left( \begin{array}{ccc} \times & \times & \times \\
\times & \times & \times \\
0 & 0 & \times \\ \end{array} \right)\,,\\
\nonumber&& \text{Case}~\mathcal{C}:~~ \left( \begin{array}{ccc} \times & \times & 0\\
\times & \times & 0 \\
\times & \times & \times \\ \end{array} \right)\,,~~\quad~~
\text{Case~}\mathcal{D}:~~\left( \begin{array}{ccc} \times & \times & \times \\
\times & \times & \times \\
\times & \times & 0 \\ \end{array} \right)\,,\\
&& \label{eq:case5_singlet}\text{Case}~\mathcal{E}:~~
\left( \begin{array}{ccc} \times & \times & 0 \\
\times & \times & \times \\
0 & 0 & \times \\ \end{array} \right)\,,
\end{eqnarray}
up to row and column permutations. The symbol "$\times$" denote nonzero matrix element, and we have neglected the quark mass matrices with zero determinant in Eq.~\eqref{eq:case5_singlet}. If the we interchange the assignments for the left-handed and right-handed quark fields in Eq.~\eqref{eq:assign_2nd}, another new texture of quark mass matrix can be obtained
\begin{equation}
\label{eq:case6_singlet}\text{Case}~\mathcal{F}:~~ \left( \begin{array}{ccc} \times & \times & 0 \\
\times & \times & 0 \\
0 & \times & \times \\ \end{array} \right)\,.
\end{equation}

\section{\label{sec:model_quark} Phenomenologically viable models for quark masses and CKM mixing}

As shown in section~\ref{sec:QMM_ds} and section~\ref{sec:QMM_s}, both up quark and down quark mass matrices can take six possible textures with zero elements: $\text{Case}~\mathcal{A}$, $\text{Case}~\mathcal{B}$, $\text{Case}~\mathcal{C}$, $\text{Case}~\mathcal{D}$, $\text{Case}~\mathcal{E}$ and $\text{Case}~\mathcal{F}$ given in Eqs.~(\ref{eq:texture1}, \ref{eq:case5_singlet}, \ref{eq:case6_singlet}), if we properly assign the modular weights of the quark fields and their transformation properties under $T'$. Combining the up quark sector with the down quark sector, we can obtain the possible up quark and down quark mass matrices predicted by $T'$ modular symmetry. We find that many cases can accommodate the experimental data on quark masses and CKM mixing matrix, and the resulting predictions for quark mass matrices can be classified according to the number of zero elements and the number of involved free parameters. In order to show concrete examples, we shall present five interesting models for quarks in the following.

\begin{description}[labelindent=-0.8em, leftmargin=0.3em]

\begin{table}[t!]
\centering
\resizebox{1.0\textwidth}{!}{
\begin{tabular}{|c|c|c|c|c|c|c|c|c|c|c|}
\hline  \hline
\multicolumn{2}{|c|}{} & $Q_{D}$ & $Q_{3}$ & $u^{c}$ & $c^{c}$ & $t^{c}$ & \multicolumn{2}{c|}{$d^{c}_{D}\equiv(d^c, s^c)$} & $b^{c}$  \\ \hline
  \multicolumn{2}{|c|}{$SU(2)_{L}\times U(1)_{Y}$} & $(2,1/6)$ & $(2,1/6)$ & $(1,-2/3)$ & $(1,-2/3)$ & $(1,-2/3)$ & \multicolumn{2}{c|}{$(1,1/3)$} & $(1,1/3)$ \\ \hline \hline
\multirow{2}{*}{\texttt{Model I}} & $T'$ & $\mathbf{2}$ & $\mathbf{1}$ & $\mathbf{1}''$ & $\mathbf{1'}$ & $\mathbf{1}''$ & \multicolumn{2}{c|}{$\mathbf{2}'$} & $\mathbf{1}$  \\ \cline{2-10}
& $k_{I}$ & $0$ & $-1$ & $3$ & $5$ & $5$ & \multicolumn{2}{c|}{$4$} & $1$ \\ \hline \hline
  \multirow{2}{*}{\texttt{Model II}}&  $T'$ & $\mathbf{2}$ & $\mathbf{1}$ & $\mathbf{1}$ & $\mathbf{1}''$ & $\mathbf{1}''$ & \multicolumn{2}{c|}{$\mathbf{2}'$} & $\mathbf{1}$ \\
\cline{2-10}
& $k_{I}$ & $0$ & $-1$ & $3$ & $3$ & $5$ & \multicolumn{2}{c|}{$4$} & $1$  \\ \hline \hline
  \multirow{2}{*}{\texttt{Model III}}&  $T'$ & $\mathbf{2}$ & $\mathbf{1}$ & $\mathbf{1}''$ & $\mathbf{1}'$ & $\mathbf{1}$ & \multicolumn{2}{c|}{$\mathbf{2}'$} & $\mathbf{1}''$ \\
\cline{2-10}
& $k_{I}$ & $0$ & $-1$ & $3$ & $5$ & $1$ & \multicolumn{2}{c|}{$4$} & $5$  \\ \hline \hline
  \multirow{2}{*}{\texttt{Model IV}} &   $T'$ & $\mathbf{2}$ & $\mathbf{1}$ & $\mathbf{1}$ & $\mathbf{1}''$ & $\mathbf{1}''$ & ~ $\mathbf{1}$ ~ & $\mathbf{1}''$ & $\mathbf{1}$  \\
\cline{2-10}
  & $k_{I}$ & $0$ & $-1$ & $5$ & $5$ & $1$ & ~$3$~ & $3$ & $1$ \\
\hline \hline
  \multirow{2}{*}{\texttt{Model V}} &   $T'$ & $\mathbf{2}$ & $\mathbf{1}$ & $\mathbf{1}$ & $\mathbf{1}''$ & $\mathbf{1}$ & ~ $\mathbf{1}$ ~ & $\mathbf{1}''$ & $\mathbf{1}$  \\
\cline{2-10}
  & $k_{I}$ & $0$ & $-1$ & $5$ & $5$ & $1$ & ~$3$~ & $3$ & $1$ \\
\hline \hline
\end{tabular}}
\caption{\label{Tab:quark_summary} The transformation properties of the quark fields under the Standard Model gauge group $SU(2)_{L}\times U(1)_{Y}$ and under $T'$ modular symmetry for different models, where $-k_{I}$ refers to the modular weights. The two Higgs doublets $H_{u,d}$ are invariant under $T'$ and their modular weights are assumed to be vanishing.}
\end{table}

\item[~~Model I: ]\textbf{6 zero elements and 10 free parameters }

The classification of the quark fields under the standard model gauge symmetry and $T'$ modular symmetry are listed in table~\ref{Tab:quark_summary}. The quark $SU(2)$ doublets are assigned to doublet and singlet $Q_{D}\sim\mathbf{2}$, $Q_3\sim\mathbf{1}$, the up type quark $SU(2)$ singlets $u^c$, $c^c$ and $t^c$ are assigned to $\mathbf{1}''$, $\mathbf{1'}$ and $\mathbf{1}''$ respectively, and the right-handed down type quarks are assumed to transform as doublet and singlet $d^{c}_D\equiv(d^c, s^c)\sim\mathbf{2}'$, $b^c\sim\mathbf{1}'$.
Then we can read out the modular invariant superpotentials for up and down sectors as follow,
\begin{equation}
\begin{aligned}
&\mathcal{W}_{u}=y_{1}^{u}u^{c}Q_{D}Y^{(3)}_{\mathbf{2}}H_{u}+y_{2}^{u}c^{c}Q_{D}Y^{(5)}_{\mathbf{2}'}H_{u}+y_{3}^{u}t^{c}Q_{D}Y^{(5)}_{\mathbf{2}}H_{u}+y_{4}^{u}t^{c}Q_{3}Y^{(4)}_{\mathbf{1}}H_{u}\,,\\
&\mathcal{W}_{d}=y_{1}^{d}d^{c}_{D}Q_{D}Y^{(4)}_{\mathbf{3}}H_{d}+y_{2}^{d}d^{c}_{D}Q_{D}Y^{(4)}_{\mathbf{1}'}H_{d}+y_{3}^{d}b^{c}Q_{3}H_{d}\,,
\end{aligned}
\end{equation}
where the coupling constant $y^{u}_{1,2,3,4}$ and $y^{d}_{1,3}$ can be taken to be real positive by rephasing the quark fields without loss of generality, while the phase of $y^{d}_{2}$ can not be removed. Applying the decomposition rules of the $T'$ tensor products in Appendix~\ref{app:Tp_group}, we find the quark mass matrices are given by
\begin{eqnarray}
\nonumber& M_{u}=\left( \begin{array}{ccc} y_{1}^{u}Y^{(3)}_{\mathbf{2},2} ~& -y_{1}^{u}Y^{(3)}_{\mathbf{2},1} ~& 0 \\
y_{2}^{u}Y^{(5)}_{\mathbf{2}',2} ~& -y_{2}^{u}Y^{(5)}_{\mathbf{2}',1} ~& 0 \\
y_{3}^{u}Y^{(5)}_{\mathbf{2},2} ~& -y_{3}^{u}Y^{(5)}_{\mathbf{2},1} ~& y_{4}^{u}Y^{(4)}_{\mathbf{1}} \\ \end{array} \right)v_u\,,\\
\label{eq:mq_Mod1}&M_{d}=\left( \begin{array}{ccc} \sqrt{2}e^{\frac{5i\pi}{12}} y^{d}_{1} Y^{(4)}_{\mathbf{3}, 1} ~&~ -y^{d}_{1}Y^{(4)}_{\mathbf{3}, 2}+y^{d}_{2}Y^{(4)}_{\mathbf{1}'} ~&~ 0 \\
-y^{d}_{1}Y^{(4)}_{\mathbf{3}, 2}-y^{d}_{2}Y^{(4)}_{\mathbf{1}'} ~&~  \sqrt{2}e^{\frac{7i\pi}{12}} y^{d}_{1} Y^{(4)}_{\mathbf{3}, 3} ~&~ 0 \\
0 ~&~ 0 ~&~ y^{d}_{3} \\ \end{array}\right)v_d\,.
\end{eqnarray}
We see that there are totally six zero entries in $M_u$ and $M_d$, and the down quark mass matrix $M_d$ is block diagonal. Apart form the dependence of $M_u$ and $M_d$ on the VEV of the complex modulus $\tau$, we have six  real input parameters $y^{u}_{1,2,3,4}$, $y^{d}_{1,3}$ and one complex parameter $y^{d}_{2}$ to describe the quark masses, mixing angles and phases.

\item[~~Model II: ]\textbf{6 zero elements and 10 free parameters }

The transformation rules of the left-handed quarks and the right-handed down quarks under the $T'$ modular symmetry are identical with those of Model I while the assignments of the right-handed up quarks are different, as shown in table~\ref{Tab:quark_summary}. The superpotential for the quark Yukawa interactions is given by
\begin{equation}
\label{eq:wq_Mod2}\begin{aligned}
&\mathcal{W}_{u}=y_{1}^{u}u^{c}Q_{D}Y^{(3)}_{\mathbf{2}''}H_{u}+y_{2}^{u}c^{c}Q_{D}Y^{(3)}_{\mathbf{2}}H_{u}+y_{3}^{u}t^{c}Q_{D}Y^{(5)}_{\mathbf{2}}H_{u}+y_{4}^{u}t^{c}Q_{3}Y^{(4)}_{\mathbf{1'}}H_{u}\,,\\
&\mathcal{W}_{d}=y_{1}^{d}d^{c}_{D}Q_{D}Y^{(4)}_{\mathbf{3}}H_{d}+y_{2}^{d}d^{c}_{D}Q_{D}Y^{(4)}_{\mathbf{1}'}H_{d}+y_{3}^{d}b^{c}Q_{3}H_{d}\,,
\end{aligned}
\end{equation}
where the phases of the couplings $y_{1, 2, 3, 4}^{u}$ and $y^{d}_{1, 3}$ can be absorbed into the quark fields, while the phase of $y_{2}^{d}$ can not be eliminated by field redefinition. The superpotential in Eq.~\eqref{eq:wq_Mod2} leads to the following up and down quark mass matrices,
\begin{eqnarray}
\nonumber& M_{u}=\left( \begin{array}{ccc} y_{1}^{u}Y^{(3)}_{\mathbf{2}'',2} ~& -y_{1}^{u}Y^{(3)}_{\mathbf{2}'',1} ~& 0 \\
y_{2}^{u}Y^{(3)}_{\mathbf{2},2} ~& -y_{2}^{u}Y^{(3)}_{\mathbf{2},1} ~& 0 \\
y_{3}^{u}Y^{(5)}_{\mathbf{2},2} ~& -y_{3}^{u}Y^{(5)}_{\mathbf{2},1} ~& y_{4}^{u}Y^{(4)}_{\mathbf{1'}} \\ \end{array} \right)v_u\,,\\
\label{eq:mq_Mod2}&M_{d}=\left( \begin{array}{ccc} \sqrt{2}e^{\frac{5i\pi}{12}} y^{d}_{1} Y^{(4)}_{\mathbf{3}, 1} ~&~ -y^{d}_{1}Y^{(4)}_{\mathbf{3}, 2}+y^{d}_{2}Y^{(4)}_{\mathbf{1}'} ~&~ 0 \\
-y^{d}_{1}Y^{(4)}_{\mathbf{3}, 2}-y^{d}_{2}Y^{(4)}_{\mathbf{1}'} ~&~  \sqrt{2}e^{\frac{7i\pi}{12}} y^{d}_{1} Y^{(4)}_{\mathbf{3}, 3} ~&~ 0 \\
0 ~&~ 0 ~&~ y^{d}_{3} \\ \end{array}\right)v_d\,.
\end{eqnarray}
which depends on ten real input parameters including the real and imaginary part of the modulus $\tau$.

\item[~~Model III: ]\textbf{6 zero elements and 10 free parameters }

  In this model, we assign the three generations of left-handed quark fields to doublet $\mathbf{2}$ and singlet $\mathbf{1}$, the right-handed up type quarks $u^c$, $c^c$ and $t^c$ transform as $\mathbf{1}''$, $\mathbf{1}'$ and $\mathbf{1}$ respectively while the three right-handed down type quarks transform as $\mathbf{2}'\oplus \mathbf{1}''$. The quark masses are described by
\begin{equation}
\begin{aligned}
&\mathcal{W}_{u}=y_{1}^{u}u^{c}Q_{D}Y^{(3)}_{\mathbf{2}}H_{u}+y_{2}^{u}c^{c}Q_{D}Y^{(5)}_{\mathbf{2}'}H_{u}+y_{3}^{u}t^{c}Q_{3}H_{u}\,,\\
&\mathcal{W}_{d}=y_{1}^{d}d^{c}_{D}Q_{D}Y^{(4)}_{\mathbf{3}}H_{d}+y_{2}^{d}d^{c}_{D}Q_{D}Y^{(4)}_{\mathbf{1}'}H_{d}+y_{3}^{d}b^{c}Q_{D}Y^{(5)}_{\mathbf{2}}H_{d}+y_{4}^{d}b^{c}Q_{3}Y^{(4)}_{\mathbf{1}'}H_{d}\,,
\end{aligned}
\end{equation}
where all coupling constants except $y^d_2$ can be taken to be real by using the freedom of field redefinition. We can read out the up and down quark mass matrices as follow,
\begin{equation}
\label{eq:mq_Mod3}M_{u}=\left(\begin{array}{ccc} y_{1}^{u}Y^{(3)}_{\mathbf{2},2} ~&~ -y_{1}^{u}Y^{(3)}_{\mathbf{2},1} ~&~ 0 \\
y_{2}^{u}Y^{(5)}_{\mathbf{2}',2} ~&~ -y_{2}^{u}Y^{(5)}_{\mathbf{2}',1}
~&~ 0 \\
0 ~&~ 0 ~&~ y_{3}^{u} \end{array} \right)v_u\,,~~
M_{d}=\left( \begin{array}{ccc} \sqrt{2}e^{\frac{5i\pi}{12}} y^{d}_{1} Y^{(4)}_{\mathbf{3}, 1} ~&~ -y^{d}_{1}Y^{(4)}_{\mathbf{3}, 2}+y^{d}_{2}Y^{(4)}_{\mathbf{1}'} ~&~ 0 \\
-y^{d}_{1}Y^{(4)}_{\mathbf{3}, 2}-y^{d}_{2}Y^{(4)}_{\mathbf{1}'} ~&~  \sqrt{2}e^{\frac{7i\pi}{12}} y^{d}_{1} Y^{(4)}_{\mathbf{3}, 3} ~&~ 0 \\
y^{d}_{3}Y^{(5)}_{\mathbf{2},2} ~&~ -y^{d}_{3}Y^{(5)}_{\mathbf{2},1} ~&~ y^{d}_{4}Y^{(4)}_{\mathbf{1}'} \\ \end{array}\right)v_d\,.
\end{equation}
We see that $M_u$ and $M_d$ are expressed in terms of eleven free input parameters: $y^{u}_{1,2,3}$, $y^{d}_{1, 3, 4}$, $|y^d_2|$, $\text{arg}(y^d_2)$, $\text{Re}\tau$, $\text{Im}\tau$. The up quark mass matrix is block diagonal in this model.

\item[~~Model IV: ]\textbf{5 zero elements and 11 free parameters }

The classification of the left-handed quark fields under the $T'$ modular symmetry is the same as that of Model III, while both up type and down type right-handed quark fields are assigned to transform as singlets of $T'$ , i.e., $u^{c}\sim \mathbf{1}$, $c^{c}\sim \mathbf{1}''$, $t^{c}\sim \mathbf{1}''$, $d^{c}\sim \mathbf{1}$, $s^{c}\sim \mathbf{1}''$ and $b^{c}\sim \mathbf{1}$ . The superpotential for quark masses is of the following form
\begin{eqnarray}
\nonumber\mathcal{W}_{u}&=&y_{1}^{u}u^{c}Q_{D}Y^{(5)}_{\mathbf{2}''}H_{u}+y_{2}^{u}c^{c}Q_{D}Y^{(5)}_{\mathbf{2}}H_{u}+ y_{3}^{u}t^{c}Q_{D}Y^{(1)}_{\mathbf{2}}H_{u}+ y_{4}^{u}u^{c}Q_{3}Y^{(4)}_{\mathbf{1}}H_{u}+y_{5}^{u}c^{c}Q_{3}Y^{(4)}_{\mathbf{1}'}H_{u}\,,\\
\mathcal{W}_{d}&=&y_{1}^{d}d^{c}Q_{D}Y^{(3)}_{\mathbf{2}''}H_{d}+y_{2}^{d}s^{c}Q_{D}Y^{(3)}_{\mathbf{2}}H_{d}+y_{3}^{d}b^{c}Q_{3}H_{d}\,,
\end{eqnarray}
where all coefficients except $y^u_5$ can be taken to be real positive. Using the decomposition rules of $T'$ group in Appendix~\ref{app:Tp_group}, we obtain
\begin{equation}
\label{eq:mq_Mod4}M_{u}=\left( \begin{array}{ccc} y_{1}^{u}Y^{(5)}_{\mathbf{2}'',2} ~&~ -y_{1}^{u}Y^{(5)}_{\mathbf{2}'',1} ~&~ y_{4}^{u}Y^{(4)}_{\mathbf{1}} \\
y_{2}^{u}Y^{(5)}_{\mathbf{2},2} ~&~ -y_{2}^{u}Y^{(5)}_{\mathbf{2},1} ~&~ y_{5}^{u}Y^{(4)}_{\mathbf{1}'} \\
y_{3}^{u}Y^{(1)}_{\mathbf{2}, 2} ~&~ -y_{3}^{u}Y^{(1)}_{\mathbf{2}, 1} ~&~ 0 \\ \end{array} \right)v_u\,,~~~~
M_{d}=\left( \begin{array}{ccc} y^{d}_{1} Y^{(3)}_{\mathbf{2}'', 2} ~&~ -y^{d}_{1}Y^{(3)}_{\mathbf{2}'', 1} ~&~ 0 \\
y^{d}_{2}Y^{(3)}_{\mathbf{2},2} ~&~  -y^{d}_{2} Y^{(3)}_{\mathbf{2},1} ~&~ 0 \\
0 ~&~ 0 ~&~ y^{d}_{3} \\ \end{array}\right)v_d\,,
\end{equation}
which involve eleven free real parameters including the real and imaginary parts of the modulus $\tau$.

\item[~~Model V: ]\textbf{4 zero elements and 11 free parameters}

  In this model, the assignments of quark fields are almost same as those of Model IV except for the transformation rule of $t^{c}$, as shown in table~\ref{Tab:quark_summary}. The modular invariant superpotential in the quark sector is
\begin{eqnarray}
\nonumber\mathcal{W}_{u}&=&y_{1}^{u}u^{c}Q_{D}Y^{(5)}_{\mathbf{2}''}H_{u}+y_{2}^{u}c^{c}Q_{D}Y^{(5)}_{\mathbf{2}}H_{u}+ y_{3}^{u}u^{c}Q_{3}Y^{(4)}_{\mathbf{1}}H_{u}+y_{4}^{u}c^{c}Q_{3}Y^{(4)}_{\mathbf{1}'}H_{u}+ y_{5}^{u}t^{c}Q_{3}H_{u}\,,\\
\mathcal{W}_{d}&=&y_{1}^{d}d^{c}Q_{D}Y^{(3)}_{\mathbf{2}''}H_{d}+y_{2}^{d}s^{c}Q_{D}Y^{(3)}_{\mathbf{2}}H_{d}+y_{3}^{d}b^{c}Q_{3}H_{d}\,,
\end{eqnarray}
where the coupling constants $y^{u}_{1, 2, 3,5}$ and $y^d_{1, 2,3}$ can be taken to be real and positive without loss of generality while the phase of $y^u_4$ can not be absorbed into quark fields. The up and down quark mass matrices are given as
\begin{equation}
\label{eq:mq_Mod5}M_{u}=\left( \begin{array}{ccc} y_{1}^{u}Y^{(5)}_{\mathbf{2}'',2} ~&~ -y_{1}^{u}Y^{(5)}_{\mathbf{2}'',1} ~&~ y_{3}^{u}Y^{(4)}_{\mathbf{1}} \\
y_{2}^{u}Y^{(5)}_{\mathbf{2},2} ~&~ -y_{2}^{u}Y^{(5)}_{\mathbf{2},1} ~&~ y_{4}^{u}Y^{(4)}_{\mathbf{1}'} \\
0 ~&~ 0 ~&~ y_{5}^{u} \\ \end{array} \right)v_u\,,~~~~
M_{d}=\left( \begin{array}{ccc} y^{d}_{1} Y^{(3)}_{\mathbf{2}'', 2} ~&~ -y^{d}_{1}Y^{(3)}_{\mathbf{2}'', 1} ~&~ 0 \\
y^{d}_{2}Y^{(3)}_{\mathbf{2},2} ~&~  -y^{d}_{2} Y^{(3)}_{\mathbf{2},1} ~&~ 0 \\
0 ~&~ 0 ~&~ y^{d}_{3} \\ \end{array}\right)v_d\,.
\end{equation}

\end{description}

\subsection{\label{subsec:numerical_quark}Numerical results }

\begin{table}[t!]
\centering
\begin{tabular}{|c|c|} \hline  \hline
Parameters & $\mu_i\pm1\sigma$ \\ \hline
$y_{u}/10^{-6}$ & $2.73325\pm 0.84731$ \\
$y_{c}/10^{-3}$ & $1.41719\pm 0.04960$ \\
$y_{t}$ & $0.50232 \pm 0.01200$ \\
$y_{d}/10^{-6}$ & $5.12495 \pm 0.56374$ \\
$y_{s}/10^{-4}$ & $1.01438 \pm 0.05478$ \\
$y_{b}/10^{-3}$ & $5.56096 \pm 0.06103$ \\
$y_{e}/10^{-6}$ & $2.07526 \pm 0.01245$ \\
$y_{\mu}/10^{-4}$ & $4.38107 \pm 0.02629$ \\
$y_{\tau}/10^{-3}$ & $7.48026 \pm 0.03898$ \\ \hline
$\theta_{12}^{q}$ & $0.22736 \pm 0.00073$ \\
$\theta_{13}^{q}/10^{-2}$ & $0.34938 \pm 0.01258$ \\
$\theta_{23}^{q}$ & $0.04015 \pm 0.00064 $ \\
$\delta_{CP}^{q}/^{\circ}$ & $69.21330 \pm 3.11460 $ \\\hline
$\sin^{2}\theta_{12}^{l}$ & $0.310_{-0.012}^{+0.013}$ \\
$\sin^{2}\theta_{23}^{l}$ & $0.563_{-0.024}^{+0.018}$ \\
$\sin^{2}\theta_{13}^{l}$ & $0.02237_{-0.00065}^{+0.00066}$ \\
$\delta_{CP}^{l}/^{\circ}$ & $221_{-28}^{+39}$ \\
$\frac{\Delta m_{21}^{2}}{10^{-5} \text{eV}^{2}}$ & $7.39_{-0.20}^{+0.21}$ \\
$\frac{\Delta m_{31}^{2}}{10^{-3} \text{eV}^{2}}$ & $2.528_{-0.031}^{+0.029}$ \\
\hline \hline
\end{tabular}
\caption{\label{Tab:parameter_values} The best fit values $\mu_i$ and $1\sigma$ uncertainties of the quark and lepton parameters when evolve to the GUT scale as calculated in~\cite{Antusch:2013jca}, with the SUSY breaking scale $M_{\text{SUSY}}=10$ TeV and $\tan\beta=10$, where the error widths represent $1\sigma$ intervals. The quark masses are given as $m_{u,c,t}=y_{u,c,t}v_u$ and  $m_{d, s, b}=y_{d, s, b}v_d$. The values of lepton mixing angles, leptonic Dirac CP violation phases $\delta^{l}_{CP}$ and the neutrino mass squared difference are taken from NuFIT 4.1~\cite{Esteban:2018azc}.}
\end{table}

In previous work, the texture zero structure of the quark mass matrices is an assumption or the zero entry is highly suppressed when imposing certain flavor symmetry, and the mass matrix is assumed to be symmetric or hermitian. In the present formalism, the texture zero is exactly dictated by the modular symmetry, and the nonvaninshing entries are correlated. Therefore modular symmetry provides a natural framework to realize texture zero. Moreover, it is remarkable that the down quark mass matrix $M_d$ is block diagonal, the (13), (23), (31) and (32) elements of $M_d$ are vanishing exactly for all the five models summarized in table~\ref{Tab:quark_summary}. As a consequence, the small off-diagonal entries $V_{ub}$, $V_{cb}$, $V_{td}$ and $V_{ts}$ of the CKM matrix completely arise from the up type quark sector.

The above predictions for up and down quark mass matrices $M_u$ and $M_d$ in Eqs.~(\ref{eq:mq_Mod1}, \ref{eq:mq_Mod2}, \ref{eq:mq_Mod3}, \ref{eq:mq_Mod4}, \ref{eq:mq_Mod5}) are given at the scale where the modulus $\tau$ obtains the vacuum expectation value. Here we assume that the flavor symmetry breaking scale is very large around the grand unified theory (GUT) scale $2\times10^{16}$ GeV. For each set of given values of the coupling constants and the modulus $\tau$, we numerically diagonalize $M_u$ and $M_d$, subsequently we extract the predictions for quark masses and CKM mixing matrix which may be compared with their GUT scale values. It is well-known that the quark masses and mixing parameters at GUT scale can be obtained from the measured values at low energy experiments by performing renormalization group evolution (RGE). The running of best-fit and error values to the GUT scale are generally dependent on SUSY parameters, primarily depends on the SUSY breaking scale $M_{\mathrm{SUSY}}$ and $\tan\beta$. In our numerical analysis, we shall extract the GUT scale values of all Yukawa couplings and the CKM parameters from~\cite{Antusch:2013jca}, assuming a SUSY breaking scale $M_{\mathrm{SUSY}}=10$ TeV and $\tan\beta=10$. The central values as well as the $1\sigma$ uncertainty ranges of different observables used in our analysis are listed in table~\ref{Tab:parameter_values}.

Following Ref.~\cite{Feruglio:2017spp} and other papers in the literature, we shall treat the VEV of the complex modulus $\tau$ as a free parameter. The modulus value is expected to be dynamically fixed as a minimum of scalar potential in supergravity theory. The modulus stabilization could be achieved in $SL(2,Z)$ modular invariant theories after the non-perturbative superpotential in supergravity theory is taken into account~\cite{Ferrara:1990ei,Cvetic:1991qm}. It has been shown that modulus stabilization can also be realized in the bottom-up modular symmetry models in a similar manner~\cite{Kobayashi:2019xvz}.

In order to determine the optimum values of the input parameters for which the experimental data can be accommodated best, we define a $\chi^2$ function
\begin{equation}
\chi^2=\sum^{n}_{i=1}\left(\frac{P_i(x_1, x_2,\ldots, x_m)-\mu_i}{\sigma_i}\right)^2\,,
\end{equation}
where $P_i$ are predictions for the physical observables derived from the up type and down type quark mass matrices $M_u$ and $M_d$ as complex nonlinear
functions of the free parameters of the model, $\mu_i$ and $\sigma_i$ are the GUT scale central values and $1\sigma$ deviations respectively of the corresponding quantities listed in table~\ref{Tab:parameter_values}. The measured values of the top and bottom quark masses can be reproduced exactly by properly choosing the values of the overall parameters $y^u_1v_u$ and $y^d_1v_d$. Hence we include the mass ratios $m_u/m_c$, $m_c/m_t$, $m_d/m_s$ and $m_s/m_b$ instead of quark masses $m_{u,c,t}$ and $m_{d,s,b}$ individually. We take the complex modulus $\tau$ as random complex number in the fundamental domain $\{\tau: |\text{Re}\tau|\leq1/2, \text{Im}\tau>0, |\tau|\geq1\}$. The absolute values of all coupling constants are scanned in the region $[0, 10^6]$ while the phases are freely varied in the range $[0, 2\pi]$. The function $\chi^2$ is numerically minimized by using the minimization algorithms incorporated in the package MINUIT developed by CERN to determine the best fit values of the input parameters. From the best fit values of the input parameters, one can obtain the corresponding values of quark masses and CKM mixing matrix.

\begin{table}
\centering
\resizebox{1.0\textwidth}{!}{
\begin{tabular}{|c|c|c|}
  \hline  \hline
  \multirow{5}*{\texttt{Model I } } & \multirow{3}*{\texttt{Input }}  &
\begin{tabular}{ccccc}
  $\text{Re}\tau$ & $\text{Im}\tau$ & $y^{u}_{2}/y^{u}_{1}$ & $y^{u}_{3}/y^{u}_{1}$ & $y^{u}_{4}/y^{u}_{1}$  \\ \hline
0.03428&2.49711&3.05239&125.82400&759.06706 \\
\end{tabular}\\
\cline{3-3}
&  &
\begin{tabular}{ccccc}
 $|y^{d}_{2}/y^{d}_{1}| $ & $\text{arg}(y^{d}_{2}/y^{d}_{1})/\pi$ & $y^{d}_{3}/y^{d}_{1}$ & $y_{1}^{u}v_{u}/\text{GeV}$ & $y_{1}^{d}v_{d}/\text{GeV}$ \\ \hline
 3.59552&1.13410&1.00599&6.59371&0.96242  \\
\end{tabular}\\
  \cline{2-3}
&  \texttt{Obs } &
\begin{tabular}{cccccccc}
$m_{u}/m_{c}$ & $m_{c}/m_{t}$ & $m_{d}/m_{s}$ & $m_{s}/m_{b}$ &$\theta_{12}^{q}$ & $\theta_{13}^{q}$  & $\theta_{23}^{q}$ & $\delta_{CP}^{q}/^{\circ}$  \\
\hline
0.00197&0.00280&0.05007&0.01824&0.22732&0.00354&0.03890&70.20896  \\
\end{tabular}\\
  \hline  \hline
 \multirow{5}*{\texttt{Model II } } & \multirow{3}*{\texttt{Input }} &
\begin{tabular}{ccccc}
  $\text{Re}\tau$ & $\text{Im}\tau$ & $y^{u}_{2}/y^{u}_{1}$ & $y^{u}_{3}/y^{u}_{1}$ & $y^{u}_{4}/y^{u}_{1}$  \\
 \hline
0.03504&2.45660&712.70240&82120.92932&$7.67910\times 10^{6}$ \\
\end{tabular}\\
\cline{3-3}
& &
\begin{tabular}{ccccc}
 $|y^{d}_{2}/y^{d}_{1}| $ & $\text{arg}(y^{d}_{2}/y^{d}_{1})/\pi$ & $y^{d}_{3}/y^{d}_{1}$ & $y_{1}^{u}v_{u}/\text{GeV}$ & $y_{1}^{d}v_{d}/\text{GeV}$  \\
 \hline
3.27938&1.13321&1.00207&0.00932&0.96618 \\
\end{tabular}\\
  \cline{2-3}
    &  \texttt{Obs} &
\begin{tabular}{cccccccc}
 $m_{u}/m_{c}$ & $m_{c}/m_{t}$ & $m_{d}/m_{s}$ & $m_{s}/m_{b}$ & $\theta_{12}^{q}$ & $\theta_{13}^{q}$  & $\theta_{23}^{q}$ & $\delta_{CP}^{q}/^{\circ}$  \\
\hline
0.00197&0.00282&0.04977&0.01830&0.22718&0.00356&0.03586&69.37967  \\
\end{tabular}\\
  \hline
  \hline
  \multirow{5}*{\texttt{Model III }  } & \multirow{3}*{\texttt{Input }} &
\begin{tabular}{ccccc}
$\text{Re}\tau$ & $\text{Im}\tau$ & $y^{u}_{2}/y^{u}_{1}$ & $y^{u}_{3}/y^{u}_{1}$ & $|y^{d}_{2}/y^{d}_{1}|$  \\
 \hline
 0.03527&2.45148&2.85554&13.51464&3.23841   \\
\end{tabular}\\
  \cline{3-3}
& &
\begin{tabular}{ccccc}
  $\text{arg}(y^{d}_{2}/y^{d}_{1})/\pi $ & $y^{d}_{3}/y^{d}_{1}$ & $y^{d}_{4}/y^{d}_{1}$ & $y_{1}^{u}v_{u}/\text{GeV}$ & $y_{1}^{d}v_{d}/\text{GeV}$ \\
 \hline
 1.13453 & 8.91712&831.75015&6.47117&0.94256 \\
\end{tabular}\\
  \cline{2-3}
& \texttt{Obs } &
 \begin{tabular}{cccccccc}
$m_{u}/m_{c}$ & $m_{c}/m_{t}$ & $m_{d}/m_{s}$ & $m_{s}/m_{b}$ & $\theta_{12}^{q}$ & $\theta_{13}^{q}$  & $\theta_{23}^{q}$ & $\delta_{CP}^{q}/^{\circ}$  \\
\hline
0.00223&0.00275&0.04968&0.01784&0.22729&0.00356&0.03558&69.02269 \\
\end{tabular}\\
 \hline  \hline
  \multirow{5}*{\texttt{Model IV }} & \multirow{3}*{\texttt{Input }} &
\begin{tabular}{cccccc}
$\text{Re}\tau$ & $\text{Im}\tau$ & $y^{u}_{2}/y^{u}_{1}$ & $y^{u}_{3}/y^{u}_{1}$ & $y^{u}_{4}/y^{u}_{1}$ & $|y^{u}_{5}/y^{u}_{1}|$  \\
 \hline
 0.04374&2.08683&235.73307&0.00453&94.95425&9006.45768  \\
\end{tabular}\\
  \cline{3-3}
 & &
\begin{tabular}{ccccc}
$\text{arg}(y^{u}_{5}/y^{u}_{1})/\pi $ & $y^{d}_{2}/y^{d}_{1}$ & $y^{d}_{3}/y^{d}_{1}$ & $y_{1}^{u}v_{u}/\text{GeV}$ & $y_{1}^{d}v_{d}/\text{GeV}$ \\
 \hline
1.75886&27.41769&56.40063&3.65402&0.01717 \\
\end{tabular}\\
  \cline{2-3}
& \texttt{Obs }&
\begin{tabular}{cccccccc}
$m_{u}/m_{c}$ & $m_{c}/m_{t}$ & $m_{d}/m_{s}$ & $m_{s}/m_{b}$ &  $\theta_{12}^{q}$ & $\theta_{13}^{q}$  & $\theta_{23}^{q}$ & $\delta_{CP}^{q}/^{\circ}$ \\
\hline
0.00191&0.00282&0.05024&0.01823&0.22740&0.00349&0.04037&70.55234 \\
\end{tabular}\\
 \hline  \hline
  \multirow{5}*{\texttt{Model V } } & \multirow{3}*{\texttt{Input }} &
\begin{tabular}{cccccc}
 $\text{Re}\tau$ & $\text{Im}\tau$ & $y^{u}_{2}/y^{u}_{1}$ & $y^{u}_{3}/y^{u}_{1}$ & $|y^{u}_{4}/y^{u}_{1}|$ & $\text{arg}(y^{u}_{4}/y^{u}_{1})/\pi$  \\
 \hline
0.04369&2.08551&233.40867&94.45134&8951.18611&1.76267  \\
\end{tabular}\\
  \cline{3-3}
& &
\begin{tabular}{ccccc}
$y^{u}_{5}/y^{u}_{1} $ & $y^{d}_{2}/y^{d}_{1}$ & $y^{d}_{3}/y^{d}_{1}$ & $y_{1}^{u}v_{u}/\text{GeV}$ & $y_{1}^{d}v_{d}/\text{GeV}$ \\
 \hline
0.03654&27.38334&56.40088&3.66649&0.01717 \\
\end{tabular}\\
  \cline{2-3}
&  \texttt{Obs } &
\begin{tabular}{cccccccc}
 $m_{u}/m_{c}$ & $m_{c}/m_{t}$ & $m_{d}/m_{s}$ & $m_{s}/m_{b}$ & $\theta_{12}^{q}$ & $\theta_{13}^{q}$  & $\theta_{23}^{q}$ & $\delta_{CP}^{q}/^{\circ}$ \\
\hline
0.00192&0.00280&0.05030&0.01821&0.22731&0.00351&0.04011&70.97865 \\
\end{tabular}\\
\hline  \hline
\end{tabular}}
\caption{\label{Tab:predictions_quark_only} The best fit values of the input parameters, quark mass ratios and the CKM parameters for the five models summarized in table~\ref{Tab:quark_summary}, where \texttt{Obs} is the abbreviation of observable. The fit is performed for the GUT scale extrapolated data given in table~\ref{Tab:parameter_values}. Notice that the measured values of the top quark mass $m_t$ and bottom quark mass $m_b$ are reproduced exactly here. }
\end{table}

As shown in section~\ref{sec:model_quark}, \texttt{Model I}, \texttt{Model II} and \texttt{Model III} have ten free parameters and the remaining \texttt{Model IV} and \texttt{Model V} have eleven free parameters. We shall fit all the ten or eleven free parameters of each model using the ten observables including six quark masses $m_{u,c,t}$, $m_{d,s,b}$, three quark mixing angles $\theta^{q}_{12}$, $\theta^q_{13}$, $\theta^q_{23}$ and one quark CP violation phase $\delta^q_{CP}$. The results of the fitting are shown in table~\ref{Tab:predictions_quark_only}. We see that all the five models can give very good fit to the data, almost all observables in quark sector lie within their $1\sigma$ ranges. When extending these models to include the lepton sector in the following section, we fix the complex modulus $\tau$ to be the best fit values in table~\ref{Tab:predictions_quark_only}  which are determined from the precisely measured quark masses and CKM mixing matrix.

\section{\label{sec:QLU}Quark lepton unification }

Inspired by the success of $T'$ modular symmetry in explaining the quark masses and CKM mixing matrix, we shall extend this framework to the lepton sector. Since the solar mixing angle $\theta^l_{12}$ and atmospheric mixing angle $\theta^l_{23}$ are large, and the reactor mixing angle $\theta^l_{13}$ is of the same order as the Cabibbo angle, the lepton mixing matrix doesn't have a hierarchical structure. Therefore we shall not use the doublet representations of $T'$ to distinguish the first two generation leptons from the third generation. The three generations of left-handed lepton doublets $L=(L_1, L_2, L_3)^T$ are assigned to be in a triplet $\mathbf{3}$ under $T'$, while the right-handed charged leptons $e^{c}$, $\mu^c$ and $\tau^c$ are assumed to transform as singlet representations $\mathbf{1}$, $\mathbf{1}'$ or $\mathbf{1}''$ of $T'$ modular group. For the irreducible representations $\mathbf{3}$, $\mathbf{1}$, $\mathbf{1}'$ and $\mathbf{1}''$, two distinct elements of $T'$ group are described by the same matrices that represent the elements in $A_4$. Therefore the group $T'$ can not be distinguished from $A_4$ when working with these representations. As a consequence, the flavor symmetry is essentially the $A_4$ modular group in the lepton sector. A systematical classification of lepton models with $A_4$ modular symmetry has been performed in~\cite{Ding:2019zxk}. In the present work, we shall focus on two economical models which are named as $\mathcal{L}_1$ and $\mathcal{L}_2$, the model $\mathcal{L}_2$ is identical to the lepton model $\mathcal{D}_6$ in~\cite{Ding:2019zxk}. The transformation properties
of the lepton chiral superfields and the right-handed neutrinos $N^c=(N^c_1, N^c_2, N^c_3)^T$ under $T'$ and their modular weights are summarized in table~\ref{Tab:lepton_assignment_D6D9}. Note that the Higgs doublets $H_{u,d}$ transform trivially under $T'$ with zero modular weight.

\begin{table}[t!]
\centering
\begin{tabular}{|c|c|c|c|c|c|c|}\hline  \hline
\multicolumn{2}{|c|}{} & $L$ & $e^{c}$ & $\mu^{c}$ & $\tau^{c}$ & $N^{c}$ \\ \hline
\multicolumn{2}{|c|}{$SU(2)_{L}\times U(1)_{Y}$} & $(2, -1/2)$ & $(1, 1)$ & $(1, 1)$ & $(1, 1)$ & $(1, 0)$  \\ \hline  \hline
\multirow{2}{*}{$\mathcal{L}_1$} & $T'$ & $\mathbf{3}$ & $\mathbf{1}$ & $\mathbf{1}'$ & $\mathbf{1}''$ & $\mathbf{3}$  \\ \cline{2-7}
 &  $k_{I}$ & $1$ & $1$ & $1$ & $3$ & $1$  \\ \hline  \hline
\multirow{2}{*}{$\mathcal{L}_2$} & $T'$ & $\mathbf{3}$ & $\mathbf{1}'$ & $\mathbf{1}'$ & $\mathbf{1}$ & $\mathbf{3}$  \\ \cline{2-7}
& $k_{I}$ & $1$ & $1$ & $3$ & $1$ & $1$  \\
\hline \hline
\end{tabular}
\caption{\label{Tab:lepton_assignment_D6D9}
The transformation properties of the lepton chiral superfields under the Standard Model gauge group $SU(2)_{L}\times U(1)_{Y}$ and under $T'$ modular symmetry for the models $\mathcal{D}_6$ and $\mathcal{D}_9$ of~\cite{Ding:2019zxk}, where $-k_{I}$ refers to the modular weights. }
\end{table}

\subsection{Lepton sector}

In the following, we shall present the neutrino and charged lepton mass terms and the corresponding lepton mass matrices for the models $\mathcal{L}_1$ and $\mathcal{L}_2$. The neutrino masses are assumed to be generated from the type I seesaw mechanism. As listed in table~\ref{Tab:lepton_assignment_D6D9}, the right-handed neutrinos $N^c$ compose a $T'$ triplet $\mathbf{3}$.

\begin{itemize}[labelindent=-0.8em, leftmargin=1.2em]
\item{$\mathcal{L}_{1}$ lepton model}

In this case, the modular invariant superpotential in the lepton sector reads
\begin{subequations}
\begin{align}
\label{subeq:We}\mathcal{W}_{e}&=\alpha e^{c}(LY^{(2)}_{\mathbf{3}})_{\mathbf{1}}H_{d}+\beta \mu^{c}(LY^{(2)}_{\mathbf{3}})_{\mathbf{1}''}H_{d}+\gamma \tau^{c}(LY^{(4)}_{\mathbf{3}})_{\mathbf{1}'}H_{d}\,,\\
\label{subeq:Wnu}\mathcal{W}_{\nu}&=g_{1}((N^{c}L)_{\mathbf{3}_{S}}Y^{(2)}_{\mathbf{3}})_{\mathbf{1}}H_{u}+g_{2}((N^{c}L)_{\mathbf{3}_{A}}Y^{(2)}_{\mathbf{3}})_{\mathbf{1}}H_{u}+\Lambda ((N^{c}N^{c})_{\mathbf{3}_{S}}Y^{(2)}_{\mathbf{3}})_{\mathbf{1}}\,.
\end{align}
\end{subequations}
The resulting charged lepton mass matrix $M_e$, the Dirac neutrino
mass matrix $M_D$ as well as the Majorana mass matrix $M_N$ for heavy neutrinos are given by
\begin{eqnarray}
\nonumber &M_{e}=\left( \begin{array}{ccc}  \alpha Y^{(2)}_{\mathbf{3},1} ~&~
\alpha Y^{(2)}_{\mathbf{3},3} ~&~  \alpha Y^{(2)}_{\mathbf{3},2} \\
\beta  Y^{(2)}_{\mathbf{3}, 3} & \beta Y^{(2)}_{\mathbf{3}, 2}  ~&~ \beta Y^{(2)}_{\mathbf{3}, 1} \\
\gamma Y^{(4)}_{\mathbf{3},2} ~&~  \gamma Y^{(4)}_{\mathbf{3},1} ~&~  \gamma Y^{(4)}_{\mathbf{3},3} \\ \end{array} \right)v_d\,,~~~
M_{N}= \left( \begin{array}{ccc} 2Y^{(2)}_{\mathbf{3},1} ~&~ -Y^{(2)}_{\mathbf{3},3} ~&~ -Y^{(2)}_{\mathbf{3},2} \\
-Y^{(2)}_{\mathbf{3},3} ~&~ 2Y^{(2)}_{\mathbf{3},2} ~&~ -Y^{(2)}_{\mathbf{3},1} \\
-Y^{(2)}_{\mathbf{3},2} ~&~ -Y^{(2)}_{\mathbf{3},1} ~&~ 2Y^{(2)}_{\mathbf{3},3} \\ \end{array} \right)\Lambda \,,\\
\label{eq:Me_MD_MN_D6}&M_{D}=\left( \begin{array}{ccc} 2g_{1}Y^{(2)}_{\mathbf{3},1} ~&~ (-g_{1}+g_{2}) Y^{(2)}_{\mathbf{3},3}  ~&~  (-g_{1}-g_{2})Y^{(2)}_{\mathbf{3},2} \\
(-g_{1}-g_{2}) Y^{(2)}_{\mathbf{3},3} ~&~ 2g_{1}Y^{(2)}_{\mathbf{3},2} ~&~ (-g_{1}+g_{2}) Y^{(2)}_{\mathbf{3},1} \\
(-g_{1}+g_{2}) Y^{(2)}_{\mathbf{3},2} ~&~ (-g_{1}-g_{2}) Y^{(2)}_{\mathbf{3},1} ~&~ 2g_{1}Y^{(2)}_{\mathbf{3},3}
\end{array} \right)v_u\,.
\end{eqnarray}
The light neutrino mass matrix is given by the seesaw formula $m_{\nu}=-M^T_DM^{-1}_NM_D$. The phases of the parameters $\alpha$, $\beta$ and $\gamma$ can be absorbed into the right-handed charged
lepton fields, and the measured charged lepton masses can be reproduced by adjusting their values. The light neutrino mass matrix $m_{\nu}$ only depends on the complex parameter $g_2/g_1$ and modulus $\tau$ besides the overall scale $g^2_1v^2_u/\Lambda$. It is found that the modulus $\tau$ lies in narrow regions in order to be compatible with the experimental data on lepton mixing angles and neutrino masses~\cite{Ding:2019zxk}.

\item{$\mathcal{L}_{2}$ lepton model}

From table~\ref{Tab:lepton_assignment_D6D9}, we see that the models $\mathcal{L}_1$ and $\mathcal{L}_2$ differ in the representation assignments for the right-handed charged leptons, while the transformation properties of $L$ and $N^c$ are exactly the same in the two models. The superpotential for the charged lepton masses is given by
\begin{equation}
\mathcal{W}_{e}=\alpha e^{c}(LY^{(2)}_{\mathbf{3}})_{\mathbf{1}''}H_{d}+\beta \mu^{c}(LY^{(4)}_{\mathbf{3}})_{\mathbf{1}''}H_{d}+\gamma \tau^{c}(LY^{(2)}_{\mathbf{3}})_{\mathbf{1}}H_{d}\,,
\end{equation}
which leads to the following charged lepton mass matrix
\begin{equation}
M_{e}=\left( \begin{array}{ccc}  \alpha Y^{(2)}_{\mathbf{3},3} ~&~  \alpha Y^{(2)}_{\mathbf{3},2} ~&~  \alpha Y^{(2)}_{\mathbf{3},1} \\
\beta  Y^{(4)}_{\mathbf{3}, 3} ~&~ \beta Y^{(4)}_{\mathbf{3}, 2}  ~&~ \beta Y^{(4)}_{\mathbf{3}, 1} \\
\gamma Y^{(2)}_{\mathbf{3},1} ~&~  \gamma Y^{(2)}_{\mathbf{3},3}  ~&~  \gamma Y^{(2)}_{\mathbf{3},2} \end{array} \right)\,.
\end{equation}
The superpotential $\mathcal{W}_{\nu}$ for neutrino sector coincides with that of Eq.~\eqref{subeq:Wnu}, and the relevant neutrino mass matrices $M_D$ and $M_{N}$ are given by Eq.~\eqref{eq:Me_MD_MN_D6}.

\end{itemize}

\subsection{Numerical results }

We can combine the benchmark quark models \texttt{Model I}, \texttt{Model II}, \texttt{Model III}, \texttt{Model IV} and \texttt{Model V} presented in section~\ref{sec:model_quark} with the $\mathcal{L}_1$ and $\mathcal{L}_2$ lepton models to give a unified description of both quark and lepton sectors. Notice that the modulus $\tau$ in the quark and lepton mass matrices should be identical, and we transmit the best fit value of $\tau$ in table~\ref{Tab:predictions_quark_only} obtained from quark sector to the lepton sector. We perform a global fit to the complete models including both quark and lepton sectors, the models would be viable if the observed quark and lepton masses and mixing parameters can be accommodated for certain values of input parameters. It is remarkable that the resulting models have less free input parameters than the number of observable quantities including quark and lepton masses and mixing parameters. Hence it is highly nontrivial that the model can successfully fit the data. In particular, the light neutrino mass matrix would only depend on a complex parameter $g_2/g_1$ and a overall factor $g^2_1v^2_u/\Lambda$. Consequently three light neutrino masses, three lepton mixing angles and three CP violation phases are pinned down by three real input parameters $|g_2/g_1|$, $\text{arg}(g_2/g_1)$ and $g^2_1v^2_u/\Lambda$. Thus the models could make six non-trivial predictions which agree well with the present data. The extrapolated values of the charged lepton masses at the GUT scale are listed in table~\ref{Tab:parameter_values} for $\tan\beta=10$ and $M_{\text{SUSY}}=10$ TeV. For the neutrino masses and mixing angles, we use the latest results of the global data analysis from NuFIT 4.1~\cite{Esteban:2018azc}, the RGE effects from the low scale to the GUT scale are neglected, since the running of the neutrino masses and mixing angles in the MSSM is known to be negligible in the case of $\tan\beta\leq30$. It is found that the RGE dependence can really be safely neglected in a sizable region of the $\tan\beta-M_{SUSY}$ plane~\cite{Criado:2018thu}.

Since the global analysis of neutrino oscillation data favors normal mass ordering over inverted ordering~\cite{Esteban:2018azc}, we shall assume normal ordering neutrino masses in the $\chi^2$ analysis. Similar to section~\ref{subsec:numerical_quark}, the absolute value of each coupling constant freely varies between $0$ and $10^6$, all phases are treated as random numbers in the range of $0$ and $2\pi$, and the complex modulus $\tau$ is scanned in the fundamental domain of the modular group. The results of the fit are shown in table~\ref{Tab:QL_fit}. We see that very good fit to the data is obtained, and all observables from the model fall well into the experimentally allowed regions. From the fitted values of the input parameters, we can further obtain predictions for the unmeasured observables such as the Dirac CP phase $\delta^l_{CP}$ and Majorana CP phase $\alpha_{21}$ and $\alpha_{31}$ in the lepton sector, the lightest neutrino mass $m_1$, and the effective  Majorana mass $|m_{ee}|$ in neutrinoless double beta decay, as shown in table~\ref{Tab:QL_fit}.

\begin{figure}[t!]
\begin{center}
\includegraphics[width=0.95\linewidth]{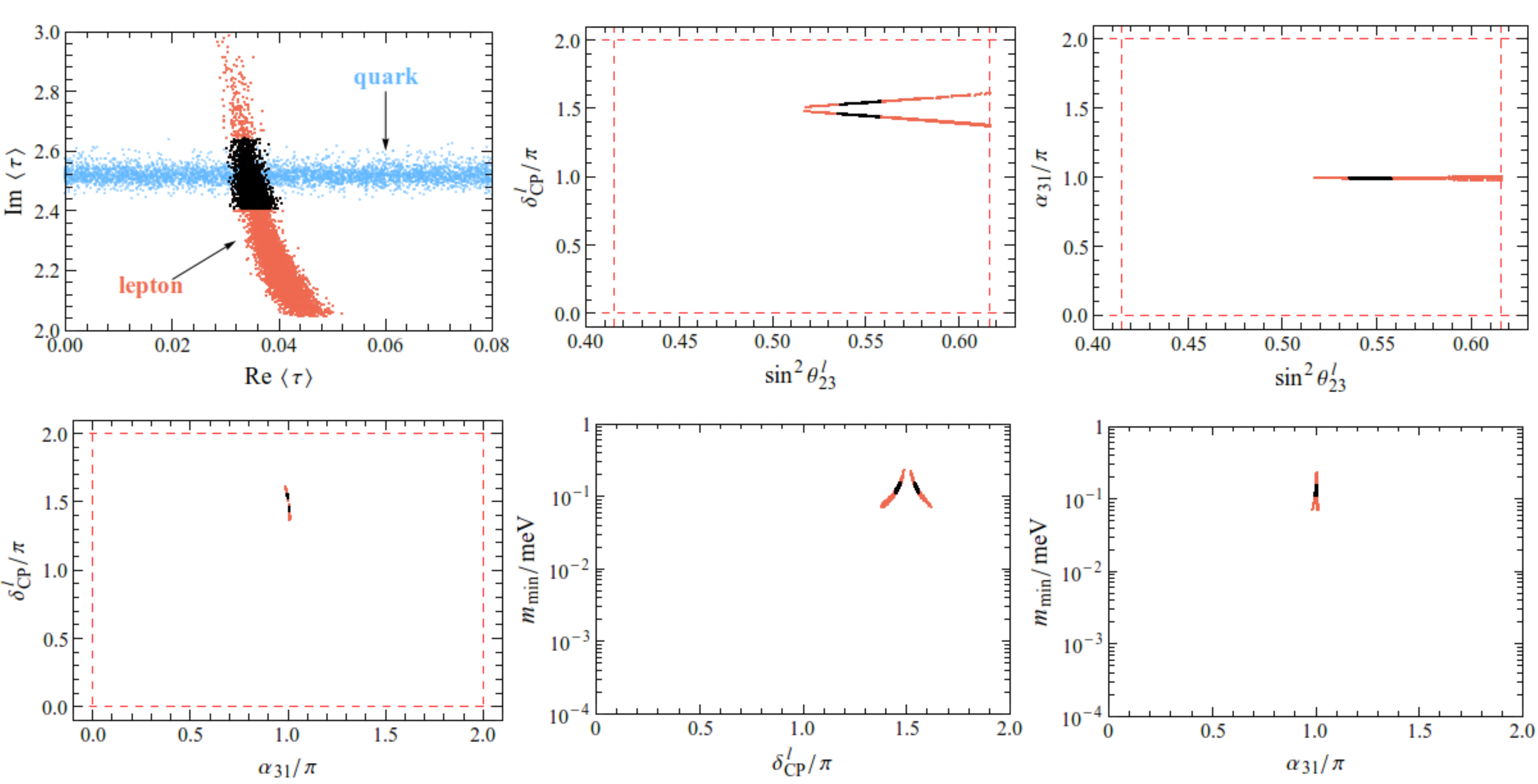}
\end{center}
\caption{\label{fig:QL_ModelID6}The allowed region of $\tau$ by the experimental data in the $\text{Re}\tau-\text{Im}\tau$ plane and the correlation between different observables. The quark masses and mixing parameters can be obtained in the blue region while the lepton masses and  parameters can be obtained in the orange region. The red dashed lines denote the experimentally preferred $3\sigma$ ranges adapted from~\cite{Esteban:2018azc}.}
\end{figure}

As an example, we consider the scenario that the \texttt{Model I} in quark sector is combined with the lepton model $\mathcal{D}_6$. We have 16 real input parameters $|y^{u}_{1,2,3,4}|$, $|y^{d}_{1,3}|$, $|y^{d}_{2}|$, $\text{arg}(y^{d}_{2})$, $|\alpha|$, $|\beta|$, $|\gamma|$, $|g_1|$, $|g_2|$, $\mathrm{arg}(g_2)$, $\mathrm{Re}\tau$ and $\mathrm{Im}\tau$ to describe the 18 measured quantities including the quark masses $m_{u,c,t}$ and $m_{d,s,b}$, the CKM mixing parameters $\theta^{q}_{12}$, $\theta^q_{13}$, $\theta^q_{23}$ and $\delta^q_{CP}$, the charged lepton masses $m_{e,\mu,\tau}$, the neutrino mass-squared splittings $\Delta m^2_{21}$ and $\Delta m^2_{31}$, and the lepton mixing angles $\theta^l_{12}$, $\theta^l_{13}$ and $\theta^{l}_{23}$.  From table~\ref{Tab:parameter_values}, we see that the charged fermion masses and the CKM mixing matrix has been measured very precisely and their $1\sigma$ ranges are quite narrow. Hence the input parameters are also constrained to be in very narrow regions to accommodate the experimental data. We find that the allowed regions of $\tau$ by the experimental data of quark and lepton sectors have overlapping area and excellent agreement with the measured values of the observables in table~\ref{Tab:parameter_values} can be achieved, as shown in figure~\ref{fig:QL_ModelID6}. In order to make the common region of $\tau$ visible, we require that the quark mass ratios $y_c/y_t$, $y_d/y_s$, $y_s/y_b$ and the CP violating phase $\delta^q_{CP}$ lie in the $2\sigma$ intervals, the quark mixing angles $\theta^q_{12}$, $\theta^q_{13}$, $\theta^q_{23}$ and the charged lepton mass ratio $y_e/y_{\mu}$, $y_{\mu}/y_{\tau}$ in the $5\sigma$ intervals, and the three neutrino mixing angles and neutrino masses squared differences in their $3\sigma$ ranges. It is notable that all the quark and lepton mixing parameters are predicted to lie in narrow regions. In particular, the atmospheric mixing angle $\theta_{23}$ is in the second octant, and the lepton Dirac CP phase $\delta^{l}_{CP}$ is determined to be around $3\pi/2$.

\begin{table}[t!]
\centering
\resizebox{1.0\textwidth}{!}{
\begin{tabular}{|c|c|c|}
\hline  \hline

\multirow{5}*{\texttt{Model I+}$\mathcal{L}_1$} & \texttt{Input } &
\begin{tabular}{cccccc}
$\beta/\alpha$ & $\gamma/ \alpha$ & $|g_{2}/g_{1}|$ & $\text{arg}(g_{2}/g_{1})/\pi$  & $\alpha v_{d}/\text{GeV}$ & $(g_{1}^{2}v_{u}^{2}/\Lambda)/\text{eV}$ \\\hline
3600.66055&1898.95134&1.11486&0.37185&0.00325&0.58722 \\
\end{tabular}\\ \cline{2-3}
& \multirow{3}*{\texttt{Obs } }&
\begin{tabular}{cccccc}
$m_{e}/m_{\mu}$ & $m_{\mu}/m_{\tau}$ & $m_{1}/\text{eV}$ & $m_{2}/\text{eV}$ & $m_{3}/\text{eV}$ & $|m_{ee}|/\text{eV}$\\
\hline
 0.00474&0.05857&0.13019&0.13048&0.13955&0.13035 \\
\end{tabular}\\ \cline{3-3}
& &
\begin{tabular}{cccccc}
$\sin^{2}\theta_{12}^{l}$ & $\sin^{2}\theta_{13}^{l}$ & $\sin^{2}\theta_{23}^{l}$ & $\delta_{CP}^{l}/^{\circ}$ & $\alpha_{21}/^{\circ}$ & $\alpha_{31}/^{\circ}$ \\ \hline
0.30390&0.02228&0.54711&261.66339&0.54808&180.62051 \\
\end{tabular}\\ \hline  \hline

\multirow{5}*{\texttt{Model II+}$\mathcal{L}_1$} &  \texttt{Input } &
\begin{tabular}{cccccc}
$\beta/\alpha$ & $\gamma/ \alpha$ & $|g_{2}/g_{1}|$ & $\text{arg}(g_{2}/g_{1})/\pi$  & $\alpha v_{d}/\text{GeV}$ & $(g_{1}^{2}v_{u}^{2}/\Lambda)/\text{eV}$ \\ \hline
3585.24816&1893.19896&1.11643&0.37294&0.00327&0.55907 \\
\end{tabular}\\
  \cline{2-3}
& \multirow{3}*{\texttt{Obs }}&
\begin{tabular}{cccccc}
  $m_{e}/m_{\mu}$ & $m_{\mu}/m_{\tau}$ & $m_{1}/\text{eV}$ & $m_{2}/\text{eV}$ & $m_{3}/\text{eV}$ &  $|m_{ee}|/\text{eV}$\\ \hline
 0.00475&0.05864&0.12393&0.12423&0.13368&0.12407 \\
\end{tabular}\\ \cline{3-3}
& &
\begin{tabular}{cccccc}
 $\sin^{2}\theta_{12}^{l}$ & $\sin^{2}\theta_{13}^{l}$ & $\sin^{2}\theta_{23}^{l}$ & $\delta_{CP}^{l}/^{\circ}$ & $\alpha_{21}/^{\circ}$ & $\alpha_{31}/^{\circ}$ \\ \hline
 0.29874&0.02190&0.55119&260.84074&0.60930&180.67191 \\
\end{tabular}\\ \hline  \hline

\multirow{5}*{\texttt{Model III+}$\mathcal{L}_1$} & \texttt{Input } &
\begin{tabular}{cccccc}
$\beta/\alpha$ & $\gamma/ \alpha$ & $|g_{2}/g_{1}|$ & $\text{arg}(g_{2}/g_{1})/\pi$  & $\alpha v_{d}/\text{GeV}$ & $(g_{1}^{2}v_{u}^{2}/\Lambda)/\text{eV}$ \\\hline
 3592.54592&1898.54478&1.11717&0.37262&0.00326&0.55611 \\
\end{tabular}\\ \cline{2-3}
& \multirow{3}*{\texttt{Obs } }&
\begin{tabular}{cccccc}
$m_{e}/m_{\mu}$ & $m_{\mu}/m_{\tau}$ & $m_{1}/\text{eV}$ & $m_{2}/\text{eV}$ & $m_{3}/\text{eV}$ & $|m_{ee}|/\text{eV}$\\
\hline
 0.00474&0.05868&0.12326&0.12356&0.13307&0.12341 \\
\end{tabular}\\ \cline{3-3}
& &
\begin{tabular}{cccccc}
$\sin^{2}\theta_{12}^{l}$ & $\sin^{2}\theta_{13}^{l}$ & $\sin^{2}\theta_{23}^{l}$ & $\delta_{CP}^{l}/^{\circ}$ & $\alpha_{21}/^{\circ}$ & $\alpha_{31}/^{\circ}$ \\ \hline
0.30802&0.02205&0.55173&260.76865&0.61384&180.67073  \\
\end{tabular}\\\hline  \hline
\multirow{5}*{\texttt{Model IV+}$\mathcal{L}_2$} & \texttt{Input } &
\begin{tabular}{cccccc}
$\beta/\alpha$ & $\gamma/ \alpha$ & $|g_{2}/g_{1}|$ & $\text{arg}(g_{2}/g_{1})/\pi$  & $\alpha v_{d}/\text{GeV}$ & $(g_{1}^{2}v_{u}^{2}/\Lambda)/\text{eV}$ \\ \hline
 4.24588&0.00031&1.15637&0.37586&0.00326&0.35546 \\
\end{tabular}\\ \cline{2-3}
& \multirow{3}*{\texttt{Obs } }&
\begin{tabular}{cccccc}
$m_{e}/m_{\mu}$ & $m_{\mu}/m_{\tau}$ & $m_{1}/\text{eV}$ & $m_{2}/\text{eV}$ & $m_{3}/\text{eV}$  & $|m_{ee}|/\text{eV}$\\
\hline
0.00474&0.05858&0.07848&0.07896&0.09315&0.07846  \\
\end{tabular}\\ \cline{3-3}
 & &
\begin{tabular}{cccccc}
 $\sin^{2}\theta_{12}^{l}$ & $\sin^{2}\theta_{13}^{l}$ & $\sin^{2}\theta_{23}^{l}$ & $\delta_{CP}^{l}/^{\circ}$ & $\alpha_{21}/^{\circ}$ & $\alpha_{31}/^{\circ}$ \\ \hline
0.30248&0.02219&0.58084&250.10633&1.49641&181.28459  \\
\end{tabular}\\ \hline  \hline
\multirow{5}*{\texttt{Model V+}$\mathcal{L}_2$ } & \texttt{Input } &
\begin{tabular}{cccccc}
$\beta/\alpha$ & $\gamma/ \alpha$ & $|g_{2}/g_{1}|$ & $\text{arg}(g_{2}/g_{1})/\pi$  & $\alpha v_{d}/\text{GeV}$ & $(g_{1}^{2}v_{u}^{2}/\Lambda)/\text{eV}$ \\\hline
4.22492&0.00031&1.15711&0.37543&0.00326&0.35468  \\
\end{tabular}\\ \cline{2-3}
& \multirow{3}*{\texttt{Obs }}&
\begin{tabular}{cccccc}
$m_{e}/m_{\mu}$ & $m_{\mu}/m_{\tau}$ & $m_{1}/\text{eV}$ & $m_{2}/\text{eV}$ & $m_{3}/\text{eV}$ & $|m_{ee}|/\text{eV}$\\ \hline
0.00474&0.05855&0.07831&0.07878&0.09299&0.07828 \\
\end{tabular}\\ \cline{3-3}
& &
\begin{tabular}{cccccc}
$\sin^{2}\theta_{12}^{l}$ & $\sin^{2}\theta_{13}^{l}$ & $\sin^{2}\theta_{23}^{l}$ & $\delta_{CP}^{l}/^{\circ}$ & $\alpha_{21}/^{\circ}$ & $\alpha_{31}/^{\circ}$ \\ \hline
0.30324&0.02235&0.58126&250.12705&1.49671&181.28533 \\
\end{tabular}\\
\hline  \hline
\end{tabular}}
\caption{\label{Tab:QL_fit} The best fit values of the input parameters, the charged lepton mass ratios, the light neutrino masses and the lepton mixing parameters for the quark-lepton unified models, where \texttt{Obs} is the abbreviation of observable. The modulus $\tau$ is taken to be the best fit values in table~\ref{Tab:predictions_quark_only} which give rise to realistic values of quark masses and CKM mixing matrix. The fit is performed for the GUT scale extrapolated data given in table~\ref{Tab:parameter_values}. Notice that the measured value tau lepton mass $m_{\tau}$ are reproduced exactly here. }
\end{table}

\section{\label{sec:non-minimal-Kahler} Non-minimal K\"ahler potential}

In the bottom-up  models with modular symmetry, the finite modular symmetry is not very restrictive for the K\"ahler potential. It is shown that the K\"ahler potential is not completely fixed by the modular symmetry~\cite{Chen:2019ewa}, and the same holds true in discrete flavor symmetry models~\cite{Chen:2012ha,Chen:2013aya}. Besides the minimal K\"ahler potential chosen in Eq.~\eqref{eq:kahler_min}, the K\"ahler potential can receive unsuppressed contributions from modular forms, and many other terms are also compatible with the modular symmetry.
In the quark sector, the most general K\"ahler potential can be written as
\begin{equation}
\mathcal{K}=\mathcal{K}_{\text{min}}+\Delta\mathcal{K}\,,
\end{equation}
with
\begin{equation}
\mathcal{K}_{\text{min}}=(-i\tau+i\bar{\tau})^{-k_{Q_i}}Q^{\dagger}_iQ_i+ (-i\tau+i\bar{\tau})^{-k_{q^c_i}}{q^c_i}^{\dagger}q^c_i\,,
\end{equation}
and
\begin{align}
\nonumber
\Delta\mathcal{K} &= \sum_{k_1,k_2,\mathbf{r}_1,\mathbf{r}_2} \lambda^{ij}_Q (-i\tau+i\bar{\tau})^{-k_{Q_i}+k_1} (Q^\dagger_i Y^{(k_1) \dagger}_{\mathbf{r}_1}Y^{(k_2)}_{\mathbf{r}_2} Q_j )_\mathbf{1}\\
&+ \sum_{k_3,k_4,\mathbf{r}_1,\mathbf{r}_2} \lambda^{ij}_{q^c} (-i\tau+i\bar{\tau})^{-k_{q^c_i}+k_3} ( q^{c \dagger}_i Y^{(k_3) \dagger}_{\mathbf{r}_1}Y^{(k_4)}_{\mathbf{r}_2} q^c_j)_\mathbf{1}+\mathrm{h.c.}~~\,.
\label{eq:Delta_Kahler}
\end{align}
Note that there are generally a few independent contractions into $T'$ singlets for each operator, and the modular weights should satisfy the following constraints
\begin{equation}
-k_{Q_i}+k_1=-k_{Q_j}+k_2,\quad -k_{q^c_i}+k_3=-k_{q^c_j}+k_4\,.
\end{equation}
The weight $k$ of the modular forms $Y^{(k)}_{\mathbf{r}}$ can run from $1$ to $+\infty$, consequently infinity possible terms could be present in the most general K\"ahler potential. The presence of free coefficients $\lambda^{ij}_Q$ and $\lambda^{ij}_{q^c}$ accompanying each operator in Eq.~\eqref{eq:Delta_Kahler} could reduce the predictivity of modular invariance models, and the impacts of these additional terms on the predictions for both masses and mixing parameters are sizable, as shown in~\cite{Chen:2019ewa}. After the complex modulus $\tau$ acquires vacuum expectation value $\langle\tau\rangle$, this additional terms $\Delta\mathcal{K}$ lead to departures from the canonical K\"ahler metric, i.e.
\begin{align}
\label{KahlerMarices}
& \mathscr{K}^{ij}_Q= \frac{\partial^2 \mathcal{K}}{\partial Q^\dagger_i \partial Q_j} = -\langle i\tau+i\bar{\tau}\rangle ^{-k_{Q_i}} \delta^{ij} + \Delta \mathcal{K}^{ij}_{Q}(\langle\tau\rangle )\,, \\
& \mathscr{K}^{ij}_{q^c}= \frac{\partial^2 \mathcal{K}}{\partial q^{c \dagger}_i \partial q^c_j} = -\langle i\tau+i\bar{\tau}\rangle ^{-k_{q^c_i}} \delta^{ij} + \Delta \mathcal{K}^{ij}_{q^c}(\langle\tau\rangle )\,.
\end{align}
Here both K\"ahlier metric matrices $\mathscr{K}_Q$ and $\mathscr{K}_{q^c}$ are hermitian and positive, and they can be rewritten as
\begin{align}
\label{eq:KQ_udc}\mathscr{K}_Q=\Sigma_Q D^2_Q \Sigma^\dagger_Q  ,\quad \mathscr{K}_{u^c}=\Sigma_{u^c}  D^2_{u^c} \Sigma^\dagger_{u^c}, \quad \mathscr{K}_{d^c}=\Sigma_{d^c}  D^2_{d^c} \Sigma^\dagger_{d^c}\,,
\end{align}
where $\Sigma_{Q, u^c, d^c}$ are unitary matrices and $D_{Q, u^c, d^c}$ are diagonal matrices with positive non-vanishing entries. In order to obtain a canonical K\"ahler metric $\mathscr{K}^{ij} \rightarrow \widehat{\mathscr{K}}^{ij}=\delta^{ij}$ such that the quark fields are canonically normalized, we should perform field redefinitions
\begin{align}
\nonumber
& Q=H_Q \hat{Q},\quad~~~~H_Q=\Sigma_Q  D^{-1}_Q \Sigma^\dagger_Q\,, \\
\nonumber
&u^c=H_{u^c} \hat{u}^c,\quad~~~ H_{u^c}=\Sigma_{u^c} D^{-1}_{u^c} \Sigma^\dagger_{u^c}\,,\\
\label{eq:field_redef_quark}&d^c=H_{d^c} \hat{d}^c,\quad~~~H_{d^c}=\Sigma_{d^c} D^{-1}_{d^c} \Sigma^\dagger_{d^c}\,,
\end{align}
where $H_{Q, u^c, d^c}$ are hermitian. If $D_{Q, u^c, d^c}$ or $\Sigma_{Q, u^c, d^c}$ are proportional to the unit matrix, we see that one should rescale the quark fields to get canonical kinetic terms, and the effect of such rescaling can be absorbed into the arbitrary coefficients of the model.
The field redefinition in Eq.~\eqref{eq:field_redef_quark} would affect the superpotential or equivalently the Yukawa coupling matrices,
\begin{equation}
\mathcal{W}= u^{c\,T} Y_u Q H_u + d^{c\,T}Y_d Q H_d\rightarrow \widehat{\mathcal{W}}=\hat{u}^{c\,T} \hat{Y}_u  \hat{Q} H_u + \hat{d}^{c\,T} \hat{Y}_d \hat{Q} H_d\,,
\end{equation}
with
\begin{equation}
\hat{Y}_u = H_{u^c}^T Y_u H_Q,~~\quad~~\hat{Y}_d = H_{d^c}^T Y_d H_Q\,.
\end{equation}
Diagonalizing the quark mass matrices in the basis of canonically normalized fields, we can obtain quark mass eigenvalues and CKM matrix,
\begin{equation}
\label{eq:CKM_new}\hat{U}_u^\dagger \hat{Y}_u^\dagger \hat{Y}_u \hat{U}_u = \text{diag}(y^2_u,~y^2_c,~y^2_t),~~~~~~\hat{U}_d^\dagger \hat{Y}_d^\dagger \hat{Y}_d \hat{U}_d = \text{diag}(y^2_d,~y^2_s,~y^2_b),~~~~~~ V_{\text{CKM}}=\hat{U}^\dagger_u \hat{U}_d\,.
\end{equation}
Since $D_{Q,u,d}$ is generally not proportional to unit matrix because of the non-diagonal terms $\Delta \mathcal{K}^{ij}_{Q}(\langle\tau\rangle )$ and $\Delta \mathcal{K}^{ij}_{q^c}(\langle\tau\rangle )$ in the K\"ahler metrics, the additional term $\Delta\mathcal{K}$ in the K\"ahler potential can
lead to nontrival effects on the observables, and it might significantly modify the predictions of a model~\cite{Chen:2019ewa,Chen:2012ha,Chen:2013aya}.

In the present work, we have focused on doublet plus singlet assignment $Q_D\sim \mathbf{2}^i, Q_3\sim \mathbf{1}^j$ for the left-handed quarks. From Eq.~\eqref{eq:Delta_Kahler}, we know the most general form of the extra K\"ahler potential is given by
\begin{align}
\nonumber
\Delta \mathcal{K}_Q &=\sum_{k_1,k_2,k_3,\mathbf{r}_1,\mathbf{r}_2} (-i\tau+i\bar{\tau})^{-k_{Q_D}+k_1} (Q_D^\dagger Y^{(k_1) \dagger}_{\mathbf{r}_1} Q_D Y^{(k_1)}_{\mathbf{r}_2})_\mathbf{1} \\
\nonumber&\qquad\qquad +(-i\tau+i\bar{\tau})^{-k_{Q_3}+k_1} Q_3^\dagger  Q_3 (Y^{(k_1) \dagger}_{\mathbf{r}_1} Y^{(k_1)}_{\mathbf{r}_2})_\mathbf{1}\\
\label{eq:DeltaK_Q}&\qquad\qquad +(-i\tau+i\bar{\tau})^{-k_{Q_D}+k_2} (Q_D^\dagger Y^{(k_2) \dagger}_{\mathbf{r}_1} Q_3 Y^{(k_3) }_{\mathbf{r}_2})_\mathbf{1}+\mathrm{h.c.}\,,
\end{align}
where we suppress all couplings in front of each operator, and the modular weights fulfill $k_2-k_3=k_{Q_D}-k_{Q_3}$. In a similar fashion, if the right-handed quarks transform as a doublet and a singlet $q^c_D \sim\mathbf{2}^k, q^c_3 \sim \mathbf{1}^l$ under $T'$, the K\"ahler potential $\Delta \mathcal{K}_{q^c}$ is of the following form
\begin{align}
\nonumber
\Delta \mathcal{K}_{q^c} &=\sum_{k_1,k_2,k_3, \mathbf{r}_1,\mathbf{r}_2} (-i\tau+i\bar{\tau})^{-k_{q^c_D}+k_1} (q^{c \dagger}_D Y^{(k_1) \dagger}_{\mathbf{r}_1} q^c_D Y^{(k_1)}_{\mathbf{r}_2})_\mathbf{1} \\
\nonumber&\qquad\qquad +(-i\tau+i\bar{\tau})^{-k_{q^c_3}+k_1} q^{c \dagger}_3 q^c_3 (Y^{(k_1) \dagger}_{\mathbf{r}_1}  Y^{(k_1)}_{\mathbf{r}_2})_\mathbf{1}\\
\label{eq:DeltaK_qc1}&\qquad\qquad +(-i\tau+i\bar{\tau})^{-k_{q^c_D}+k_2} (q^{c \dagger}_D Y^{(k_2) \dagger}_{\mathbf{r}_1} q^c_3 Y^{(k_3)}_{\mathbf{r}_2})_\mathbf{1} +\mathrm{h.c.} \,,
\end{align}
with $k_2-k_3=k_{q^c_D}-k_{q^c_3}$. For the singlet assignment $q^c_a \sim \mathbf{1}^{l_a}$ ($a=1, 2, 3$), $\Delta\mathcal{K}_{q^c}$ reads as
\begin{align}
\label{eq:DeltaK_qc2}\Delta \mathcal{K}_{q^c} &=\sum_{a,b, k_1,k_2,\mathbf{r}_1,\mathbf{r}_2} (-i\tau+i\bar{\tau})^{-k_{q^c_a}+k_1} (q^{c \dagger}_a Y^{(k_1) \dagger}_{\mathbf{r}_1} q^c_b Y^{(k_2)}_{\mathbf{r}_2})_\mathbf{1}+\mathrm{h.c.} \,,
\end{align}
with $k_1-k_2=k_{q^c_a}-k_{q^c_b}$. Since the new parameters appearing in the K\"ahler potential considerably reduce the predictability of the approach~\cite{Chen:2019ewa}, a better understanding of the K\"ahler potential is required. However, the problem of better controlling the K\"ahler potential remains an open question. In some more complete setting such as string theory, the K\"ahler potential is constrained to a better extent~\cite{Dixon:1989fj}, nevertheless the complete expressions  for higher-order terms are still unknown. Recently a new approach based on top-down model building in string theory has been proposed~\cite{Baur:2019kwi,Baur:2019iai,Nilles:2020nnc,Nilles:2020kgo}, it is a hybrid of the top-down and bottom-up approaches to modular flavor
symmetry~\cite{Nilles:2020nnc}. This new approach combines the traditional flavor symmetry and the modular flavor symmetry, and put forward the concept of eclectic flavor groups. This new scheme severely restricts the representations and modular weights of matter fields, such that both superpotential potential and K\"ahler potential are strongly constrained~\cite{Nilles:2020kgo}. In order to show the impact of the additional terms in the K\"ahler potential, we give a concrete example in the following. The transformation properties of the quark fields under the $SU(2)_{L}\times U(1)_{Y}$ gauge group and $T'$ modular symmetry are summarized in table~\ref{Tab:quark_modification}. The modular invariant superpotentials for up and down quark Yukawa couplings read as,
\begin{equation}
\begin{aligned}
&\mathcal{W}_{u}=y_{1}^{u}u^{c}Q_{D}Y^{(3)}_{\mathbf{2}}H_{u}+y_{2}^{u}c^{c}Q_{D}Y^{(1)}_{\mathbf{2}}H_{u}+y_{3}^{u}t^{c}Q_{3}H_{u}\,,\\
&\mathcal{W}_{d}=y_{1}^{d}d^{c}_{D}Q_{D}Y^{(4)}_{\mathbf{3}}H_{d}+y_{2}^{d}d^{c}_{D}Q_{D}Y^{(4)}_{\mathbf{1}'}H_{d}+y_{3}^{d}b^{c}Q_{3}H_{d}\,,
\end{aligned}
\end{equation}
\begin{table}[t!]
\centering
\resizebox{1.0\textwidth}{!}{
\begin{tabular}{|c|c|c|c|c|c|c|c|c|c|c|}
\hline  \hline
\multicolumn{2}{|c|}{} & $Q_{D}$ & $Q_{3}$ & $u^{c}$ & $c^{c}$ & $t^{c}$ & \multicolumn{2}{c|}{$d^{c}_{D}\equiv(d^c, s^c)$} & $b^{c}$  \\ \hline
\multicolumn{2}{|c|}{$SU(2)_{L}\times U(1)_{Y}$} & $(2,1/6)$ & $(2,1/6)$ & $(1,-2/3)$ & $(1,-2/3)$ & $(1,-2/3)$ & \multicolumn{2}{c|}{$(1,1/3)$} & $(1,1/3)$ \\ \hline \hline
\multirow{2}{*}{\texttt{Model VI}} & $T'$ & $\mathbf{2}$ & $\mathbf{1}$ & $\mathbf{1}''$ & $\mathbf{1}''$ & $\mathbf{1}$ & \multicolumn{2}{c|}{$\mathbf{2}'$} & $\mathbf{1}$  \\ \cline{2-10}
& $k_{I}$ & $0$ & $-1$ & $3$ & $1$ & $1$ & \multicolumn{2}{c|}{$4$} & $1$ \\
\hline \hline
\end{tabular}}
\caption{\label{Tab:quark_modification} The transformation properties of the quark fields under the Standard Model gauge group $SU(2)_{L}\times U(1)_{Y}$ and under $T'$ modular symmetry for the example model in which the contribution of non-minimal K\"ahler potential is considered, where $-k_{I}$ refers to the modular weights. The two Higgs doublets $H_{u,d}$ are invariant under $T'$ and their modular weights are vanishing. }
\end{table}
where the coupling constants $y^{u}_{1,2,3}$ and $y^{d}_{1,3}$ can be taken to be real and positive by rephasing the quark fields, while $y^{d}_{2}$ is generically a complex number. Applying the decomposition rules of the $T'$ tensor products in Appendix~\ref{app:Tp_group}, we find the quark mass matrices are given by
\begin{eqnarray}
\nonumber& M_{u}=\left( \begin{array}{ccc} y_{1}^{u}Y^{(3)}_{\mathbf{2},2} ~& -y_{1}^{u}Y^{(3)}_{\mathbf{2},1} ~& 0 \\
y_{2}^{u}Y^{(1)}_{\mathbf{2},2} ~& -y_{2}^{u}Y^{(1)}_{\mathbf{2},1} ~& 0 \\
0 ~& 0 ~& y_{3}^{u} \\ \end{array} \right)v_u\,,\\
&M_{d}=\left( \begin{array}{ccc} \sqrt{2}e^{\frac{5i\pi}{12}} y^{d}_{1} Y^{(4)}_{\mathbf{3}, 1} ~&~ -y^{d}_{1}Y^{(4)}_{\mathbf{3}, 2}+y^{d}_{2}Y^{(4)}_{\mathbf{1}'} ~&~ 0 \\
-y^{d}_{1}Y^{(4)}_{\mathbf{3}, 2}-y^{d}_{2}Y^{(4)}_{\mathbf{1}'} ~&~  \sqrt{2}e^{\frac{7i\pi}{12}} y^{d}_{1} Y^{(4)}_{\mathbf{3}, 3} ~&~ 0 \\
0 ~&~ 0 ~&~ y^{d}_{3} \\ \end{array}\right)v_d\,.
\end{eqnarray}
We see that both the up quark mass matrix $M_u$ and the down quark mass matrix $M_d$ are block diagonal. As a result, only the Cabibbo angle could be accommodated, while the other two quark mixing angles $\theta_{13}^{q}$, $\theta_{23}^{q}$ and the CP-violation phase $\delta_{CP}^{q}$ are vanishing if only the superpotential is considered. Now we proceed to consider the contribution of the K\"ahler potential. With the transformations properties of quark fields in table~\ref{Tab:quark_modification}, we can use Eqs.~(\ref{eq:DeltaK_Q},\ref{eq:DeltaK_qc1},\ref{eq:DeltaK_qc2}) and straightforwardly write down the most general K\"ahler potential $\mathcal{K}$. As an illustration example, we only consider one extra term of K\"ahler potential in right-hand up quark sector.
\begin{align}
\label{eq:Kahler_add_uc}\Delta \mathcal{K}_{u^c} &= \lambda_{u^{c}}(-i\tau+i\bar{\tau})^{1} u^{c \dagger} (Y^{(4) \dagger}_{\mathbf{3}} Y^{(2)}_{\mathbf{3}})_{\mathbf{1}''}t^c+\mathrm{h.c.} \,,
\end{align}
where $\lambda_{u^c}$ is an arbitrary complex number. The minimal K\"ahler potential $\mathcal{K}^{u^{c}}_{\text{min}}$ is
\begin{align}
\mathcal{K}^{u^c}_{\text{min}} &= (-i\tau+i\bar{\tau})^{-3} (u^{c \dagger} u^c)_\mathbf{1}+ (-i\tau+i\bar{\tau})^{-1} (c^{c \dagger} c^c)_\mathbf{1}+(-i\tau+i\bar{\tau})^{-1} (t^{c \dagger} t^c)_\mathbf{1}\,.
\end{align}
Then we can find the K\"ahler metric $\mathscr{K}_{u^{c}}$ as follow,
\begin{equation}
\mathscr{K}_{u^c}=\begin{pmatrix}
\langle-i\tau +i\bar{\tau}\rangle^{-3} & 0 & 0\\
0 & \langle-i\tau +i\bar{\tau}\rangle^{-1} & 0 \\
0 & 0 & \langle-i\tau +i\bar{\tau}\rangle^{-1}
\end{pmatrix} + \langle-i\tau +i\bar{\tau}\rangle^{1}\begin{pmatrix}
0 & 0 & \lambda_{u^{c}}Y \\
0 & 0  & 0 \\
\lambda_{u^{c}}^{*} Y^{*} & 0 & 0
\end{pmatrix}+\dots\,,
\end{equation}
where $Y=Y^{(4)*}_{\mathbf{3},3}Y^{(2)}_{\mathbf{3},2}+Y^{(4)*}_{\mathbf{3},1}Y^{(2)}_{\mathbf{3},3}+Y^{(4)*}_{\mathbf{3},2}Y^{(2)}_{\mathbf{3},1}$.
We can rescale the up type quark superfields $u^{c}\rightarrow \braket{-i\tau +i\bar{\tau}}^{3/2}u^{c}$, $c^{c}\rightarrow \braket{-i\tau +i\bar{\tau}}^{1/2}c^{c}$ and $t^{c}\rightarrow \braket{-i\tau +i\bar{\tau}}^{1/2}t^{c}$ which can be absorbed into the arbitrary coefficients in this model. Then the K\"ahler metrix $\mathcal{K}_{u^{c}}$ simplifies into
\begin{equation}
\mathscr{K}_{u^c}=\begin{pmatrix}
1 & 0 & 0\\
0 & 1 & 0 \\
0 & 0 & 1
\end{pmatrix} + \begin{pmatrix}
 0 & ~0 ~& \lambda'_{u^{c}} \\
0 & ~0  ~& 0 \\
(\lambda'_{u^{c}})^{*}  & ~0 ~& 0
\end{pmatrix}+\dots\,.
\end{equation}
where $\lambda_{u^{c}}'\equiv \langle-i\tau +i\bar{\tau}\rangle^{3}\lambda_{u^{c}}Y$.
The K\"ahler metric is not diagonal anymore because of the additional term in Eq.~\eqref{eq:Kahler_add_uc} so that mixture between the third and the first two generations can be induced. A numerical analysis is performed, we find that the correct values of quark masses and quark mixing angles and CP violation phase can be obtained. The numerical fitting results are shown in table~\ref{Tab:predictions_quark_modification}.
The non-minimal K\"ahler potential in the lepton sector is discussed in Appendix~\ref{app:non-minimal-Kahler-lepton}. 

\begin{table}
\centering
\resizebox{1.0\textwidth}{!}{
\begin{tabular}{|c|c|c|}
  \hline  \hline
  \multirow{5}*{\texttt{Model VI } } & \multirow{3}*{\texttt{Input }} &
\begin{tabular}{cccccc}
  $\text{Re}\tau$ & $\text{Im}\tau$ & $|\lambda'_{u^{c}}|$ & $\text{arg}(\lambda'_{u^{c}})/\pi$ & $y^{u}_{2}/y^{u}_{1}$ & $y^{u}_{3}/y^{u}_{1}$   \\
  \hline
  $-0.08817$& 2.51405& 1.00000 & 0.19108 & 5.01136 & 0.92896 \\
\end{tabular}\\
\cline{3-3}
& &
\begin{tabular}{ccccc}
 $|y^{d}_{2}/y^{d}_{1}| $ & $\text{arg}(y^{d}_{2}/y^{d}_{1})/\pi$ & $y^{d}_{3}/y^{d}_{1}$ & $y_{1}^{u}v_{u}/\text{GeV}$ & $y_{1}^{d}v_{d}/\text{GeV}$  \\
 \hline
 3.71694 & 0.14153 & 1.00590 & 0.14779 & 0.96250 \\
\end{tabular}\\
  \cline{2-3}
    &  \texttt{Obs} &
\begin{tabular}{cccccccc}
 $m_{u}/m_{c}$ & $m_{c}/m_{t}$ & $m_{d}/m_{s}$ & $m_{s}/m_{b}$ & $\theta_{12}^{q}$ & $\theta_{13}^{q}$  & $\theta_{23}^{q}$ & $\delta_{CP}^{q}/^{\circ}$  \\
\hline
 0.00194& 0.00282& 0.04973& 0.01824& 0.22541& 0.00349& 0.03977& 69.28416  \\
\end{tabular}\\
 \hline  \hline
\end{tabular}}
\caption{\label{Tab:predictions_quark_modification} The best fit values of the input parameters as well as quark mass ratios and flavor mixing parameters for the \texttt{Model VI}. }
\end{table}

\section{\label{sec:conclusion} Conclusion and summary}

Modular invariance as the flavor symmetry is a promising approach to understand the puzzle of quark and lepton masses and flavor mixing. The Yuakwa couplings are modular forms of a certain level $N$ in this approach.
We have generalized this formalism to modular forms of general integer weights which can be arranged into irreducible representations of the homogeneous finite modular group $\Gamma'_N$~\cite{Liu:2019khw}. Notice that $\Gamma'_N$ is the double covering of $\Gamma_N$. It is well known that $SU(2)$ is the double covering group of $SO(3)$, and two $SU(2)$ elements correspond to one $SO(3)$ element with the same Euler angles. $\Gamma'_N$ can be regarded as the inverse image of the inhomogeneous finite modular group $\Gamma_N$ under this map for smaller $N$. In addition to the irreducible representations of $\Gamma_N$, $\Gamma'_N$ has other new representations~\cite{Liu:2019khw}. The even weight modular forms of level $N$ transform in the irreducible representations of $\Gamma_N$~\cite{Feruglio:2017spp}, while the odd weight modular forms of level $N$ are arranged into the new representations of $\Gamma'_N$~\cite{Liu:2019khw}. The odd weight modular forms provide us interesting opportunity for fermion mass model building.

With the idea of reduce the number of free parameters, the ansatz of texture zero has been widely studied. It is shown that the texture zeros of the fermion mass matrices can be enforced by means of Abelian symmetries. In the present work, we show that the texture zeros of the fermion mass matrices can be naturally produced if we properly assign the representations and weights of the matter fields under the modular symmetry. As a concrete example, we impose the $\Gamma'_3\cong T'$ modular symmetry on the quark sector. We assign the first two generations of the left-handed quark doublets to a $T'$ doublet, and the third generation of the left-handed quark is a singlet of $T'$. As regards the right-handed quark fields, we have considered two scenarios that the three generations of right-handed quarks transform as a doublet and a singlet under $T'$ or they are three singlets of $T'$. We find that the quark mass matrix can take six possible forms with zero entries up to row and column permutations: $\text{Case}~\mathcal{A}$, $\text{Case}~\mathcal{B}$, $\text{Case}~\mathcal{C}$, $\text{Case}~\mathcal{D}$, $\text{Case}~\mathcal{E}$ and $\text{Case}~\mathcal{F}$ given in Eqs.~(\ref{eq:texture1}, \ref{eq:case5_singlet}, \ref{eq:case6_singlet}). Combing the up and down quark sectors together, we can obtain the possible up quark and down quark mass matrices predicted by $T'$ modular symmetry. It is usually assumed the mass matrix is symmetric or hermitian in texture zero models. In our approach, the explicit form of mass matrix is completely fixed by modular symmetry and it is generally neither symmetric nor hermitian. We present five interesting quark models summarized in table~\ref{Tab:quark_summary}. These models contain only ten or eleven independent real parameters including the complex modulus $\tau$, and the experimental data of quark masses and CKM mixing matrix can be accommodated very well.

Furthermore, we investigate the lepton sector with the $T'$ modular symmetry. The lepton fields are assumed to transform as triplet $\mathbf{3}$ or singlets $\mathbf{1}$, $\mathbf{1}'$, $\mathbf{1}''$ under $T'$ such that the $T'$ modular symmetry can not be distinguished from the $A_4$ modular symmetry in the lepton sector since doublet representations of $T'$ are not involved. A systematical classification of lepton models with $A_4$ modular symmetry has been performed in~\cite{Ding:2019zxk}. We combine the five quark models in table~\ref{Tab:quark_summary} with the $\mathcal{D}_6$ and $\mathcal{D}_9$ lepton models of~\cite{Ding:2019zxk} to give a unified description of both quark and lepton masses and flavor mixing. It is highly nontrivial that the resulting models can produce very good fit to the experimental data although they contain less number of free parameters than the observables.

In summary, the modular symmetry can naturally produce texture zeros in fermion mass matrix if odd weight modular forms are considered.
The modular invariance approach has the merits of both abelian flavor symmetry and discrete non-abelian flavor symmetry. It is interesting to investigate the possible texture zero structures of the lepton sector which can be obtained from modular symmetry.

In the present paper, we have followed the bottom-up model construction approach of modular symmetry to describe the flavor structure of the SM. It is interesting to investigate whether the benchmark quark-lepton unification models in section~\ref{sec:QLU} can be embedded in a string theory framework. It has been shown that the finite modular group $T'$ can naturally appear in  simple string constructions such as $Z_3$ orbifold~\cite{Baur:2019kwi,Baur:2019iai}. Therefore we expect that certain (not all) models of the present paper or at least texture zero structure of quark mass matrix  could be realized in the low-energy effective theory, although explicit model building in string theory is needed. The phenomenological bottom-up models based on modular invariance can explain many different aspects of the flavor puzzle in SM. However, the possible connections between this approach and the more fundamental theory such as string theory are still not clear, see~\cite{Baur:2019kwi,Baur:2019iai,Nilles:2020nnc} for recent progress toward this direction. Discrete flavor symmetries and modular invariance could appear simultaneously in top-down model building in string theory. Recently it is suggested to extend the traditional discrete flavor group by finite modular symmetries~\cite{Nilles:2020kgo}. The K\"ahler potential and superpotential are severely constrained in this new scheme, in particularly the off-diagonal contributions to the K\"ahler metric are forbidden. As a result, the modular form dependent terms in the K\"ahler potential does not considerably change the phenomenological predictions obtained by assuming a minimal K\"ahler potential. Hence it is a reasonable and good approximation to take the standard minimal K\"ahler potential in bottom-up model construction.

\section*{Acknowledgements}

This work is supported by the National Natural Science Foundation of China under Grant Nos 11975224 and 11835013.

\section*{Appendix}

\setcounter{equation}{0}
\renewcommand{\theequation}{\thesection.\arabic{equation}}

\begin{appendix}

\section{\label{app:Tp_group}Group Theory of $T^{\prime}$}

The $T'$ group is the double covering of the tetrahedral group $A_4$. All the elements of $T'$ can be generated by three generators $S$, $T$ and $\mathbb{R}$ which obey the following relations,
\begin{equation}
S^{2}=\mathbb{R},~~(ST)^{3}=T^{3}=\mathbb{R}^{2}=1,~~\mathbb{R}T = T\mathbb{R}\,.
\end{equation}
Hence the generator $\mathbb{R}$ commutes with all elements of the group.
Besides the $A_4$ representations: one triplet $\mathbf{4}$ and three singlets $\mathbf{1}$, $\mathbf{1}'$ and $\mathbf{1}''$, the $T'$ group has three two-dimensional irreducible representations $\mathbf{2}$, $\mathbf{2}'$ and $\mathbf{2}''$. In the present work, we shall adopt the same basis of~\cite{Liu:2019khw}, in particular using the following
explicit representation matrices for the generators $S$, $T$ and $\mathbb{R}$ in different irreps,
\begin{eqnarray*}
\label{eq:irr_reps} \begin{array}{cccc}
\mathbf{1:} & S=1, & T=1,& \mathbb{R}=1\,,\\
\mathbf{1}':& S=1, & T=\omega, & \mathbb{R}=1\,,\\
\mathbf{1}'': & S=1, & T=\omega^{2}, & \mathbb{R}=1\,,\\
\mathbf{2:} ~&~ S=-\dfrac{1}{\sqrt{3}}\left(\begin{array}{cc}
 i & \sqrt2e^{i\pi/12} \\
 -\sqrt2e^{-i\pi/12} & -i \\
\end{array}\right), & T=\left(\begin{array}{cc}
\omega & 0 \\
0 & 1 \\
\end{array}\right), & \mathbb{R}=-\left(\begin{array}{cc}
1 & 0 \\
0 & 1 \\
\end{array}\right)\,,\\
\mathbf{2}': ~&~ S=-\dfrac{1}{\sqrt{3}}\left(\begin{array}{cc}
 i & \sqrt2e^{i\pi/12} \\
 -\sqrt2e^{-i\pi/12} & -i \\
\end{array}\right), & T=\left(\begin{array}{cc}
\omega^{2} & 0 \\
0 & \omega \\
\end{array}\right), & \mathbb{R}=-\left(\begin{array}{cc}
1 & 0 \\
0 & 1 \\
\end{array}\right)\,,\\
\mathbf{2}'': ~&~ S=-\dfrac{1}{\sqrt{3}}\left(\begin{array}{cc}
 i & \sqrt2e^{i\pi/12} \\
 -\sqrt2e^{-i\pi/12} & -i \\
\end{array}\right), & T=\left(\begin{array}{cc}
1 & 0 \\
0 & \omega^{2} \\
\end{array}\right), & \mathbb{R}=-\left(\begin{array}{cc}
1 & 0 \\
0 & 1 \\
\end{array}\right)\,,\\
\mathbf{3:} & S=\frac{1}{3}\left(\begin{array}{ccc}
-1&2&2\\
2&-1&2\\
2&2&-1
\end{array}\right),
 & T=\left(
\begin{array}{ccc}
1~&0~&0\\
0~&\omega~&0\\
0~&0~&\omega^2
\end{array}\right), &~~ \mathbb{R}=\left(
\begin{array}{ccc}
1~&0~&0\\
0~&1~&0\\
0~&0~&1
\end{array}\right)\,,
\end{array}
\end{eqnarray*}
with $\omega=e^{i2\pi/3}$. The generator $\mathbb{R}$ is represented by an identity matrix for the odd-dimensional representations $\mathbf{1}$, $\mathbf{1}'$, $\mathbf{1}''$ and $\mathbf{3}$. Therefore the elements of $T'$ coincide two by two and can be described by the same matrices that represent the elements in $A_4$ for these representations. The Kronecker products between different irreducible representations of $T'$ are given by
\begin{eqnarray}
\nonumber&&\mathbf{1}^a\otimes \mathbf{r}^b = \mathbf{r}^b \otimes \mathbf{1}^a= \mathbf{r}^{a+b~(\text{mod}~3)},~~~~~{\rm for}~~\mathbf{r} = \mathbf{1}, \mathbf{2}\,,\\
\nonumber&&\mathbf{1}^a \otimes \mathbf{3} = \mathbf{3}\otimes \mathbf{1}^a=\mathbf{3}\,,\\
\nonumber&&\mathbf{2}^a \otimes \mathbf{2}^b=\mathbf{3}\oplus \mathbf{1}^{a+b+1~(\text{mod}~3)}\,, \\
\nonumber&&\mathbf{2}^a \otimes \mathbf{3} =\mathbf{3} \otimes \mathbf{2}^a = \mathbf{2}\oplus \mathbf{2}'\oplus \mathbf{2}''\,, \\
\label{eq:mult}&&\mathbf{3} \otimes \mathbf{3} = \mathbf{3}_S \oplus \mathbf{3}_A \oplus \mathbf{1} \oplus \mathbf{1}' \oplus \mathbf{1}''\,,
\end{eqnarray}
where $a,b=0, 1, 2$ and we have denoted $\mathbf{1}\equiv\mathbf{1}^0$, $\mathbf{1}'\equiv\mathbf{1}^{1}$, $\mathbf{1}''\equiv\mathbf{1}^{2}$ for singlet representations and $\mathbf{2}\equiv\mathbf{2}^0$, $\mathbf{2}'\equiv\mathbf{2}^{1}$, $2''\equiv\mathbf{2}^{2}$ for the doublet representations. The notations $\mathbf{3}_S$ and $\mathbf{3}_A$ stand for the symmetric and antisymmetric triplet combinations respectively. The CG coefficients in our working basis can be also be found in~\cite{Liu:2019khw}, we would like to list them in the following for completeness. We shall use $\alpha_i$ to denote the elements of the first representation, $\beta_i $ to indicate these of the second representation of the product.
\begin{eqnarray}
\mathbf{1}^a\otimes\mathbf{1}^b &=& \mathbf{1}^{a+b~(\text{mod}~3)} \sim \alpha\beta \,, \\
\mathbf{1}^a\otimes\mathbf{2}^b &=& \mathbf{2}^{a+b~(\text{mod}~3)} \sim \left(\begin{array}{c} \alpha\beta_1\\
\alpha\beta_2 \\ \end{array}\right) \,,\\
  \mathbf{1}'\otimes\mathbf{3} &=& \mathbf{3} \sim \left(\begin{array}{c} \alpha\beta_3\\
\alpha\beta_1 \\
\alpha\beta_2 \end{array}\right) \,, \\
\mathbf{1}''\otimes\mathbf{3} &=& \mathbf{3} \sim \left(\begin{array}{c} \alpha\beta_2\\
\alpha\beta_3 \\
\alpha\beta_1 \end{array}\right) \,.
\end{eqnarray}

\begin{eqnarray}
  \mathbf{2}\otimes\mathbf{2}=\mathbf{2}'\otimes \mathbf{2}''=\mathbf{3}\oplus\mathbf{1}' ~&\text{with}&~\left\{
\begin{array}{l}
\mathbf{1}'\sim \alpha_1\beta_2-\alpha_2\beta_1 \\ [0.1in]
\mathbf{3}\sim
\left(\begin{array}{c} e^{i \pi /6}\alpha_2\beta_2 \\
\frac{1}{\sqrt{2}}e^{i 7\pi /12}(\alpha_1\beta_2+\alpha_2\beta_1)  \\
\alpha_1\beta_1 \end{array}\right)
\end{array}
\right. \\
\mathbf{2}\otimes\mathbf{2}'=\mathbf{2}''\otimes\mathbf{2}''=\mathbf{3}\oplus\mathbf{1}''~&\text{with}&~\left\{
\begin{array}{l}
\mathbf{1}''\sim \alpha_1\beta_2-\alpha_2\beta_1\\ [0.1in]
\mathbf{3}\sim
 \left(\begin{array}{c} \alpha_1\beta_1\\
 e^{i \pi /6}\alpha_2\beta_2 \\
 \frac{1}{\sqrt{2}}e^{i 7\pi /12}(\alpha_1\beta_2+\alpha_2\beta_1) \end{array}\right)
\end{array}
\right. \\
\mathbf{2}\otimes\mathbf{2}''=\mathbf{2}'\otimes\mathbf{2}'=\mathbf{3}\oplus\mathbf{1} ~&\text{with}&~\left\{
\begin{array}{l}
\mathbf{1} \sim \alpha_1\beta_2-\alpha_2\beta_1 \\ [0.1in]
\mathbf{3}\sim
\left(\begin{array}{c}  \frac{1}{\sqrt{2}}e^{i 7\pi /12}(\alpha_1\beta_2+\alpha_2\beta_1)\\
 \alpha_1\beta_1\\
 e^{i \pi /6}\alpha_2\beta_2  \end{array}\right)
\end{array}
\right. \\
\mathbf{2}\otimes\mathbf{3}=\mathbf{2}\oplus\mathbf{2}'\oplus\mathbf{2}'' ~&\text{with}&~\left\{
\begin{array}{l}
\mathbf{2}\sim
 \left(\begin{array}{c} \alpha_1\beta_1-\sqrt{2}e^{i 7\pi /12}\alpha_2\beta_2 \\
 -\alpha_2\beta_1+\sqrt{2}e^{i 5\pi / 12}\alpha_1\beta_3 \end{array}\right)  \\ [0.1in]
\mathbf{2}'\sim
\left(\begin{array}{c} \alpha_1\beta_2-\sqrt{2}e^{i 7\pi /12}\alpha_2\beta_3 \\
-\alpha_2\beta_2+\sqrt{2}e^{i 5\pi / 12}\alpha_1\beta_1  \end{array}\right) \\ [0.1in]
\mathbf{2}''\sim
\left(\begin{array}{c} \alpha_1\beta_3-\sqrt{2}e^{i 7\pi /12}\alpha_2\beta_1 \\
-\alpha_2\beta_3+\sqrt{2}e^{i 5\pi / 12}\alpha_1\beta_2  \end{array}\right) \\ [0.1in]
\end{array}
\right. \\
\mathbf{2}'\otimes\mathbf{3}=\mathbf{2}\oplus\mathbf{2}'\oplus\mathbf{2}'' ~&\text{with}&~\left\{
\begin{array}{l}
\mathbf{2}\sim
\left(\begin{array}{c} \alpha_1\beta_3-\sqrt{2}e^{i 7\pi /12}\alpha_2\beta_1 \\
-\alpha_2\beta_3+\sqrt{2}e^{i 5\pi / 12}\alpha_1\beta_2  \end{array}\right)  \\ [0.1in]
\mathbf{2}'\sim
 \left(\begin{array}{c} \alpha_1\beta_1-\sqrt{2}e^{i 7\pi /12}\alpha_2\beta_2 \\
 -\alpha_2\beta_1+\sqrt{2}e^{i 5\pi / 12}\alpha_1\beta_3 \end{array}\right) \\ [0.1in]
\mathbf{2}''\sim
\left(\begin{array}{c} \alpha_1\beta_2-\sqrt{2}e^{i 7\pi /12}\alpha_2\beta_3 \\
-\alpha_2\beta_2+\sqrt{2}e^{i 5\pi / 12}\alpha_1\beta_1  \end{array}\right) \\ [0.1in]
\end{array}
\right. \\
\mathbf{2}''\otimes\mathbf{3}=\mathbf{2}\oplus\mathbf{2}'\oplus\mathbf{2}'' ~&\text{with}&~\left\{
\begin{array}{l}
\mathbf{2}\sim
\left(\begin{array}{c} \alpha_1\beta_2-\sqrt{2}e^{i 7\pi /12}\alpha_2\beta_3 \\
-\alpha_2\beta_2+\sqrt{2}e^{i 5\pi / 12}\alpha_1\beta_1  \end{array}\right)  \\ [0.1in]
\mathbf{2}'\sim
\left(\begin{array}{c} \alpha_1\beta_3-\sqrt{2}e^{i 7\pi /12}\alpha_2\beta_1 \\
-\alpha_2\beta_3+\sqrt{2}e^{i 5\pi / 12}\alpha_1\beta_2  \end{array}\right) \\ [0.1in]
\mathbf{2}''\sim
 \left(\begin{array}{c} \alpha_1\beta_1-\sqrt{2}e^{i 7\pi /12}\alpha_2\beta_2 \\
 -\alpha_2\beta_1+\sqrt{2}e^{i 5\pi / 12}\alpha_1\beta_3 \end{array}\right) \\ [0.1in]
\end{array}
\right. \\
\mathbf{3}\otimes\mathbf{3}=\mathbf{3}_S\oplus\mathbf{3}_A\oplus\mathbf{1}\oplus\mathbf{1}'\oplus\mathbf{1}'' ~&\text{with}&~\left\{
\begin{array}{l}
\mathbf{3}_S\sim
 \left(\begin{array}{c} 2\alpha_1\beta_1 - \alpha_2\beta_3 - \alpha_3\beta_2 \\
 2\alpha_3\beta_3 - \alpha_1\beta_2 - \alpha_2\beta_1  \\
 2\alpha_2\beta_2 - \alpha_1\beta_3 - \alpha_3\beta_1 \end{array}\right) \\ [0.1in]
\mathbf{3}_A\sim
 \left(\begin{array}{c} \alpha_2\beta_3 - \alpha_3\beta_2 \\
\alpha_1\beta_2 - \alpha_2\beta_1  \\
 \alpha_3\beta_1 - \alpha_1\beta_3 \end{array}\right) \\ [0.1in]
\mathbf{1} \sim \alpha_1\beta_1 + \alpha_2\beta_3 + \alpha_3\beta_2 \\ [0.1in]
\mathbf{1}' \sim \alpha_3\beta_3 + \alpha_1\beta_2 + \alpha_2\beta_1 \\ [0.1in]
\mathbf{1}'' \sim \alpha_2\beta_2 + \alpha_1\beta_3 + \alpha_3\beta_1 \\ [0.1in]
\end{array}
\right.
\end{eqnarray}

\setcounter{equation}{0}
\renewcommand{\theequation}{\thesection.\arabic{equation}}

\section{\label{app:High_MDF} Higher weight modular forms  }

Using the contraction rules of the $T'$ group in Appendix~\ref{app:Tp_group}, we can construct the modular forms of weight $k=5, 6, 7, 8$ and level 3. The linear space of modular forms of weight $k$ and level 3 has dimension $k+1$. The action of the homogeneous finite modular group $T'$ divides the space of weight 5 modular forms into three doublets transforming the irreducible representations $\mathbf{2}$, $\mathbf{2}'$ and $\mathbf{2}''$ of $T'$. They are given by
\begin{eqnarray}
\nonumber&&Y^{(5)}_{\mathbf{2}}\equiv \begin{pmatrix}
Y^{(5)}_{\mathbf{2},1}\\Y^{(5)}_{\mathbf{2},2} \end{pmatrix} = \begin{pmatrix} 2\sqrt{2}e^{i7\pi/12}Y^4_1Y_2+e^{i\pi/3}Y_1Y^4_2 \\ 2\sqrt{2}e^{i7\pi/12}Y^3_1Y^2_2+e^{i\pi/3}Y^5_2 \end{pmatrix}\,, \\
\nonumber&&Y^{(5)}_{\mathbf{2}'}\equiv \begin{pmatrix}
Y^{(5)}_{\mathbf{2}',1}\\Y^{(5)}_{\mathbf{2}',2} \end{pmatrix} =\begin{pmatrix}-Y^5_1+2(1-i)Y^2_1Y^3_2 \\ -Y^4_1Y_2+2(1-i)Y_1Y^4_2 \end{pmatrix},\\
&&Y^{(5)}_{\mathbf{2}''}\equiv \begin{pmatrix}
Y^{(5)}_{\mathbf{2}'',1}\\Y^{(5)}_{\mathbf{2}'',2} \end{pmatrix} =\begin{pmatrix} 5e^{i\pi/6}Y^3_1Y^2_2-(1-i)e^{i\pi/6}Y^5_2 \\ -\sqrt{2}e^{i5\pi/12}Y^5_1-5e^{i\pi/6}Y^2_1Y^3_2 \end{pmatrix} \,.
\end{eqnarray}
The weight 6 modular forms can be decomposed into two triplets $\mathbf{3}$
and one singlet $\mathbf{1}$ of $T'$, and they explicitly read
\begin{eqnarray}
\nonumber&&Y^{(6)}_{\mathbf{1}}=(1-i)e^{i\pi/6}Y^6_2-(1+i)e^{i\pi/6}Y^6_1-10e^{i\pi/6}Y^3_1Y^3_2\,, \\
\nonumber&&Y^{(6)}_{\mathbf{3}I} \equiv \begin{pmatrix}
Y^{(6)}_{\mathbf{3}I,1}\\Y^{(6)}_{\mathbf{3}I,2} \\ Y^{(6)}_{\mathbf{3}I,3} \end{pmatrix}=\begin{pmatrix} -2(1-i)Y^3_1Y^3_2+iY^6_2 \\ -4e^{i\pi/6}Y^4_1Y^2_2-(1-i)e^{i\pi/6}Y_1Y^5_2 \\ 2\sqrt{2}e^{i7\pi/12}Y^5_1Y_2+e^{i\pi/3}Y^2_1Y^4_2\end{pmatrix}\,,\\
&&Y^{(6)}_{\mathbf{3}II} \equiv \begin{pmatrix}
Y^{(6)}_{\mathbf{3}II,1}\\Y^{(6)}_{\mathbf{3}II,2} \\ Y^{(6)}_{\mathbf{3}II,3} \end{pmatrix}= \begin{pmatrix}-Y^6_1+2(1-i)Y^3_1Y^3_2 \\  -e^{i\pi/6}Y^4_1Y^2_2+2(1-i)e^{i\pi/6}Y_1Y^5_2 \\ 4e^{i\pi/3}Y^2_1Y^4_2-(1+i)e^{i\pi/3}Y^5_1Y_2 \end{pmatrix}\,.
\end{eqnarray}
For the modular forms of weight 7, we have
\begin{eqnarray}
\nonumber&&Y^{(7)}_{\mathbf{2}I}\equiv \begin{pmatrix}
Y^{(7)}_{\mathbf{2}I,1}\\Y^{(7)}_{\mathbf{2}I,2} \end{pmatrix} = \begin{pmatrix} 3iY_1Y_2^3[2(1 + i)Y_1^3 + Y_2^3] \\ -4Y_1^6 Y_2 + (1 - i) Y_1^3 Y_2^4 - i Y_2^7 \end{pmatrix} \\
\nonumber&&Y^{(7)}_{\mathbf{2}II}\equiv \begin{pmatrix}
Y^{(7)}_{\mathbf{2}II, 1}\\Y^{(7)}_{\mathbf{2}II, 2} \end{pmatrix} = \begin{pmatrix} -Y_1[Y_1^6 -(1-i)Y_1^3 Y_2^3 + 4 i Y_2^6] \\ 3 Y_1^3Y_2[Y_1^3-2(1-i)Y_2^3] \end{pmatrix} \\
\nonumber&&Y^{(7)}_{\mathbf{2}'}\equiv \begin{pmatrix}
Y^{(7)}_{\mathbf{2}',1}\\Y^{(7)}_{\mathbf{2}',2} \end{pmatrix} =\begin{pmatrix}-3e^{i\pi/6} Y_1^2 Y_2^2[Y_1^3-2(1-i) Y_2^3] \\ -e^{i\pi/6}Y_1[Y_1^3-2(1-i) Y_2^3][(1 + i)Y_1^3 -Y_2^3] \end{pmatrix},\\
&&Y^{(7)}_{\mathbf{2}''}\equiv \begin{pmatrix}
Y^{(7)}_{\mathbf{2}'', 1}\\Y^{(7)}_{\mathbf{2}'', 2} \end{pmatrix} =\begin{pmatrix} [2(1+i)Y_1^3 + Y_2^3](e^{i\pi/3} Y_1^3 Y_2 + e^{i\pi/12}\sqrt{2} Y_2^4) \\ -3e^{i\pi/3}Y_1^2 Y_2^2[2(1+i)Y_1^3 + Y_2^3] \end{pmatrix} \,.
\end{eqnarray}
Finally we present the expressions of the weight 8 modular forms of level 3 as follow,
\begin{eqnarray}
\nonumber&&Y^{(8)}_{\mathbf{1}}=[2(1+i) Y_1^3 + Y_2^3](4 e^{i\pi/3}Y_1^3Y_2^2 + e^{i\pi/12} \sqrt{2}Y_2^5)\,, \\
\nonumber&& Y^{(8)}_{\mathbf{1}'}= 4Y_1^7 Y_2 -7(1-i) Y_1^4 Y_2^4 +4iY_1Y_2^7\,, \\
\nonumber&&Y^{(8)}_{\mathbf{1}''}=[Y_1^4 - 2(1-i) Y_1 Y_2^3]^2\,,\\
\nonumber&&Y^{(8)}_{\mathbf{3}I} \equiv \begin{pmatrix}
Y^{(8)}_{\mathbf{3}I, 1}\\Y^{(8)}_{\mathbf{3}I, 2} \\ Y^{(8)}_{\mathbf{3}I, 3} \end{pmatrix} = \begin{pmatrix} -e^{i2\pi/3} Y_2[-4 i Y_1^6Y_2 + (1 + i) Y_1^3 Y_2^4 + Y_2^7] \\ e^{i5\pi/6} [-2(1 -i) Y_1^7 Y_2 + 5i Y_1^4 Y_2^4 + (1 + i) Y_1 Y_2^7] \\ 3i Y_1^2 Y_2^3 [2(1+i) Y_1^3 + Y_2^3]\end{pmatrix}\,, \\
&&Y^{(8)}_{\mathbf{3}II} \equiv \begin{pmatrix}
Y^{(8)}_{\mathbf{3}II, 1}\\Y^{(8)}_{\mathbf{3}II, 2} \\ Y^{(8)}_{\mathbf{3}II, 3} \end{pmatrix}= \begin{pmatrix} 3 e^{i\pi/6} Y_1^3 Y_2^2 [Y_1^3 - 2(1-i) Y_2^3] \\  e^{i\pi/3}[(1 + i) Y_1^7 Y_2 - 5 Y_1^4 Y_2^4 + 2(1-i) Y_1 Y_2^7] \\ -Y_1^2[Y_1^6 -(1 -i) Y_1^3 Y_2^3 + 4i Y_2^6] \end{pmatrix}\,.
\end{eqnarray}

\section{\label{app:non-minimal-Kahler-lepton}Non-minimal K\"ahler potential in lepton sector }

\begin{figure}[t!]
\begin{center}
\includegraphics[width=0.46\linewidth]{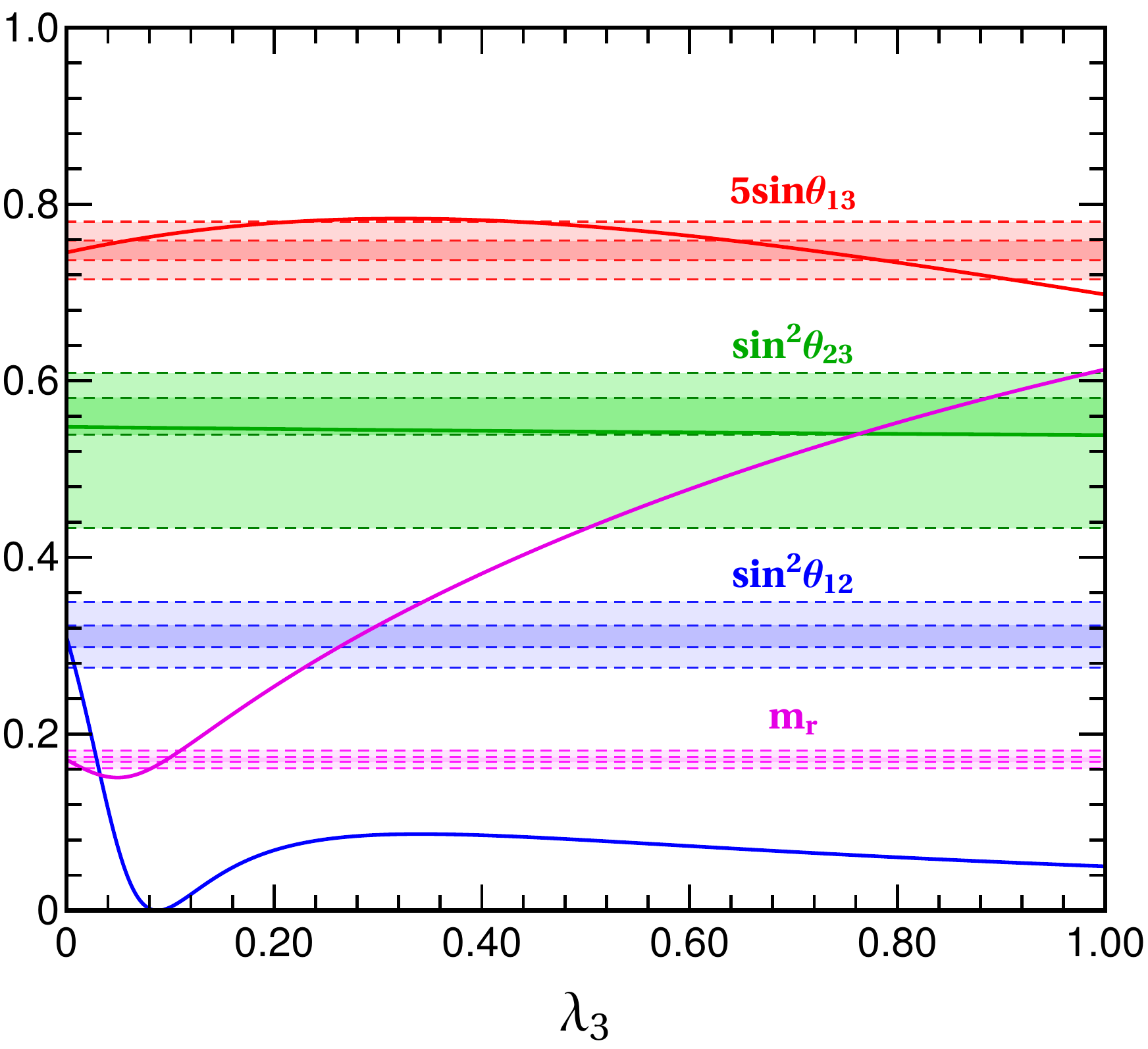}~
\includegraphics[width=0.50\linewidth]{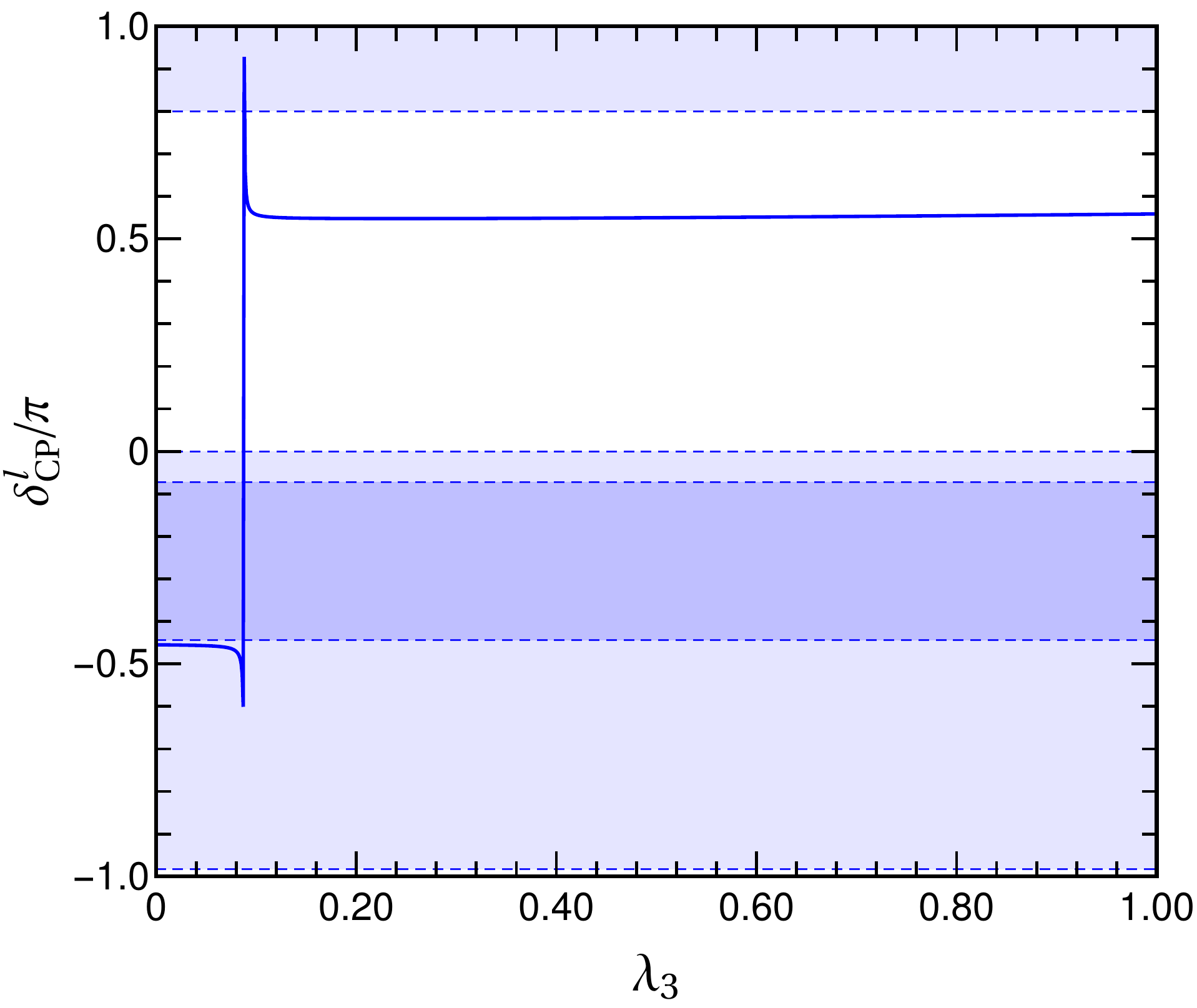}
\end{center}
\caption{\label{fig:mixingpar_kahler_D6}
Variation of the the lepton mixing angles, mass ratio $m_{r}\equiv \sqrt{\Delta m_{21}^{2}/\Delta m_{31}^{2}}$ (left panel) and Dirac CP phase $\delta^l_{CP}$ (right panel) with respect to the coupling $\lambda_3$. The shaded regions represent the $1\sigma$ and $3\sigma$ ranges of the mixing parameters~\cite{Esteban:2018azc}.}
\end{figure}

In the same manner as for quarks in section~\ref{sec:non-minimal-Kahler}, the minimal K\"ahler potential in the lepton sector is given by
\begin{equation}
\mathcal{K}_{\text{min}}=(-i\tau+i\bar{\tau})^{-k_{L_i}}L^{\dagger}_iL_i+ (-i\tau+i\bar{\tau})^{-k_{E^c_i}}{E^c_i}^{\dagger}E^c_i + (-i\tau+i\bar{\tau})^{-k_{N^c_i}}{N^c_i}^{\dagger}N^c_i\,,
\end{equation}
where $L$ and $E^c$ denote lepton doublets and singlets respectively, and $N^c$ stands for the right-handed neutrinos superfield. Depending on the weight and representation assignments of $L$, $E^c$ and $N^c$, the most general form of the additional term $\Delta\mathcal{K}$ is given by
\begin{align}
\nonumber\Delta \mathcal{K}&=\sum_{k_1,k_2,\mathbf{r}_1,\mathbf{r}_2} \lambda^{ij}_L (-i\tau+i\bar{\tau})^{-k_{L_i}+k_1} (L^\dagger_i Y^{(k_1)\dagger}_{\mathbf{r}_1}Y^{(k_2)}_{\mathbf{r}_2} L_j)_\mathbf{1}\\
\nonumber&+ \sum_{k_3,k_4,\mathbf{r}_1,\mathbf{r}_2} \lambda^{ij}_{E^c} (-i\tau+i\bar{\tau})^{-k_{E^c_i}+k_3} (E^{c\dagger}_iY^{(k_3)\dagger}_{\mathbf{r}_1}Y^{(k_4)}_{\mathbf{r}_2} E^c_j)_\mathbf{1} \\
&+ \sum_{k_5,k_6,\mathbf{r}_1,\mathbf{r}_2} \lambda^{ij}_{N^c} (-i\tau+i\bar{\tau})^{-k_{N^c_i}+k_5} ( N^{c \dagger}_iY^{(k_5)\dagger}_{\mathbf{r}_1}Y^{(k_6)}_{\mathbf{r}_2}N^c_j)_\mathbf{1}+\mathrm{h.c.}~~\,,
\end{align}
with
\begin{equation}
k_1-k_2=k_{L_i}-k_{L_j},~~~k_3-k_4=k_{E^c_i}-k_{E^c_j},~~~k_5-k_6=k_{N^c_i}-k_{N^c_j}\,.
\end{equation}
The hermitian and positive K\"ahler metrics for $L$, $E^c$ and $N^c$ can be easily read out and they can be written as
\begin{equation}
\mathscr{K}_L=\Sigma_L D^2_L \Sigma^\dagger_L,\quad \mathscr{K}_{E^c}=\Sigma_{E^c} D^2_{E^c} \Sigma^\dagger_{E^c},\quad \mathscr{K}_{N^c}=\Sigma_{N^c} D^2_{N^c} \Sigma^\dagger_{N^c} \,,
\end{equation}
where $\Sigma_{L, E^c, N^c}$ are unitary matrices, and
$D_{L, E^c, N^c}$ are diagonal with positive real entries. The K\"ahler metrics as well as the kinetic terms can be diagonalized by the field redefinition
\begin{align}
\nonumber
&L\rightarrow \hat{L}=H^{-1}_L L,~~~~~~~H_L= \Sigma_LD^{-1}_L \Sigma^\dagger _L\,,\\
\nonumber&E^c\rightarrow \hat{E}^c=H^{-1}_{E^c} E^c,~~~~H_{E^c}= \Sigma_{E^c}D^{-1}_{E^c} \Sigma^\dagger _{E^c}\,,\\
&N^c\rightarrow \hat{N}^c=H^{-1}_{N^c} N^c,~~~H_{N^c}= \Sigma_{N^c} D^{-1}_{N^c} \Sigma^\dagger_{N^c}\,.
\end{align}
This field redefinition would lead to change of charged lepton and neutrino mass matrices. If neutrino masses are described by the effective Weinberg operator, we have
\begin{align}
\mathcal{W}_l = E^{c\,T} Y_e L H_d + \frac{1}{2\Lambda} L^T Y_\nu L H^2_u\rightarrow\mathcal{\widehat{W}}_{l}=\hat{E}^{c\,T} \hat{Y}_e  \hat{L} H_d +\frac{1}{2\Lambda} \hat{L}^T  \hat{Y}_\nu \hat{L} H^2_u\,,
\end{align}
where $\hat{Y}_e=H^T_{E^c}Y_eH_L$ and $\hat{Y}_\nu=H^T_{L}Y_{\nu}H_L$.
If neutrino masses are generated by the seesaw mechanism, going to the basis of canonically normalized fields, we obtain
\begin{equation}
\mathcal{W}_l= E^{cT}Y_e L H_d + N^{cT} Y_D L H_u +\frac{1}{2} N^{cT} M_{N} N^c=\hat{E}^{cT} \hat{Y}_e  \hat{L} H_d + \hat{N}^{cT}\hat{Y}_D \hat{L} H_u +\frac{1}{2} \hat{N}^{cT} \hat{M}_{N} \hat{N}^c\,,
\end{equation}
with $\hat{Y}_e=H^T_{E^c} Y_e H_L$, $\hat{Y}_D=H^T_{N^c}Y_D H_L$ and $ \hat{M}_N=H^{T}_{N^c} M_N H_{N^c}$. The charged lepton and neutrino mass eigenvalues and the lepton mixing matrix should be calculated for canonically normalized fields. In order to be more specific, we shall take the model $\mathcal{L}_{2}$ as an example.
\begin{table}
\centering
\resizebox{1.0\textwidth}{!}{
\begin{tabular}{|c|c|c|}
  \hline  \hline
\multirow{5}*{$\mathcal{L}_2$} &  \texttt{Input } &
\begin{tabular}{cccccccc}
$\text{Re}\tau$ & $\text{Im}\tau$ & $\beta/\alpha$ & $\gamma/ \alpha$ & $|g_{2}/g_{1}|$ & $\text{arg}(g_{2}/g_{1})/\pi$  & $\alpha v_{d}/\text{GeV}$ & $(g_{1}^{2}v_{u}^{2}/\Lambda)/\text{eV}$ \\ \hline
$-0.48188$ & 0.87803 & 0.37487 & 0.00433 & 1.11456 & 1.62791 & 8.34809 & 0.41822 \\
\end{tabular}\\
  \cline{2-3}
& \multirow{3}*{\texttt{Obs }}&
\begin{tabular}{cccccc}
  $m_{e}/m_{\mu}$ & $m_{\mu}/m_{\tau}$ & $m_{1}/\text{eV}$ & $m_{2}/\text{eV}$ & $m_{3}/\text{eV}$ &  $|m_{ee}|/\text{eV}$\\ \hline
0.00474 & 0.05858 & 0.13012 & 0.13040 & 0.13948 & 0.13028 \\
\end{tabular}\\ \cline{3-3}
& &
\begin{tabular}{cccccc}
 $\sin^{2}\theta_{12}^{l}$ & $\sin^{2}\theta_{13}^{l}$ & $\sin^{2}\theta_{23}^{l}$ & $\delta_{CP}^{l}/^{\circ}$ & $\alpha_{21}/^{\circ}$ & $\alpha_{31}/^{\circ}$ \\ \hline
 0.31012 & 0.02222 & 0.54786 & 278.10936 & 359.47123 & 179.07277 \\
\end{tabular}\\ \hline  \hline
\end{tabular}}
\caption{\label{Tab:predictions_quark_modification} The best fit values of the input parameters as well as lepton mass ratios and flavor mixing parameters for the lepton model $\mathcal{L}_{2}$. }
\end{table}
Given the field assignments in table~\ref{Tab:lepton_assignment_D6D9}, we can find the most general superpotential for $L$ which is a $T'$ triplet with modular weight 1,
\begin{align}
\nonumber\mathcal{K}_{L} &=(-i\tau+i\bar{\tau})^{-1}L^{\dagger}L+ \sum_{k_1,\mathbf{r}_1,\mathbf{r}_2} (-i\tau+i\bar{\tau})^{k_1-1}(L^\dagger Y^{(k_1)\dagger}_{\mathbf{r}_1} Y^{(k_1)}_{\mathbf{r}_2} L )_\mathbf{1}+\mathrm{h.c.}~~\\
\nonumber&=(-i\tau+i\bar{\tau})^{-1}L^{\dagger}L
+\lambda_1 ((L^{\dagger} Y^{(1)\dagger}_{\mathbf{2}})_{\mathbf{2}''} (L Y^{(1)}_\mathbf{2})_\mathbf{2})_{\mathbf{1}} + \lambda_2 ((L^{\dagger} Y^{(1)\dagger}_{\mathbf{2}})_{\mathbf{2}'} (L Y^{(1)}_\mathbf{2})_{\mathbf{2}'})_{\mathbf{1}}\\
\label{eq:Kahler_L-D6}&\quad +\lambda_3 ((L^{\dagger} Y^{(1)\dagger}_{\mathbf{2}})_{\mathbf{2}} (L Y^{(1)}_\mathbf{2})_{\mathbf{2}''})_{\mathbf{1}} + \dots\,,
\end{align}
where the couplings $\lambda_{1,2,3}$ are real. After field rescaling $L\rightarrow (-i\tau+i\bar{\tau})^{1/2}L$, the K\"ahler metric matrix is
\begin{align}
\nonumber
&\mathscr{K}_L=\begin{pmatrix}
 1 & 0 & 0\\
 0 & 1 & 0\\
 0 & 0 & 1
\end{pmatrix}
+\lambda_1(-i\tau+i\bar{\tau})\begin{pmatrix}
Y^{\dagger}_1 Y_1+Y^{\dagger}_2 Y_2 ~& -\sqrt{2}e^{i\frac{7\pi}{12}}Y^{\dagger}_1 Y_2 ~& -\sqrt{2}e^{i\frac{5\pi}{12}}Y^{\dagger}_2 Y_1 \\
-\sqrt{2}e^{-i\frac{7\pi}{12}}Y^{\dagger}_2 Y_1 ~& 2 Y^{\dagger}_2 Y_2 ~& 0 \\
-\sqrt{2}e^{-i\frac{5\pi}{12}}Y^{\dagger}_1 Y_2 ~& 0 ~& 2 Y^{\dagger}_1 Y_1
\end{pmatrix} \\
\nonumber
&~~~~~+\lambda_2 (-i\tau+i\bar{\tau})\begin{pmatrix}
2 Y^{\dagger}_1 Y_1~ & -\sqrt{2}e^{-i\frac{5\pi}{12}}Y^{\dagger}_1 Y_2~ & 0 \\
-\sqrt{2}e^{i\frac{5\pi}{12}}Y^{\dagger}_2 Y_1 ~& Y^{\dagger}_1 Y_1+Y^{\dagger}_2 Y_2 ~&  -\sqrt{2}e^{i\frac{7\pi}{12}}Y^{\dagger}_1 Y_2  \\
0 ~& -\sqrt{2}e^{-i\frac{7\pi}{12}}Y^{\dagger}_2 Y_1 ~& 2 Y^{\dagger}_2 Y_2
\end{pmatrix} \\
\label{eq:kahler_metric_L}&~~~~~+\lambda_3 (-i\tau+i\bar{\tau})\begin{pmatrix}
2 Y^{\dagger}_2 Y_2 ~& 0 ~&  -\sqrt{2}e^{-i\frac{7\pi}{12}}Y^{\dagger}_2 Y_1  \\
0 ~& 2 Y^{\dagger}_1 Y_1 ~& -\sqrt{2}e^{-i\frac{5\pi}{12}}Y^{\dagger}_1 Y_2 \\
-\sqrt{2}e^{i\frac{7\pi}{12}}Y^{\dagger}_1 Y_2 ~& -\sqrt{2}e^{i\frac{5\pi}{12}}Y^{\dagger}_2 Y_1 ~& Y^{\dagger}_1 Y_1+Y^{\dagger}_2 Y_2
\end{pmatrix}+\ldots\,.
\end{align}
The right-handed neutrinos $N^c$ also transform as a triplet $N^c\sim\mathbf{3}$ under $T'$ and carry modular weight 1, hence the K\"ahler potential and K\"ahler metric of $N^c$ are of similar form as Eq.~\eqref{eq:Kahler_L-D6} and Eq.~\eqref{eq:kahler_metric_L} respectively. The right-handed charged lepton are $T'$ singlets $\mathbf{1}'$, $\mathbf{1}'$ and $\mathbf{1}$. The associated most general K\"ahler potential reads
\begin{align}
\nonumber\mathcal{K}_{E^c} &=(-i\tau+i\bar{\tau})^{-1}e^{c\dagger}e^c+(-i\tau+i\bar{\tau})^{-3}\mu^{c\dagger}\mu^c+(-i\tau+i\bar{\tau})^{-1}\tau^{c\dagger}\tau^c\\
\nonumber&\quad+\zeta_1(-i\tau+i\bar{\tau})e^{c\dagger}\tau^c(Y^{(2)\dagger}_{\mathbf{3}}Y^{(2)}_{\mathbf{3}})_{\mathbf{1}'}+\zeta^{*}_1(-i\tau+i\bar{\tau})\tau^{c\dagger}e^{c}(Y^{(2)\dagger}_{\mathbf{3}}Y^{(2)}_{\mathbf{3}})_{\mathbf{1}''}\\
\nonumber&\quad+\zeta_2e^{c\dagger}\mu^c(Y^{(1)\dagger}_{\mathbf{2}}Y^{(3)}_{\mathbf{2}})_{\mathbf{1}}+\zeta^{*}_2\mu^{c\dagger}e^{c}(Y^{(3)\dagger}_{\mathbf{2}}Y^{(1)}_{\mathbf{2}})_{\mathbf{1}}\\
&\quad+\zeta_3\mu^{c\dagger}\tau^c(Y^{(3)\dagger}_{\mathbf{2}''}Y^{(1)}_{\mathbf{2}})_{\mathbf{1}'}+\zeta^{*}_3\tau^{c\dagger}\mu^{c}(Y^{(1)\dagger}_{\mathbf{2}}Y^{(3)}_{\mathbf{2}''})_{\mathbf{1}''}+\ldots~\,.
\end{align}
We plot the dependence of the lepton mixing angles, mass ratio $m_{r}\equiv \sqrt{\Delta m_{21}^{2}/\Delta m_{31}^{2}}$ and $\delta^l_{CP}$ on the parameter $\lambda_3$ in figure~\ref{fig:mixingpar_kahler_D6}, where the values of other inputting parameters in the model $\mathcal{D}_{6}$ are adapted from table~\ref{Tab:predictions_quark_modification}. We see that the corrections from the K\"ahler potential can be sizable, this confirm the insight of~\cite{Chen:2019ewa}. This drawback of bottom-up construction with modular symmetry can be overcame by combining the traditional and modular flavor symmetries~\cite{Nilles:2020kgo}. In this scheme, both superpotential and K\"ahler potentials are strongly constrained, the off-diagonal contributions to the K\"ahler metric are forbidden by all symmetries of the theory. Consequently the additional terms depending on modular forms in the K\"ahler potential don't significantly change the phenomenological predictions obtained by assuming a minimal K\"ahler potential.

\end{appendix}


\providecommand{\href}[2]{#2}\begingroup\raggedright\endgroup

\end{document}